

\magnification=\magstep1


\vsize=24.0 truecm
\hsize=15.0 truecm


\vsize=22.5 truecm
\hsize=16.2 truecm


\tolerance=5000
\vskip 0.5truecm

\rightline{ DIAS-STP-91-29 \break}
\rightline{ UdeM-LPN-TH-71/91 \break}
\vskip 1.0truecm
\baselineskip=15pt

\centerline{\bf ON THE GENERAL STRUCTURE OF HAMILTONIAN REDUCTIONS }
\vskip 0.4truecm
\centerline{\bf OF THE WZNW THEORY}

\vskip 1.0truecm
\centerline{
L. Feh\'er\footnote*{Present address: Laboratoire de Physique Nucl\'eaire,
Universit\'e de Montr\'eal, Montr\'eal, Canada H3C 3J7.
e-mail: feher@lps.umontreal.ca.bitnet
},
L. O'Raifeartaigh, P. Ruelle and I. Tsutsui}
\medskip
\centerline{\it Dublin Institute for Advanced Studies}
\centerline{\it 10 Burlington Road, Dublin 4, Ireland}
\vskip 0.8truecm
\centerline{A. Wipf}
\medskip
\centerline{\it Institut f\"ur Theoretische Physik}
\centerline{\it Eidgen\"ossische Technische Hochschule}
\centerline{\it H\"onggerberg, Z\"urich CH-8093, Switzerland}

\vskip 1.0truecm
\centerline{\bf Abstract}
\vskip 0.4truecm

The structure  of Hamiltonian reductions of the
Wess-Zumino-Novikov-Witten (WZNW) theory
 by first class Kac-Moody constraints is analyzed in detail.
Lie algebraic conditions are given for  ensuring
the presence of exact integrability, conformal invariance and
$\cal W$-symmetry in the reduced theories.
A Lagrangean, gauged WZNW implementation of the reduction
is established in the general case and  thereby
the path integral as well as the BRST formalism  are
 set up for studying the quantum version of the reduction.
The general results are applied to a number of examples.
In particular, a ${\cal W}$-algebra is associated to each
embedding of $sl(2)$ into the simple Lie
algebras by using purely first class constraints.
The importance of these $sl(2)$ systems is demonstrated
by showing that they underlie the $W_n^l$-algebras as well.
New generalized Toda theories are found whose chiral algebras are
the  ${\cal W}$-algebras belonging to the half-integral $sl(2)$ embeddings,
and the ${\cal W}$-symmetry of the effective action of those generalized
Toda theories associated with the integral gradings is exhibited explicitly.

\vfill\eject

\baselineskip=16.5pt

\def\d{\dotfill}
\def\i{\item{}}

\centerline{\bf Contents}
\vskip 7mm

\item{1.}
Introduction
\dotfill 3

\vskip 3mm
\item{2.}
General structure of KM and WZNW reductions
\dotfill 11
\vskip 1mm
\item{}
2.1. First class and conformally invariant KM constraints
\dotfill 11
\item{}
2.2. Lagrangean realization of the Hamiltonian reduction
\dotfill 15
\item{}
2.3. Effective field theories from left-right dual reductions
\dotfill 18

\vskip 3mm
\item{3.}
Polynomiality in KM reductions and the ${\cal W}_{\cal S}^{\cal G}$-algebras
\dotfill 26
\vskip 1mm
\item{}
3.1. A sufficient condition for polynomiality
\dotfill 26
\item{}
3.2. The polynomiality of the Dirac bracket
\dotfill 31
\item{}
3.3. First class constraints for the  ${\cal W}^{\cal G}_{\cal S}$-algebras
\dotfill 34
\item{}
3.4. The ${\cal W}_{\cal S}^{\cal G}$ interpretation of the $W_n^l$-algebras
\dotfill 41

\vskip 3mm
\item{4.}
Generalized Toda theories
\dotfill 46
\vskip 1mm
\item{}
4.1. Generalized Toda theories associated with integral gradings
\dotfill 46
\item{}
4.2. Generalized Toda theories for half-integral $sl(2)$ embeddings
\dotfill 48
\item{}
4.3. Two examples of generalized Toda theories
\dotfill 53

\vskip 3mm
\item{5.}
Quantum framework for WZNW reductions
\dotfill 57
\vskip 1mm
\item{}
5.1. Path-integral for constrained WZNW theory
\dotfill 57
\item{}
5.2. Effective theory in the physical gauge
\dotfill 60
\item{}
5.3. The ${\cal W}$-symmetry of the generalized Toda action $I^H_{\rm eff}(b)$
\dotfill 62
\item{}
5.4. BRST formalism for WZNW reductions
\dotfill 64
\item{}
5.5. The Virasoro centre in two examples
\dotfill 67

\vskip 3mm
\item{6.}
Discussion
\dotfill 70

\vskip 3mm
\noindent
\hskip 2mm
Appendices
\item{}
A: A solvable but not nilpotent gauge algebra
\dotfill 73
\item{}
B: $H$-compatible $sl(2)$ and the non-degeneracy condition
\dotfill 77
\item{}
C: $H$-compatible $sl(2)$ embeddings and halvings
\dotfill 80

\vskip 3mm
\noindent
\hskip 2mm
References
\dotfill 87

\vfill\eject

\baselineskip=15pt
\parskip=5pt

\centerline{\bf 1. Introduction }

\vskip 0.8truecm

Due to their intimate relationship with Lie algebras,
the various one- and two-dimensional Toda systems are among the most
important models of the theory of integrable non-linear equations
[1-19].
In particular, the standard conformal Toda field theories, which are
given by the Lagrangean
$$
{\cal L}_{\rm Toda}(\varphi )={\kappa \over 2}
\biggl( \sum_{i,j=1}^l{1\over {2\vert \alpha_i\vert^2}}K_{ij}
\partial_\mu\varphi^i\partial^\mu \varphi^j
-\sum_{i=1}^l m_i^2\exp \Bigl\{ {1\over 2}\sum_{j=1}^lK_{ij}\varphi^j
\Bigl\} \biggr) \ ,
\eqno(1.1)$$
where
 $\kappa$ is a coupling constant,
 $K_{ij}$ is the Cartan matrix and
the $\alpha_i$ are the simple roots of a simple Lie algebra
 of rank
$l$,
have been the subject of many studies [1,3,4,8-13,19].
It has been
first shown by Leznov and Saveliev [1,3]
that the Euler-Lagrange equations of (1.1)
can be written as a zero curvature condition,
are exactly integrable, and possess interesting non-linear symmetry
algebras [3,4,10,11,13,19].
These symmetry algebras are
generated by chiral
conserved currents, and are
 polynomial extensions of the
chiral Virasoro algebras generated by the traceless energy-momentum
tensor.
The chiral currents in question are conformal primary fields,
whose
conformal weights are given by the orders of the independent Casimirs of
the corresponding simple Lie algebra.
Polynomial extensions of the Virasoro algebra by chiral primary
fields are generally known as
$\cal W$-algebras [20], which are expected to play an important
role
in the classification of conformal field
theories and are in the focus of current investigations [20-29].
The importance of Toda systems
in two-dimensional conformal field theory is in fact
greatly enhanced by their
realizing the ${\cal W}$-algebra symmetries.

It has been discovered recently that the conformal Toda field theories
can be naturally viewed as Hamiltonian reductions of the
Wess-Zumino-Novikov-Witten (WZNW) theory [12,13].
The main feature of the WZNW theory is its affine Kac-Moody (KM)
symmetry, which underlies its integrability [30,31].
The WZNW theory provides the most \lq economical'
realization of the KM symmetry in the sense that its
 phase space is essentially
a direct product of the ${\rm left}\times {\rm right}$ KM phase
spaces.
The WZNW $\rightarrow$ Toda
Hamiltonian reduction is achieved by imposing certain first class,
conformally invariant constraints on the KM currents,
which reduce the chiral KM
phase spaces to phase spaces carrying the
chiral ${\cal W}$-algebras as their Poisson bracket structure
[12,13].
Thus
the ${\cal W}$-algebra is related to the phase space of the Toda
theory in the same way as the KM algebra is related to the
phase space of the WZNW
theory.
In the above manner,
the $\cal W$-symmetry of the Toda theories becomes manifest
by describing these theories as reduced  WZNW
theories.
This way of looking at
Toda theories has also  numerous other advantages,
described in detail in [13].

The constrained WZNW (KM) setting of
the standard Toda theories ($\cal W$-algebras) allows for
generalizations, some of which have already been investigated
[14-18,26-29].
An important recent
development is the realization that
it is possible to associate a generalized ${\cal W}$-algebra
to every embedding of the Lie algebra $sl(2)$ into the
simple Lie algebras [16-18].
The standard ${\cal W}$-algebra, occurring in Toda theory,
corresponds to the so called principal $sl(2)$.
In fact, these generalized ${\cal W}$-algebras
can be obtained from the KM algebra by constraining the current
to the highest weight gauge, which has been originally
introduced in [13] for describing the standard case.
Another interesting development is the $W_n^l$-algebras introduced by
Bershadsky [26] and further studied in [28].  It is known
that the simplest non-trivial case $W_3^2$, which was originally
proposed by Polyakov [27], falls into a special case of the ${\cal
W}$-algebras obtained by the $sl(2)$ embeddings mentioned above.
It has not been clear, however, as to whether the two classes
of ${\cal W}$-algebras are related in general,
or to what extent one
can further generalize the KM reduction to achieve new ${\cal
W}$-algebras.

In the present paper, we undertake the first systematic study of
the Hamiltonian reductions of the WZNW theory, aiming at
uncovering the general structure of the reduction and,
at the same time, try to
answer the above question.
Various different questions arising from this main problem are
also addressed (see Contents),
and some of them can be examined on its own right.
As this provides our motivation and in fact
most of the later developments originate from it,
we wish to recall here the main points of
the WZNW $\rightarrow$ Toda reduction
before giving a more detailed outline
of the content.

To make contact with the
Toda theories, we consider the WZNW theory\footnote*{
The KM level $k$ is $-4\pi\kappa$.
The space-time conventions are:
$\eta_{00}=-\eta_{11}=1$ and $x^{\pm}={1\over 2}(x^0\pm x^1)$.
The WZNW field $g$ is periodic in $x^1$ with period $2\pi r$.}
$$
S_{\rm WZ}(g)={\kappa \over 2}\int\,d^2x\,{\eta}^{\mu\nu}\,{\rm Tr}\,
(g^{-1}\partial_\mu g)(g^{-1}\partial_\nu g)
-{\kappa \over 3}\int_{B_3}\,{\rm Tr}\, (g^{-1}dg)^3 \ ,
\eqno(1.2)$$
for a simple,
maximally non-compact, connected real Lie group $G$.
In other words, we assume that the simple Lie algebra,
$\cal G$, corresponding to $G$ allows for a Cartan decomposition
over the field
of real numbers.
The field equation of the
WZNW theory
 can be written in the equivalent
forms
$$
\partial_-J=0
\qquad
{\rm or}
\qquad
\partial_+\tilde J=0 \ ,
\eqno(1.3)$$
where
$$
J=\kappa\partial_+g\cdot g^{-1}\ ,
\quad {\rm and}\quad
\tilde J=-\kappa g^{-1}\partial_-g \ .
\eqno(1.4)$$
These equations express the conservation of the left- and right KM
currents, $J$ and $\tilde J$, respectively.
The general solution of the WZNW field equation is
given by the simple formula
$$
g(x^+,x^-)=g_L(x^+)\cdot g_R(x^-)\ ,
\eqno(1.5)$$
where $g_L$ and $g_R$ are arbitrary $G$-valued functions, i.e.,
constrained only by the boundary condition imposed on $g$.

Let now $M_-$, $M_0$ and $M_+$ be the standard generators of the
principal $sl(2)$ subalgebra of ${\cal G}$ [32].
By considering the eigenspaces ${\cal G}_m$ of $M_0$ in the adjoint
of ${\cal G}$, ${\rm ad}_{M_0} = [M_0,\,\,]$,
one can define a grading of ${\cal G}$ by the
eigenvalues $m$.
Under the principal $sl(2)$
this grading is an integral grading, in fact the spins
occurring in the decomposition of the adjoint of ${\cal G}$
are the exponents of ${\cal G}$, which are related to
the orders of the independent Casimirs by a shift by $1$.
It is also worth noting that the grade $0$ part of
$$
{\cal G}={\cal G}_+ + {\cal G}_0 + {\cal G}_-\ ,
\qquad\qquad
{\cal G}_\pm =\sum_{m=1}^N{\cal G}_{\pm m} \ ,
\eqno(1.6)$$
is a Cartan subalgebra, and (by using some automorphism
of the Lie algebra)
 one can assume that the
generator $M_0$ is given by the formula
$M_0={1\over 2}\sum_{\alpha >0} H_{\alpha}$, where $H_{\alpha}$
is the standard Cartan generator corresponding to the positive root
$\alpha$,
and the generators $M_\pm$ are certain linear combinations of
the step operators $E_{\pm {\alpha_i}}$ corresponding
 to the simple roots $ \alpha_i$,
$i=1,\ldots ,{\rm rank}\,{\cal G}$.

The basic observation of [12,13] has been
that the standard Toda theory can be  obtained from the WZNW theory
by imposing first class constraints which restrict the
currents to take the following form:
$$
J(x) = \kappa M_- + j(x),
\qquad {\rm with}\qquad j(x) \in ({\cal G}_0+{\cal G}_+)\,,
\eqno(1.7{\rm a})$$
and
$$
\tilde J(x) =-\kappa M_+ +\tilde j(x),
\qquad {\rm with}\qquad \tilde j(x) \in ({\cal G}_0+{\cal G}_-)\,.
\eqno(1.7{\rm b})$$
(For clarity, we note that
one should in principle include some dimensional
constants in $M_\pm$ which are dimensionless,
but such constants are always put to unity
in this paper, for simplicity.)
To derive the Toda theory (1.1) from the WZNW
theory (1.2),
one uses the generalized Gauss decomposition
$g=g_+\cdot g_0\cdot g_-$ of the WZNW field $g$, where $g_{0,\pm }$
are from the subgroups $G_{0,\pm }$ of $G$
corresponding to the  Lie subalgebras
${\cal G}_{0,\pm}$, respectively.
In this framework the Toda fields $\varphi_i$  are given
by the middle-piece of the Gauss
decomposition,
$g_0=\exp [{1\over 2}\sum_{i=1}^l\varphi_i H_i]$,
which is invariant under the triangular KM gauge transformations
belonging to the first class constraints (1.7).
Note that here the
elements $H_i\in {\cal G}_0$ are the standard Cartan generators
associated to the simple roots.
In fact,
the Toda field equation can be derived directly
from the WZNW field equation by inserting the
Gauss decomposition of $g$ into (1.3)
and using the constraints (1.7).
The effective action of the reduced theory, (1.1), can also
be obtained
in a natural way, by using the
Lagrangean, gauged WZNW implementation of the Hamiltonian
reduction [13].

In their pioneering
work [1,3], Leznov and Saveliev proved
the exact integrability of the conformal Toda systems by
exhibiting chiral quantities by using
the field equation and the special
graded structure of the
Lax potential ${\cal A}_{\pm}$, in terms of which the Toda equation
takes the zero curvature form
$$
[\partial_+-{\cal A}_+\,,\, \partial_--{\cal A}_-]=0\ .
\eqno(1.8)$$
In our framework the exact integrability of Toda systems
is seen as an immediate consequence of the
obvious
integrability of the WZNW theory,
which survives
the reduction to Toda theory.
In other words, the chiral fields
underlying the integrability of the Toda equation are available from
the
very beginning, that is,
they come
from the fields entering
the ${\rm left}\times {\rm right}$
decomposition of the general WZNW solution (1.5).
Furthermore,
the Toda Lax potential itself emerges naturally
from the trivial,
chiral Lax potential of the WZNW theory.
To see this one first observes
that the WZNW field equation is a zero curvature condition,
since one can write for example the first equation
in (1.3) as
$$
[\partial_+ -J\, ,\, \partial_- - 0]=0\ .
\eqno(1.9)$$
Using the constraints of the reduction,
the Toda zero curvature condition
(1.8) of [1,3]  arises from (1.9) by conjugating
this equation by
$g_+^{-1}(x^+,x^-)$, namely
by the inverse of the upper triangular piece of the
generalized Gauss decomposition of the WZNW field $g$ [18].

The  ${\cal W}$-symmetry of the Toda theory appears
in the WZNW setting in a very direct and natural way.
Namely, one can interpret the ${\cal W}$-algebra as the KM Poisson
bracket algebra of the gauge invariant differential polynomials of
the constrained currents in (1.7).
Concentrating on the left  sector,
the  gauge transformations
act on the current according to
$$
J(x)\rightarrow e^{a(x^+)}\,J(x)\,e^{-a(x^+)}
+\kappa (e^{a(x^+)})^{\prime}\,e^{-a(x^+)},
\eqno(1.10)$$
where $a(x^+)\in {\cal G}_+$ is an arbitrary
chiral parameter
function.\footnote*{Throughout the paper, the notation
$f^\prime = 2\partial_1 f$ is used for every function $f$, including the
spatial $\delta$-functions. For a chiral function $f(x^+)$ one
has then $f^\prime =\partial_+ f$.}
The constraints (1.7) are chosen in such a way that the following
Virasoro generator
$$
L_{M_0}(x)\equiv L_{\rm KM}(x)-{\rm Tr\,}(M_0 J^{\prime}(x) ) ,
\quad{\rm where}\quad
L_{\rm KM}(x)={1\over {2\kappa}}{\rm Tr}(J^2(x)) ,
\eqno(1.11)$$
is gauge invariant,
which ensures the conformal invariance of the reduced theory.

One obtains an equivalent interpretation
of the ${\cal W}$-algebra by identifying it
with the Dirac bracket algebra of the
differential polynomials of the current components in certain gauges,
which are such that a basis of the gauge invariant
differential
polynomials reduces to the independent current components after the
gauge fixing.
We call the gauges in question Drinfeld-Sokolov (DS)
gauges [13], since such gauges has been used also in [5].
They have the nice property that any constrained current $J(x)$
can be brought to the gauge fixed form by a unique gauge
transformation depending  on $J(x)$ in a differential polynomial
way.
The most important DS gauge
is the highest weight gauge [13], which is defined by requiring the
gauge fixed current to be of the following form:
$$
J_{\rm red}(x)=\kappa M_-+j_{\rm red}(x)\ ,
\qquad
j_{\rm red}(x)\in {\rm Ker}({\rm ad}_{M_+})\ ,
\eqno(1.12)$$
where ${\rm Ker(ad}_{M_+})$ is the kernel of the adjoint of $M_+$.
In other words, $j_{\rm red}(x)$ is restricted to be an
arbitrary linear combination of the highest weight vectors of
the $sl(2)$ subalgebra in the adjoint of $\cal G$.
The special  property of the highest weight gauge
is that in this gauge the conformal properties become
manifest.
Of course,
the quantity  $L_{\rm red}(x)$ obtained by
restricting $L_{M_0}(x)$ in (1.11) to the highest weight gauge
generates a Virasoro algebra under Dirac bracket.
(Note that in our case $L_{\rm red}(x)$  is proportional to
the $M_+$-component of $j_{\rm red}(x)$.)
The important point is that, with the
exception of the $M_+$-component, the spin $s$ component of
$j_{\rm red}(x)$ is in fact
a primary field
of conformal weight $(s+1)$ with respect to $L_{\rm red}(x)$
under the Dirac bracket.
Thus
{\it the highest weight gauge automatically yields a primary field
basis of the ${\cal W}$-algebra},
from which one sees that the spectrum of conformal weights is fixed
by the $sl(2)$ content of the adjoint of ${\cal G}$ [13].

In the above we arrived at the description of the ${\cal W}$-algebra
as a Dirac bracket algebra by gauge fixing the first class system
of constraints corresponding to (1.7).
However, it is clear now that it
would have been possible  to define the ${\cal W}$-algebra
as the Dirac bracket algebra of the components of
$j_{\rm red}$ in (1.12)
in the first place.
Once this point is realized,
a natural generalization arises immediately [16-18].
Namely, one can associate a classical ${\cal W}$-algebra
to any $sl(2)$ subalgebra  ${\cal S}=\{M_-,\ M_0,\ M_+\}$
of any simple Lie algebra ${\cal G}$, by defining it to be
the Dirac bracket algebra of the components of $j_{\rm red}$
in (1.12),
where one simply substitutes the generators $M_\pm$ of
the {\it arbitrary} $sl(2)$ subalgebra ${\cal S}$ for
those of the {\it principal} $sl(2)$.
As we shall see in this paper, this Dirac bracket algebra
is a polynomial extension of the Virasoro algebra by primary
fields, whose conformal weights are related to the
spins occurring in the decomposition of the adjoint
of ${\cal G}$ under ${\cal S}$ by a shift by $1$,
in  complete analogy with the case of the principal $sl(2)$.
We shall designate the generalized
${\cal W}$-algebra associated to the $sl(2)$ embedding
${\cal S}$ as ${\cal W}_{\cal S}^{\cal G}$.

With the main features of the WZNW $\rightarrow$ Toda
reduction and the above definition of the
${\cal W}_{\cal S}^{\cal G}$-algebras at our disposal, now
we sketch the philosophy and the outline
of the present paper.
We start by giving the most important assumption
underlying our investigations, which is that
we consider those reductions which can be
obtained by imposing {\it first class} KM constraints
generalizing
the ones in (1.7).
To be more precise,
our most general
constraints restrict the current to take the
following form:
$$
J(x) =\kappa M + j(x),
\qquad {\rm with} \qquad
j(x) \in \Gamma^\perp\ ,
\eqno(1.13)$$
where $M$ is {\it some} constant element of the underlying simple
Lie algebra ${\cal G}$,
and $\Gamma^\perp$ is the subspace consisting of the
Lie algebra elements trace orthogonal to {\it some}
subspace $\Gamma$ of ${\cal G}$.
We note that earlier in (1.7a)
we have chosen $\Gamma ={\cal G}_+$ and $M = M_-$,
but we do not need any $sl(2)$ structure here.
The whole analysis is based on requiring
the first-classness of the system of linear KM
constraints corresponding the pair
$(\Gamma , M)$ according to (1.13).
However, this first-classness assumption
is not as restrictive as one perhaps
might think at first sight.
In fact, as far as we know,
our first class method is capable of
covering all Hamiltonian reductions of the WZNW theory
considered to date.
The many technical advantages of
using
purely first class KM constraints will
be apparent.

The investigations in this paper are organized
according to three distinct levels of generality.
At the most general level we only make the
first-classness assumption
and deduce the following results.
First, we give a complete Lie algebraic analysis
of the conditions on the pair $(\Gamma , M)$ imposed
by the first-classness of the constraints.
We shall see that
$\Gamma$ in (1.13)
has to be a subalgebra of ${\cal G}$ on which the
Cartan-Killing form vanishes, and that every
such subalgebra is solvable.
The Lie subalgebra $\Gamma$  will be referred to as
the \lq gauge algebra' of
the reduction.
For a given ${\Gamma}$, the first-classness
imposes a further condition on the element $M$,
and we shall describe  the space of the allowed $M$'s.
Second, we establish a gauged WZNW implementation of the
reduction, generalizing the one found previously in the
standard case [13].
This gauged WZNW setting of the reduction will be
first seen classically, but
it will be also established  in the quantum
theory by considering the phase space path integral of
the constrained WZNW theory.
Third, the gauged WZNW framework will be used to
set up the BRST formalism for the
quantum Hamiltonian reduction in the general case.
Fourth, by making the additional assumption that
the left and right gauge algebras
are dual to each
other with respect to the Cartan-Killing form,
we will be able to give a detailed local analysis of
the effective theories resulting from the reduction.
This duality assumption will also
be related to the parity invariance of the effective
theories, which is satisfied in the standard Toda
case where the left and right gauge algebras are ${\cal G}_+$
and ${\cal G}_-$ in (1.6), respectively.
In general, the WZNW reduction not only allows
us to make contact with known theories, like the Toda theory
in (1.1), where the simplicity
and the large symmetry of the \lq parent' WZNW theory
are fully exploited for analyzing them,
but also leads to new theories
which are \lq integrable by construction'.

At the next level of generality, we
study the conformally invariant reductions.
The basic idea here is that one can
guarantee the conformal invariance of the
reduced theory by exhibiting a Virasoro
density
such that the corresponding conformal action
preserves the constraints in (1.13).
Generalizing (1.11),
we assume that this Virasoro
density is of the form
$$
L_H (x) = L_{\rm KM}(x) - {\rm Tr\,}(H J^\prime (x))\ ,
\eqno(1.14)$$
where $H$ is {\it some} Lie algebra element,
to be determined from the
condition that $L_H$ weakly
commutes with the first class constraints.
We shall describe  the relations
which are imposed on the triple of quantities
$(\Gamma , M , H)$ by this requirement, and thereby
obtain a Lie algebraic sufficient condition
for conformal invariance.

At the third level of generality, we deal
with polynomial reductions and ${\cal W}$-algebras.
The above mentioned
sufficient condition for
conformal invariance
is a guarantee for $L_H$ being a gauge invariant differential
polynomial.
We shall provide an additional condition on
the triple of quantities $(\Gamma , M, H)$
which allows one to  construct out of the current
in (1.13) a
complete set of gauge invariant differential polynomials
by means of a polynomial gauge fixing algorithm.
The KM Poisson bracket algebra of the gauge invariant
differential polynomials yields  a polynomial
extension of the Virasoro algebra generated by $L_H$.
The most important application of our sufficient
condition  for polynomiality concerns the
${\cal W}_{\cal S}^{\cal G}$-algebras
mentioned previously.

Let us remember that,
for an arbitrary $sl(2)$ subalgebra
${\cal S}$ of ${\cal G}$,
the ${\cal W}_{\cal S}^{\cal G}$-algebra
can be defined as the
Dirac bracket algebra of the highest weight
current in (1.12)
realized by purely second
class constraints.
However, we shall see in this paper that these second class
constraints can be replaced by purely first class constraints
even in the case of arbitrary,
{\it integral or half-integral}, $sl(2)$ embeddings.
Since the first class constraints satisfy our
sufficient condition for polynomiality,
we can realize the
${\cal W}_{\cal S}^{\cal G}$-algebra as the KM
Poisson bracket
algebra of the corresponding gauge invariant
differential polynomials.
After having our hands on first class
KM constraints leading
to the ${\cal W}_{\cal S}^{\cal G}$-algebras,
we shall immediately apply our general construction
to exhibiting reduced WZNW theories realizing
these ${\cal W}$-algebras as their chiral algebras
for arbitrary $sl(2)$-embeddings.
In the non-trivial case
of half-integral $sl(2)$-embeddings,
these generalized Toda theories
represent
a new class of integrable models,
which will be studied in some detail.
It is also  worth noting that realizing
the ${\cal W}_{\cal S}^{\cal G}$-algebra  as
a KM Poisson bracket
algebra of  gauge invariant
differential polynomials
should in principle allow
for quantizing it
through the KM representation theory, for
example by using the general BRST formalism which will be
set up in this paper.
As a first step, we shall give a concise
formula for the Virasoro centre of this algebra in terms
of the level of the underlying KM algebra.

The existence of purely first class KM  constraints
leading to the ${\cal W}_{\cal S}^{\cal G}$-algebra
might be perhaps surprizing
to the reader, since earlier
in [16] it was claimed to be inevitably
necessary to use
at least some second class constraints
from the very beginning, when
reducing the KM algebra to
${\cal W}_{\cal S}^{\cal G}$ in the
case of a half-integral $sl(2)$ embedding.
Contrary to their claim,
we will demonstrate that it is possible and in fact easy
to obtain
the appropriate first class constraints
which lead to ${\cal W}_{\cal S}^{\cal G}$.
Roughly speaking, this will be
achieved by discarding
\lq half' of those constraints which
form the second class part in the mixed system of the
constraints imposed in [16].
The mixed system of
constraints can be recovered by a partial
gauge fixing of our purely first class KM
constraints.
Similarly, Bershadsky's
constraints [26], used to define the $W_n^l$-algebra,
are also a mixed system in the above sense,
i.e., it contains both first and second class parts.
We can also replace these
constraints by purely first class ones
without changing the
final reduced phase space.
In this procedure we shall
uncover the hidden $sl(2)$ structure of the $W_n^l$-algebras,
namely, we shall identify them in general as
further reductions of particular ${\cal W}_{\cal S}^{\cal G}$-algebras.

The study of WZNW reductions embraces
various subjects, such as integrable models, ${\cal W}$-algebras and
their field theoretic realizations.  We hope that the readers with
different interests will find relevant results throughout this
paper, and find an interplay of general considerations and
investigations of numerous examples.

\vfill\eject

\centerline{\bf 2. General structure of KM and WZNW reductions}

\vskip 0.8truecm
The purpose of this chapter is to investigate the general structure of those
reductions of the KM phase space and
corresponding reductions of the
full WZNW theory which can be defined
by imposing {\it first class constraints setting certain current
components to constant values}.
In the rest of the paper, we assume that the WZNW group, $G$, is a
connected real Lie group whose Lie algebra, ${\cal G}$, is a
non-compact real form of a complex simple Lie algebra, ${\cal G}_c$.
We shall first uncover the  Lie algebraic implications  of the
constraints being first class, and also discuss  a sufficient
 condition which may be used to ensure their conformal invariance.
In particular, we shall see why the compact real form is outside
our framework.
We then set up a gauged WZNW theory which
provides  a Lagrangean realization of the
WZNW reduction, for the case of general first class constraints.
Finally, we shall describe the effective field theories
 resulting from the reduction in some detail in an important special
case, namely when the left and right KM currents
are constrained for such
subalgebras of ${\cal G}$ which are dual to each other with respect to the
Cartan-Killing form.

\vskip 1.2truecm

\noindent
{\bf 2.1. First class and  conformally invariant KM constraints}

\vskip 0.8truecm

Here we analyze the general form of the KM
constraints which will be used subsequently to reduce the WZNW
theory.
The analysis applies to
each current $J$ and $\tilde J$ separately so we choose
one of them, $J$ say, for definiteness.
To fix the conventions, we first note that the
KM Poisson bracket reads
$$
\{
\langle u,J(x)\rangle \,,\,\langle v ,J(y)\rangle \}_{\vert x^0=y^0}
=\langle [u,v],J(x)\rangle\delta (x^1-y^1)
+\kappa\langle u,v\rangle \delta^{\prime}(x^1-y^1),
\eqno(2.1)$$
where $u$ and $v$ are arbitrary generators of ${\cal G}$
and the inner product
$\langle u\,,\,v\rangle ={\rm Tr\,}(u\cdot v)$ is normalized
so that the long roots of ${\cal G}_c$ have
length squared $2$.
This normalization means that in terms of the adjoint representation one has
$\langle u\,,\,v\rangle ={1\over 2g}{\rm tr\,}({\rm ad}_u\cdot
{\rm ad}_v)$, where $g$ is the dual Coxeter number.
It is worth noting that $\langle u\,,\,v\rangle$ is the usual
matrix trace in the defining, vector representation for the classical Lie
algebras $A_l$ and $C_l$, and it is ${1\over 2}\times {\rm trace}$ in
the defining representation for the $B_l$ and $D_l$ series.
We also wish to point out that the KM Poisson bracket
together with all the subsequent
relations which follow from it hold {\it in the same form} both on the usual
canonical phase space and on the space of the classical solutions
of the theory.
This is the advantage of using equal time Poisson brackets and
spatial $\delta$-functions even on the latter space, where $J(x)$
depends on $x=(x^0,x^1)$ only through $x^+$
(see the footnote on page 7).

The KM reduction we consider
is defined by requiring the constrained current to be of
the following special form:
$$
J(x)=\kappa M+j(x)\ ,
\qquad {\rm with}\qquad j(x)\in \Gamma^\perp\ ,
\eqno(2.2)$$
where $\Gamma$ is some linear subspace and
$M$ is some element of ${\cal G}$.
Equivalently, the constraints can be given as
$$
\phi_{\gamma}(x) =
\langle \gamma\,,\,J(x)\rangle -\kappa\langle \gamma\,,\, M\rangle =0\ ,
\qquad\qquad \forall\ \gamma\in \Gamma\ .
\eqno(2.3)$$
In words, our constraints set the current components corresponding
to $\Gamma$ to constant values.
It is clear both from (2.2) and (2.3) that
$M$ can be shifted by an arbitrary   element from
the space $\Gamma^\perp$ without changing the actual content of
the constraints.
This  ambiguity is unessential, since
one can fix $M$, for example, by requiring
that it is from some given linear complement
of $\Gamma^\perp$ in ${\cal G}$,
which can be chosen by convention.

In our method we assume that the above system of
constraints is {\it first class}, and now we analyze the
content of this condition.
Immediately from (2.1), we have\footnote*{For simplicity, we set
$\kappa$ to
$1$ in the rest of the paper,
except in Chapter 5, where $\kappa$ occurs in
 the formula of the
Virasoro centre.}
$$
\{ \phi_\alpha (x) ,
 \phi_\beta (y)\}=
\phi_{[\alpha ,\beta ]} (x) \delta(x^1-y^1) +
\omega_M(\alpha ,\beta )\delta(x^1-y^1)
+\langle \alpha ,\beta\rangle\delta^\prime (x^1-y^1),
\eqno(2.4)$$
where the second term contains the restriction to $\Gamma$ of
the following anti-symmetric 2-form of ${\cal G}$:
$$
\omega_M(u,v) \equiv \langle M\,,\,[u\,,\,v]\rangle\ ,
\qquad\quad \forall u\,,\, v\in {\cal G}\ .
\eqno(2.5)$$
It is evident from (2.4) that the constraints are first class if, and
only if,
we have
$$
[\alpha\,,\,\beta]\in \Gamma , \qquad
\langle\alpha\,,\,\beta \rangle =0
\quad {\rm and}\quad
 \omega_M(\alpha\,,\,\beta)=0,
\quad {\rm for}\quad \forall \alpha\,,\,\beta\in \Gamma .
\eqno(2.6)$$
This means that
the linear subspace $\Gamma$  has to be {\it a
subalgebra on which the Cartan-Killing form and $\omega_M$ vanish}.
It is easy to see that the three conditions in (2.6) can be
equivalently written as
$$
[\Gamma\,,\,\Gamma^{\perp}] \subset \Gamma^{\perp}, \qquad
 \Gamma \subset \Gamma^{\perp}
\qquad {\rm and}\qquad
[M\,,\,\Gamma] \subset \Gamma^{\perp}\ ,
 \eqno(2.7)$$
respectively.
Subalgebras $\Gamma$ satisfying $\Gamma\subset \Gamma^\perp$ exist in
every
real form of the complex simple Lie algebras
except the compact one, since for the compact real form the
Cartan-Killing inner product is (negative) definite.

We note that for a given $\Gamma$
the third condition and the ambiguity in choosing $M$
can be concisely summarized by the (equivalent) statement that
$$
M\in [\Gamma\,,\,\Gamma]^\perp/\Gamma^\perp\ .
\eqno(2.8)$$
The constraints defined by the zero element of this factor-space
are in a sense trivial.
It is clear that, for a subalgebra $\Gamma$ such that
$\Gamma\subset \Gamma^\perp$, the above factor-space
contains non-zero elements if and only if
$[\Gamma ,\Gamma ]\neq \Gamma$.
Actually this is always so because $\Gamma\subset \Gamma^\perp$
implies that $\Gamma$ is a {\it solvable}
 subalgebra
of ${\cal G}$.
To prove this, we first note that if $\Gamma$ is not solvable then,
by Levi's theorem [33],
it contains a semi-simple subalgebra, in which one can find
 either an $so(3,R)$ or an $sl(2,R)$ subalgebra. From this one sees
that there exists at least one generator
$\lambda$
of $\Gamma$ for which the operator
${\rm ad}_\lambda$ is diagonalizable with real
eigenvalues.
It cannot be that all eigenvalues of ${\rm ad}_\lambda$ are $0$ since
${\cal G}$ is a simple Lie algebra,
and from this one gets that
$\langle \lambda\,,\,\lambda\rangle \neq 0$,
which contradicts $\Gamma \subset \Gamma^\perp$.
Therefore one can conclude that $\Gamma$
 is necessarily a solvable subalgebra of ${\cal G}$.

The
second condition in (2.6) can be satisfied for example by
assuming that  every $\gamma \in\Gamma$ is a nilpotent element
of $\cal G$.
This is true in the concrete instances of the reduction
studied in Chapters 3 and 4.
We note that in this case $\Gamma$ is actually a nilpotent
Lie algebra, by Engel's theorem [33].
However, the nilpotency of ${\Gamma}$ is not necessary
for satisfying $\Gamma\subset \Gamma^\perp$.
In fact, a solvable but not nilpotent
 $\Gamma$ can be found in Appendix A.

The current
components constrained in (2.3) are the infinitesimal generators of
the KM transformations corresponding to the subalgebra $\Gamma$,
which act on the KM phase space as
$$
J(x)\longrightarrow
e^{a^i(x^+) \gamma_i}\,J(x)\,e^{-a^i(x^+)\gamma_i}+
( e^{a^i(x^+) \gamma_i})^\prime\,
e^{-a^i(x^+)\gamma_i}\ ,
\eqno(2.9)$$
where the $a^i(x^+)$ are parameter functions and there is a summation
over some basis $\gamma_i$ of $\Gamma$.
Of course, the first class conditions are equivalent to the
statement that the constraint
surface, consisting of currents of the form (2.2),
is left invariant by the above transformations. From the point of
 view of the reduced theory, these transformations
are to be regarded as gauge transformations, which means that
the reduced phase space can be identified as the
space of gauge orbits
 in the constraint surface.
Taking this into account, we shall often refer to $\Gamma$ as the
gauge algebra of the reduction.

We next discuss a sufficient condition for the conformal invariance
of the constraints.
We assume that $M\notin \Gamma^\perp$ from now on.
The standard conformal symmetry  generated by the
Sugawara Virasoro density
$L_{\rm KM}(x)$ is then broken by the constraints (2.3), since they
set some component of the current, which has spin 1, to a non-zero
constant.
The idea is to circumvent
this apparent violation of conformal invariance
by changing the
standard
action of the conformal group
on the KM phase space to one which does leave the constraint surface
invariant.
One can try to generate the new
conformal action by  changing the usual KM
Virasoro density to the new Virasoro density
$$
L_H(x)=L_{\rm KM}(x)-\langle H,J^{\prime}(x) \rangle,
\eqno(2.10)$$
where $H$ is some element of
${\cal G}$.
The
 conformal action
 generated by $L_H(x)$ operates on the KM phase space as
$$\eqalign{
\delta_{f,H}\,
J(x)\equiv & -\int dy^1\, f(y^+)\,\{L_H(y) \,,\,J(x)\}\cr
=&\,f(x^+)J^\prime (x)+
f^\prime (x^+)\bigl( J(x)+[H,J(x)] \bigr)
+f^{\prime\prime}(x^+)H\ ,\cr}
\eqno(2.11)$$
for any parameter function $f(x^+)$,
corresponding to the conformal coordinate transformation
$\delta_{f}\,x^+ =-f(x^+)$.
In particular,
 $j(x)$ in (2.2) transforms
under this new conformal action according to
$$
\delta_{f,H}\, j(x)=f(x^+)j^\prime (x)+f^{\prime\prime}(x^+)H+
f^\prime (x^+)\bigl( j(x)+[H,j(x)]+([H,M]+M)\bigr),
\eqno(2.12)$$
and our condition is that this variation should be in
$\Gamma^\perp$, which means that
this
conformal action
preserves  the constraint surface. From (2.12),
 one sees that this is equivalent
to having
the following relations:
$$
H\in \Gamma^\perp ,\qquad
[H,\Gamma^\perp] \subset \Gamma^\perp
\qquad {\rm and}\qquad
([H,M]+M)\in \Gamma^\perp\ .
\eqno(2.13)$$
In conclusion, the existence of an operator $H$ satisfying these relations
is a {\it sufficient
 condition} for the conformal invariance of the KM reduction
obtained by imposing (2.3).
The conditions in (2.13) are equivalent to
$L_H(x)$ being a gauge invariant quantity,
inducing a corresponding conformal action on the reduced
phase space.
Obviously,
the second relation
in (2.13)
is  equivalent to
$$
[H,\Gamma]\subset\Gamma\ .
\eqno(2.14)$$

An element $H\in {\cal G}$ is called
{\it diagonalizable}
if the linear operator ${\rm ad}_H$
possesses a
complete set of
eigenvectors in ${\cal G}$.
By the eigenspaces of ${\rm ad}_H$, such an element defines
a grading of ${\cal G}$, and below we shall
refer to a diagonalizable element as a {\it grading operator}
of ${\cal G}$.
In the examples we study later, conformal
invariance will be ensured by the existence of a grading operator
subject to (2.13).

If $H$ is a grading operator satisfying (2.13)
then it is always possible to shift  $M$
by some
element of $\Gamma^\perp$ (i.e., without
changing the physics) so that the new $M$ satisfies
$$
[H,M]=-M\ ,
\eqno(2.15)$$
instead of the last condition in (2.13).
It is also clear that if $H$ is a grading operator then
 one can take graded bases in
$\Gamma$ and $\Gamma^\perp$, since these are
invariant subspaces under ${\rm ad}_H$.
On re-inserting (2.15) into (2.12) it then follows
that
all components of $j(x)$ are primary fields with respect
to the conformal action
generated by $L_H(x)$, with the exception of  the $H$-component,
which also survives the constraints
according to the first condition in (2.13).

As an example, let us now consider some arbitrary
grading operator $H$ and denote by ${\cal G}_m$ the eigensubspace
corresponding to the eigenvalue $m$ of ${\rm ad}_H$.
Then the graded subalgebra ${\cal G}_{\geq n}$,
which is defined to be the direct sum
of the subspaces ${\cal G}_m$ for all $m\geq n$,
will qualify as a
gauge algebra $\Gamma$ for any $n> 0$ from the spectrum of
${\rm ad}_H$.
In this case $\Gamma^\perp ={\cal G}_{> -n}$ and
the factor space $[\Gamma ,\Gamma]^\perp /\Gamma^\perp$,
which is the space of the allowed $M$'s,
can be represented as the direct sum of
${\cal G}_{-n}$ and that graded
subspace of ${\cal G}_{<-n}$ which is orthogonal
to $[\Gamma ,\Gamma ]$.
It is easy to see that
one obtains  {\it conformally invariant} first class constraints
by
choosing $M$ to be any {\it graded element} from this factor space.
Indeed, if the grade of $M$ is $-m$ then $L_{H/m}$
yields a Virasoro density
weakly commuting with the
corresponding constraints.

In summary, in this section we have seen that one
can associate a first class system
of KM constraints  to any pair
($\Gamma$,$M$)
subject to (2.6) by requiring the constrained current to take the form
(2.2), and that the conformal invariance of this system of constraints
is guaranteed if one can find
an  operator $H$ such that
the triple
($\Gamma$,$M$,$H$) satisfies the conditions
in (2.13).

\vskip 1.2truecm

\noindent
{\bf 2.2. Lagrangean realization of the Hamiltonian reduction}

\vskip 0.8 truecm
We shall exhibit here
a gauged WZNW theory providing
the Lagrangean realization of those Hamiltonian reductions
of the WZNW theory which can be
defined by imposing first class  constraints of the type (2.3)
on the KM  currents $J$ and $\tilde J$ of the theory.
It should be noted that,
in the rest of this chapter, we do not assume that the constraints
are conformally invariant.

To define the WZNW reduction, we can choose left and right constraints
completely independently.
We shall denote the
pairs consisting of an appropriate subalgebra and a
constant matrix
corresponding  to the left and right constraints
as $(\Gamma ,M)$ and $(\tilde \Gamma ,-\tilde M)$, respectively.
The reduced theory is obtained by first constraining
the WZNW phase space by setting
$$
\phi_i=\langle \gamma_i\,,\,J\rangle -\langle \gamma_i\,,\,M\rangle =0,
\qquad {\rm and}\qquad
\tilde\phi_i =
- \langle \tilde\gamma_i\,,\,\tilde J\rangle
- \langle \tilde\gamma_i\,,\,\tilde M\rangle =0,
\eqno(2.16)
$$
where the $\gamma_i$ and the $\tilde \gamma_i$ form bases
of $\Gamma$ and $\tilde\Gamma$, respectively,
and then factorizing the constraint surface by the canonical
transformations generated by these constraints.
One can apply this  reduction  either to
the usual
canonical phase space or to the space of solutions of the classical field
equation.
These are equivalent procedures since the two spaces in question are
isomorphic.
For later purpose we note that the constraints generate
the following {\it chiral} gauge transformations
on the space of solutions:
$$
g(x^+,x^-)\longrightarrow e^{\gamma (x^+)}\cdot g(x^+,x^-)\cdot
e^{-\tilde \gamma (x^-)}\ ,
\eqno(2.17)$$
where $\gamma (x^+)$ and $\tilde \gamma (x^-)$ are arbitrary
$\Gamma$ and $\tilde \Gamma$ valued functions.

For completeness, we wish to mention here how the above way of reducing
the WZNW theory fits into the general theory
of Hamiltonian (symplectic) reduction of symmetries [34].
In general, the Hamiltonian reduction is obtained by setting the
phase space functions generating the
symmetry transformations through Poisson bracket
 (in other words, the components
of the momentum map) to some constant values.
The reduced phase space results by factorizing this constraint
surface by the subgroup of the symmetry group respecting
the constraints.
The symmetry group we consider
 is the ${\rm left}\times {\rm right}$ KM group
generated by $\Gamma\times\tilde\Gamma$ and our Hamiltonian
reduction  is special in the sense that the full symmetry group
preserves the constraints.
Of course, the latter fact is just a reformulation of the
first-classness of our constraints.

We now come to the main point of the section,
which is that the reduced WZNW theory,
defined in the above by using the Hamiltonian picture,
 can be identified as
the gauge invariant content of a
corresponding  gauged WZNW theory.
This gauged WZNW interpretation of the reduction was
pointed out in the concrete case of the
${\rm WZNW}\rightarrow {\rm standard\ Toda}$ reduction in [13],
and we below generalize that construction
to the present situation.

The gauged WZNW theory we are interested in is
given by the
following action functional:
$$\eqalign{
I(g,A_-,A_+)\equiv S_{\rm WZ}(g)
+&\int d^2 x\, \bigl(\langle A_-,\partial_+g g^{-1}-M\rangle\cr
+&\langle A_+ ,g^{-1}\partial_-g  -\tilde M\rangle
+\langle A_- ,gA_+g^{-1}\rangle \bigr)\,,\cr}
\eqno(2.18)$$
where the
gauge fields
$A_-(x)$ and $A_+(x)$ vary in $\Gamma$ and $\tilde \Gamma$,
respectively.
The main property of this action is that it is invariant under
the following {\it non-chiral} gauge transformations:
$$
g\rightarrow\alpha g{\tilde \alpha}^{-1};\ \
A_-\rightarrow\alpha A_-\alpha ^{-1} +
\alpha\partial_-\,\alpha^{-1};\ \
A_+\rightarrow\tilde \alpha A_+
{\tilde \alpha}^{-1} +
 (\partial_+\tilde \alpha){\tilde \alpha}^{-1}\ ,
\eqno(2.19{\rm a})$$
where
$$
\alpha =e^{\gamma (x^+,x^-)} \qquad{\rm and}\qquad
\tilde\alpha =e^{{\tilde \gamma}(x^+,x^-)}\ ,
\eqno(2.19{\rm b})$$
for any $\gamma (x^+,x^-)\in\Gamma$ and
$\tilde\gamma (x^+,x^-)\in\tilde\Gamma$.
The proof of the invariance of (2.18) under (2.19) can proceed
along the same lines as for the special case in [13].
In the proof one
rewrites $S_{\rm WZ}(\alpha g {\tilde\alpha}^{-1})$
by using the
well-known Polyakov-Wiegmann identity [35],
and in this step one uses the fact that the
WZNW action {\it vanishes} for fields in the subgroups of $G$
with Lie algebras $\Gamma$ or $\tilde \Gamma$.
This is an obvious consequence of the relations
 $\Gamma\subset \Gamma^\perp$ and
$\tilde\Gamma\subset\tilde \Gamma^\perp$.
The other crucial
point is that the terms in (2.18) containing
the constant matrices $M$ and $\tilde M$ are
separately invariant
under (2.19).
It is easy to see that this follows from the third condition in
(2.6).
For example,
 under an infinitesimal gauge
transformation belonging to $\alpha \simeq 1+\gamma$,
the term  $\langle A_- ,M\rangle$ changes by
$$
\delta\,\langle A_- , M\rangle =
- \langle \partial_-\gamma ,M\rangle
+ \omega_M(\gamma, A_-)\ ,
\eqno(2.20)$$
which is a total divergence since the second term vanishes,
as both $A_-$ and $\gamma$ are from $\Gamma$.

The Euler-Lagrange equation derived from (2.18)
by varying $g$ can be written equivalently as
$$
\partial_-(\partial_+g g^{-1}+gA_+g^{-1})+
[A_- , \partial_+g g^{-1}+gA_+g^{-1}]
+\partial_+A_-=0\ ,
\eqno(2.21{\rm a})$$
or
$$
\partial_+(g^{-1}\partial_-g+g^{-1}A_-g)-[A_+,g^{-1}\partial_-g+g^{-1}A_-g]
+ \partial_-A_+=0 \ ,
\eqno(2.21{\rm b})$$
and the field equations obtained by varying $A_-$ and $A_+$ are given
by
$$
\langle \gamma\, ,\,\partial_+g g^{-1}+gA_+g^{-1}-M\rangle =0,
\quad\qquad
\forall\ \gamma\in \Gamma\,,
\eqno(2.21{\rm c})$$
and
$$
\langle \tilde\gamma\, ,\, g^{-1}\partial_-g+
g^{-1}A_-g-\tilde M\rangle=0,
\quad\qquad
\forall\ \tilde\gamma\in\tilde\Gamma\,,
\eqno(2.21{\rm d})$$
respectively.
We now note that
by making use of the gauge invariance, $A_+$ and $A_-$ can be set equal
to zero simultaneously.
The important point for us is that,
as is easy to see, in the
$A_\pm =0$ gauge one
recovers from (2.21) both the
field equations (1.3)
of the WZNW theory and  the constraints (2.16).
Furthermore, one sees that
setting $A_\pm $ to zero is not a complete gauge fixing,
the residual
gauge transformations
are exactly the chiral gauge transformations of equation (2.17).

The above arguments tell us that
 the space of gauge orbits in
the space
of classical solutions of the  gauged WZNW theory (2.18)
can be naturally identified with the reduced phase
space belonging to the
Hamiltonian reduction of the WZNW theory
determined by the first class constraints (2.16).
It can be also shown that the Poisson bracket induced
on the reduced phase space by the Hamiltonian reduction is
the same as the one determined by the gauged WZNW action (2.18).
In summary, we see that
the gauged WZNW theory (2.18) provides a
natural Lagrangean
implementation of the WZNW reduction.

\vskip 1.2truecm

 at 14.4truept
\def\d{\delta}

\def\pa{\partial}
\def\i{\int\!d^2x\,}

\def\gi{g^{-1}}

\def\rr{\rangle}
\def\ll{\langle}

\noindent
{\bf 2.3. Effective field
theories from left-right dual reductions}

\vskip 0.8truecm

The aim of this section is to describe the effective field
equations and action functionals for an important class of
the reduced WZNW theories.
This class of theories is obtained by making the assumption that
the left and right
gauge algebras $\Gamma$ and $\tilde \Gamma$ are {\it dual to each
other} with respect to the Cartan-Killing form, which means that
one can choose bases $\gamma_i\in \Gamma$ and $\tilde\gamma_j\in\tilde\Gamma$
so that
$$
\langle \gamma_i , \tilde\gamma_j\rangle =\delta_{ij}\ .
\eqno(2.22)$$
This {\it technical} assumption
allows for having  a simple general
algorithm for disentangling the constraints:
$$
\phi_i = \ll\gamma_i,\,
 \pa_+ g\, \gi   - M \rr  = 0,\quad{\rm and}\quad
\tilde\phi_i = \ll\tilde\gamma_i,\,  \gi
\pa_- g  - \tilde M \rr  = 0,
\eqno(2.23)
$$
which define the  reduction.
We shall comment on the physical meaning of the assumption at the
end of the section, here we only point out that it holds,
e.g., if one  chooses
$\Gamma$ and $\tilde \Gamma$ to be the images of each other under
a Cartan involution\footnote{*}{A Cartan involution $\sigma$ of
the simple Lie algebra ${\cal G}$ is an automorphism for which
$\sigma^2=1$ and $\langle v,\sigma (v)\rangle <0$ for any non-zero element
$v$ of ${\cal G}$.}
 of the underlying simple Lie algebra.

For concreteness, let us consider the
maximally non-compact real form  which
can be defined as the  real span of a Chevalley basis
$H_i$, $E_{\pm\alpha}$  of the corresponding complex Lie algebra
${\cal G}_c$, and in the case
 of the classical series $A_n$, $B_n$, $C_n$ and $D_n$
is given by
$sl(n+1,R)$, $so(n,n+1,R)$, $sp(2n,R)$ and $so(n,n,R)$,
respectively.
In this case the Cartan involution
is $(-1)\times {\rm transpose}$, operating on the Chevalley
basis according to
$$
H_i\longrightarrow -H_i\,
\qquad
E_{\pm \alpha}\longrightarrow -E_{\mp \alpha}\ .
\eqno(2.24)$$
It is obvious
that $\langle v\,,\,v^t\rangle >0$ for any non-zero $v\in {\cal G}$
 and from this one
sees that
$\Gamma^t$ is dual to $\Gamma$ with respect to the Cartan-Killing
form, i.e., (2.22) holds for
$\tilde \Gamma =\Gamma^t$.
It should also be mentioned that
there  is a Cartan involution for every non-compact real form
of the complex simple Lie algebras, as explained in detail
in [36].

Equation (2.22)  implies that
the left and right gauge algebras do not intersect, and
thus we can consider a  direct sum
decomposition of ${\cal G}$ of the form
$$
{\cal G}=\Gamma +{\cal B}+\tilde\Gamma \ ,
\eqno(2.25{\rm a})$$
where ${\cal B}$ is some linear subspace of ${\cal G}$.
Here ${\cal B}$ is in principle an arbitrary
complementary space to $(\Gamma +\tilde \Gamma)$ in ${\cal G}$,
but one can always make the choice
$$
{\cal B}=(\Gamma +\tilde \Gamma )^\perp\ ,
\eqno(2.25{\rm b})$$
which is natural in the sense that the Cartan-Killing form is
non-degenerate on this ${\cal B}$.
Choosing ${\cal B}$ according to  (2.25b) is especially well-suited
in the case of the parity invariant effective theories
discussed at the end of the section.
We note that it might also be convenient if one can take the space
${\cal B}$ to be a subalgebra of ${\cal G}$, but this is not necessary
for our arguments  and is not always possible either.

We can associate a \lq generalized Gauss decomposition'
of the group $G$ to the direct sum decomposition (2.25),
which is the main tool of our analysis.
By \lq Gauss decomposing' an element $g\in G$
according to (2.25), we mean  writing it
in the form
$$
g=a\cdot b\cdot c\,,
\quad {\rm with}{\quad}
a=e^{\gamma},\quad
b=e^{\beta}
\quad{\rm and}\quad
c=e^{\tilde \gamma},
\eqno(2.26)$$
where $\gamma$, $\beta$ and $\tilde\gamma$ are from the respective
subspaces in (2.25).

There is a neighbourhood of the identity in $G$ consisting of
elements which allow a unique decomposition of this sort, and
in this neighbourhood the pieces $a$, $b$ and $c$ can be
extracted from $g$ by algebraic operations.
(Actually it is also possible to define $b$ as a
product of exponentials corresponding to subspaces of
${\cal B}$, and we shall make use of this freedom later,
in Chapter 4.)
We make the assumption that every $G$-valued
field we encounter is decomposable as $g$ in (2.26).
It is easily seen
that in this \lq Gauss decomposable
sector' the components of  $b(x^+,x^-)$ provide
a complete set of gauge invariant local fields,
which are the local fields of the reduced theory we are
after.
Below we explain how to  solve the constraints (2.23)
in the Gauss decomposable sector of
the WZNW theory.
More exactly, for our method to work, we
restrict ourselves to considering those fields
which vary in such a Gauss decomposable neighbourhood of the
identity where the matrix
$$
V_{ij}(b)=\langle \gamma_i ,b\tilde\gamma_j b^{-1}\rangle
\eqno(2.27)$$
is invertible.
Due to the assumptions, the analysis given in the following
yields a {\it local} description of the reduced theories.
It is clear that for a global description one should use patches
on $G$ obtained by multiplying out the Gauss decomposable
neighbourhood of the identity, but we do not deal with this
issue here.

First we derive the field equation of the
reduced theory  by implementing the constraints
directly in the WZNW field equation $\partial_-(\partial_+g g^{-1})=0$.
(This is allowed since the WZNW dynamics leaves the constraint surface
invariant, i.e., the WZNW Hamiltonian weakly commutes with the constraints.)
By inserting the Gauss decomposition of $g$ into (2.23) and
making use of the constraints being first class,
the constraint equations  can be rewritten as
$$
\eqalign{
  \ll\gamma_i,\, \pa_+ b b^{-1} + b(\pa_+c c^{-1})b^{-1} - M \rr & = 0\,, \cr
  \ll\tilde\gamma_i,\, b^{-1}\pa_-b + b^{-1}(a^{-1}\pa_-a)b - \tilde M \rr & =
0
   \,.
}
\eqno(2.28)
$$
With the help of the inverse of $V_{ij}(b)$ in (2.27),
one can solve these
equations for $\pa_+c c^{-1}$ and
$a^{-1}\pa_-a$ in terms of $b$,
$$
\pa_+c c^{-1} = b^{-1}T(b)b,
\qquad {\rm and}\qquad
a^{-1}\pa_-a = b \tilde T(b) b^{-1},
\eqno(2.29{\rm a})
$$
where
$$\eqalign{
T(b)&=\sum_{ij} V^{-1}_{ij}(b)
  \ll\gamma_j,\,  M - \pa_+ b b^{-1} \rr b \tilde \gamma_i b^{-1} ,\cr
{\tilde T}(b)&=\sum_{ij} V^{-1}_{ij}(b)
 \ll\tilde\gamma_i,\, \tilde M - b^{-1}\pa_-b \rr b^{-1} \gamma_j b
\,.\cr}
\eqno(2.29{\rm b})$$
It is easy to obtain the effective field equation
for the field $b(x^+,x^-)$ by using this explicit form of
the constraints.
This can be achieved for example by noting that,
by applying the operator ${\rm Ad}_{a^{-1}}$
to equation (1.9) (i.e., by conjugating it by $a^{-1}$)
the WZNW  field equation
can be written in the form
$$
[\partial_+-{\cal A}_+\,,\,\partial_--{\cal A}_-]=0
\eqno(2.30)$$
with
$$
{\cal A}_+=\partial_+b \, b^{-1}+b(\partial_+c c^{-1})b^{-1}
\quad{\rm and}\quad
{\cal A}_-=-a^{-1}\partial_-a\ .
\eqno(2.31)$$
Thus, by inserting the constraints (2.29) into the
above form of the WZNW equation,
we see that the field equation of the reduced theory
is the zero curvature condition of the following Lax potential:
$$
{\cal A}_+(b)=\partial_+b\, b^{-1}+T(b)
\quad{\rm and}\quad
{\cal A}_-(b)=-b\tilde T(b)b^{-1}\ .
\eqno(2.32)$$
More explicitly, the effective field equation reads
$$
\pa_-(\pa_+b b^{-1}) + [ b\tilde T(b)b^{-1}, \, T(b) ]
+ \pa_- T(b)+ b(\pa_+ \tilde T(b)) b^{-1} = 0.
\eqno(2.33)
$$

The expression on the left-hand-side
of (2.33) in general varies in
the full space ${\cal G}$, but
not all the components represent independent equations.
The number of the independent equations is the number of the
independent components of the WZNW field equation
minus the number of the constraints in (2.23),
since the constraints
automatically imply the corresponding components of the
WZNW  equation.
Thus there are exactly as many independent equations in (2.33)
as the number of the reduced degrees of freedom.
In fact,
the {\it independent field equations} can be obtained by taking the
Cartan-Killing inner product of (2.33)  with a basis
of the linear space ${\cal B}$ in (2.25), and
the inner product of
(2.33) with the $\gamma_i$ and
the $\tilde\gamma_i$ vanishes as a consequence of
the constraints in (2.23) together with the
independent field equations.
To see this one first recalls that the
left-hand-side of (2.33) is, upon imposing the constaints,
equivalent to $a^{-1}(\partial_- J)a$.
Thus the inner product of this with $\Gamma$,
and similarly that of $c(\partial_+ \tilde J)c^{-1}$ with $\tilde \Gamma$,
vanishes as a consequence of the constraints. From this,
 by using the identity
$a^{-1}(\partial_- J)a=-bc(\partial_+ \tilde J)c^{-1}b^{-1}$,
one can conclude that the inner product of
$a^{-1}(\partial_-J)a$ with $\tilde \Gamma$ also vanishes as
a consequence of the constraints and the independent field equations.

At this point we would like to mention certain special cases when the
above equations simplify.
First we note that if one has
$$
[{\cal B}\,,\,\Gamma ]\subset \Gamma
\qquad {\rm and}\qquad
[{\cal B}\,,\,\tilde \Gamma ]\subset \tilde\Gamma\ ,
\eqno(2.34)$$
then
$$
T(b)=M-\pi_{\tilde \Gamma}(\partial_+b b^{-1})
\quad{\rm and}\quad
\tilde T(b)=\tilde M-\pi_{\Gamma}(b^{-1} \partial_-b )\ ,
\eqno(2.35)$$
where we introduced the operators
$$
\pi_{\Gamma}=\sum_i\, \vert \gamma_i\rangle\!\langle \tilde \gamma_i\vert
\qquad{\rm and}\qquad
\pi_{\tilde \Gamma}=\sum_i\,\vert\tilde\gamma_i\rangle\!\langle \gamma_i\vert
\ ,\eqno(2.36)$$
which project onto the spaces $\Gamma$ and $\tilde \Gamma$,
and assumed that $M\in \tilde \Gamma$ and $\tilde M\in \Gamma$.
(The latter assumption can be done without loss of generality due to
the duality condition (2.22)).
One obtains (2.35) from (2.29)  by taking into account that in this case
$V_{ij}(b)$ in (2.27) is the matrix of the operator ${\rm Ad}_b$ acting
on $\tilde \Gamma$,
and thus the inverse is given by ${\rm Ad}_{b^{-1}}$.
The nicest possible situation occurs when
${\cal B}=(\Gamma +\tilde \Gamma)^\perp$ is a {\it subalgebra} of
${\cal G}$ and also satisfies (2.34).
In this case one simply has $T=M$ and $\tilde T=\tilde M$
and thus (2.33) simplifies to
$$
\partial_-(\partial_+b b^{-1})+[b\tilde M b^{-1}\,,\,M]=0\ .
\eqno(2.37)$$
The  derivative term is now an element of ${\cal B}$ and
by combining the above assumptions
with the first class conditions
$[M,\Gamma ]\subset \Gamma^\perp$ and
$[\tilde M,\tilde \Gamma ]\subset \tilde \Gamma^\perp$
one sees that the commutator term in (2.37)
also varies in ${\cal B}$, which ensures the consistency
of this equation.

The effective field equation (2.33) is in general a non-linear equation
for the field $b(x^+,x^-)$, and we can give a procedure
which can in principle  be used for producing its  {\it general solution}.
We are going to do this by making use of the fact that the
space of solutions of the reduced theory is the space of
the constrained WZNW solutions factorized by
the chiral gauge transformations, according to equation (2.17).
Thus the idea is to find the general solution of the effective field
equation by first parametrizing, in terms of
arbitrary chiral functions, those WZNW solutions which satisfy
the constraints (2.23), and then  extracting the $b$-part of those
WZNW solutions by algebraic operations.
In other words, we propose to derive the general solution of (2.33)
by looking at the origin of this equation, instead of its explicit form.

To be more concrete,
one can start the construction of the general solution  by
first Gauss-decomposing the chiral factors of the general WZNW solution
$g(x^+,x^-)=g_L(x^+)\cdot g_R(x^-)$ as
$$
g_L(x^+)=a_L(x^+)\cdot b_L(x^+)\cdot c_L(x^+),
\quad
g_R(x^-)=a_R(x^-)\cdot b_R(x^-)\cdot c_R(x^-).
\eqno(2.38)$$
Then the constraint equations (2.23) become
$$
\partial_+c_L c_L^{-1}=b_L^{-1}T(b_L)b_L
\quad{\rm and}\quad
a_R^{-1}\partial_- a_R=b_R\tilde T(b_R)b_R^{-1}\ .
\eqno(2.39)$$
In addition to the the purely algebraic problems of
computing the quantities $T$ and $\tilde T$ and
extracting  $b$ from $g=g_L\cdot g_R=a\cdot b\cdot c$,
these first order systems of ordinary differential
equations are all one has to solve to produce
the general solution of the effective field equation.
If this can be done by quadrature then the
effective field equation is also integrable by quadrature.
In general, one can proceed  by trying to
solve (2.39) for the functions $c_L(x^+)$ and $a_R(x^-)$
in terms of the arbitrary \lq input functions' $b_L(x^+)$ and $b_R(x^-)$.
Clearly, this involves only a {\it finite number of integrations
whenever the gauge algebras} $\Gamma$ {\it and} $\tilde \Gamma$ {\it consist
of nilpotent elements of} ${\cal G}$.
Thus in this case (2.33) is exactly integrable, i.e., its  general
solution can be obtained by quadrature.

We note that in concrete cases some other choice  of
input functions, instead of the chiral $b$'s,
might prove more convenient for finding the general solutions
of the systems of first order equations
on $g_L$ and $g_R$ given in (2.39) (see for instance
the derivation of the general solution of the Liouville
equation given in [12]).

It is natural  to ask for the action functional underlying
the effective field theory obtained by imposing
the constraints (2.23) on the WZNW theory.
In fact, the effective action
is given by the following formula:
$$
I_{\rm eff}(b) = S_{\rm WZ}(b) -  \i \ll
b\tilde T(b) b^{-1}\,,\, T(b) \rr.
\eqno(2.40)
$$
One can  derive the following condition for the extremum of
this action:
$$
\ll \delta b b^{-1}, \pa_-(\pa_+b b^{-1}) + [ b\tilde T(b) b^{-1}, \,T(b) ]
       + \pa_- T(b) + b(\pa_+ \tilde T(b)) b^{-1} \rr = 0.
\eqno(2.41)
$$
It is straightforward to compute this, the only thing to remember
is that the objects $b\tilde Tb^{-1}$ and $b^{-1}Tb$ introduced in
(2.29) vary in the gauge algebras $\Gamma$ and $\tilde \Gamma$.
The arbitrary variation of $b(x)$ is determined by the arbitrary variation
of $\beta (x)\in {\cal B}$, according to $b(x)=e^{\beta (x)}$,
and thus we see from (2.41) that the Euler-Lagrange equation of the action
(2.40) yields exactly the independent components of the effective field
equation (2.33), which we obtained previously
by imposing the constraints  directly in the WZNW field equation.

The effective action  given above can be derived from the gauged
WZNW action $I(g,A_-,A_+)$ given in (2.18), by eliminating the
gauge fields $A_\pm$ by means of their Euler-Lagrange
equations (2.21c-d).
By using the Gauss decomposition, these Euler-Lagrange equations
become equivalent to the relations
 $$
a^{-1}D_-a=b\tilde T(b)b^{-1}\ ,
\qquad {\rm and}\qquad
cD_+c^{-1}=-b^{-1}T(b)b\ ,
\eqno(2.42)$$
where the quantities $T(b)$ and $\tilde T(b)$ are given by the expressions
in (2.29b) and $D_\pm$ denotes the gauge covariant derivatives,
$D_\pm =\partial_\pm \mp A_\pm$.
Now we show that
$I_{\rm eff}(b)$ in (2.40) can indeed be  obtained
by substituting the solution of (2.42) for $A_\pm$  back into
$I(g,A_-,A_+)$ with  $g=abc$.
To this first we rewrite $I(abc,A_-,A_+)$ by using the
Polyakov-Wiegmann identity [35] as
$$\eqalign{
I&(abc\,,\,A_-,A_+)=S_{\rm WZ}(b)- \int d^2x\,
\Bigl(\langle a^{-1}D_-a\,,\,b(cD_+c^{-1})b^{-1}\rangle \cr
&+\langle b^{-1}\partial_-b\,,\,cD_+c^{-1}\rangle
-\langle \partial_+b b^{-1}\,,\,a^{-1}D_-a\rangle
+\langle A_-,M\rangle +\langle A_+,\tilde M\rangle \Bigr).\cr}
\eqno(2.43)$$
This equation can be regarded as the gauge covariant form of the
Polyakov-Wiegmann identity,
and all but the last two terms are
manifestly gauge invariant.
The effective action
(2.40) is derived from (2.43) together with (2.42)
by noting, for example,
that
$\langle \partial_-a a^{-1}\,,\,M\rangle$ is a total derivative,
which follows from the facts that
$a(x)\in e^\Gamma$ and $M\in [\Gamma\,,\,\Gamma]^\perp$, by
(2.8).

Above we have used the field
equations to eliminate the gauge fields from the gauged WZNW action
(2.18)
on the ground
that $A_-$ and $A_+$
are not dynamical fields,
but  \lq Lagrange multiplier fields' implementing the constraints.
However, it should be noted that without further assumptions the
Euler-Lagrange equation of the  action
resulting from (2.18) by means of this
elimination procedure  {\it does not} always give the effective
field equation, which can always be obtained directly from the WZNW
field equation.
One can see this on an example in which one imposes constraints
{\it only on one of the chiral sectors} of the WZNW theory. From this
 point of view, the role of our assumption on the duality of the
left and right gauge algebras is that it guarantees that the effective
action underlying the effective field equation can be derived from
$I(g,A_-,A_+)$ in the above manner.
To end this discussion, we note that for
$g=abc$ the non-degeneracy of $V_{ij}(b)$
in (2.27) is  equivalent to the non-degeneracy of the quadratic
expression $\langle A_-\,,\,gA_+g^{-1}\rangle$  in the components of
$A_-=A_-^i \gamma_i$ and $A_+=A_+^i \tilde\gamma_i$.
This quadratic term enters into the gauged WZNW action given by
(2.18), and its non-degeneracy is clearly important in the quantum theory,
which we consider in Chapter 5.

We mentioned at the beginning of the section that,
considering a maximally non-compact ${\cal G}$,
one can make sure that the duality assumption expressed by (2.22)
holds by choosing $\Gamma$ and $\tilde \Gamma$ to be the {\it transposes}
of each other.
Here we point out that this particular
left-right related choice of the gauge algebras
 can also be used to ensure the {\it parity
invariance} of the effective field theory.
To this first we notice that, in the case of a maximally non-compact
connected Lie group $G$,
the WZNW action $S_{\rm WZ}(g)$ is invariant under any of
the following two \lq parity transformations' $g\longrightarrow Pg$:
$$
(P_1g)(x^0,x^1)\equiv g^t(x^0,-x^1)\ ,
\quad{\rm and}\quad
(P_2g)(x^0,x^1)\equiv g^{-1}(x^0,-x^1) .
\eqno(2.44)$$
If one chooses  $\tilde \Gamma =\Gamma^t$ and
$\tilde M =M^t$ to define the WZNW reduction
then the parity  transformation  $P_1$
simply interchanges the
left and right constraints, $\phi$ and $\tilde \phi$ in
(2.23), and thus
the corresponding
effective field theory is invariant under the parity $P_1$.
The space ${\cal B}=(\Gamma +\tilde\Gamma )^\perp$,
i.e., the choice in (2.25b),
is invariant under the transpose in this case,
and thus the gauge invariant field $b$ transforms
in the same way under $P_1$ as $g$ does in (2.44).
Of course, the
parity invariance can also be seen on the level of the gauged action
$I(g,A_-,A_+)$.
Namely, $I(g,A_-,A_+)$
is invariant under $P_1$
if one  extends the definition in (2.44) to include the
following parity transformation
of the gauge fields:
$$
(P_1 A_\pm )(x^0,x^1)\equiv A_\mp^t(x^0,-x^1)\ .
\eqno(2.45)$$
The $P_1$-invariant reduction procedure does not preserve
the parity symmetry $P_2$,
but it is possible to consider reductions preserving
just $P_2$ instead of $P_1$.
In fact, such reductions can be obtained by taking
$\tilde \Gamma =\Gamma$ and $\tilde M=M$.

Finally, it is obvious that to construct parity invariant WZNW reductions
in general, for some arbitrary but non-compact
real form ${\cal G}$
of the complex simple Lie algebras,
one can use $-\sigma$ instead of the transpose, where $\sigma$
is a Cartan involution of ${\cal G}$.

\vfill\eject

\def \mtxt#1{\quad\hbox{{#1}}\quad}

\centerline{{\bf 3. Polynomiality in KM reductions and the}
 ${\cal W}_{\cal S}^{\cal G}${\bf-algebras}}

\vskip 0.8truecm

In the previous chapter we described
the conditions for (2.2) defining first class constraints
and for $L_H(J)$ in (2.10) being a gauge invariant quantity
on this constraint surface.
It is clear that the KM Poisson brackets of the gauge
invariant differential polynomials of the current
always close on such polynomials and $\delta$-distributions.
The algebra of the gauge invariant differential polynomials is
of special interest in the conformally invariant case when it
is a polynomial extension of the Virasoro algebra.
In Section 3.1 we shall give an additional
condition on the triple $(\Gamma , M ,H)$ which allows one
to construct out of the current in (2.2) a complete set of
gauge invariant differential polynomials by means of a
differential polynomial gauge fixing algorithm.
We call the KM reduction polynomial if such a polynomial
gauge fixing algorithm is available, and also call the
corresponding gauges Drinfeld-Sokolov (DS) gauges, since our construction
is a generalization of the one given in [5].
The KM Poisson bracket algebra of the gauge invariant differential
polynomials becomes the Dirac bracket algebra of the
current components in the DS gauges, which we consider in Section 3.2.
The extended conformal algebra ${\cal W}^{\cal G}_{\cal S}$ mentioned
in the Introduction is especially interesting in that its primary field
basis is manifest and given by the $sl(2)$ structure,
as we shall see in Section 3.3.
One of our main results is that we shall find here
first class KM constraints underlying this
algebra, such that they satisfy our
sufficient condition for polynomiality.
Thus we can represent ${\cal W}_{\cal S}^{\cal G}$ as a KM Poisson
bracket algebra of gauge invariant differential polynomials,
which in principle allows for its quantization through the KM representation
theory.
The importance of the ${\cal W}_{\cal S}^{\cal G}$-algebras
is clearly demonstrated by the result of Section 3.4, where
we show that the $W_n^l$-algebras of [26] can be interpreted as
further reductions of particular ${\cal W}_{\cal S}^{\cal G}$-algebras.
This makes it possible to exhibit primary fields for the $W_n^l$-algebras
and to describe their structure in detail in terms of the
corresponding ${\cal W}_{\cal S}^{\cal G}$-algebras,
which is the subject of [37].

\vskip 1.2 truecm

\noindent
{\bf 3.1. A sufficient condition for polynomiality}

\vskip 0.8truecm

Let us suppose that $(\Gamma ,M,H)$ satisfy the
previously given conditions, (2.6) and (2.13), for
$$
J(x)=M+j(x)\ ,
\qquad\qquad
j(x)\in \Gamma^\perp
\eqno(3.1)$$
describing the constraint surface of
conformally invariant first class constraints,
where $H$ is a {\it grading operator} and $M$
is subject to
$$
[H,M]=-M\,,
\qquad\qquad
M\notin \Gamma^\perp\ .
\eqno(3.2)$$
Then, as we shall show, the following two additional conditions:
$$
\Gamma \cap {\cal K}_M =\{ 0\}\ ,
\qquad {\rm where}\qquad
{\cal K}_M={\rm Ker}({\rm ad}_M)\ ,
\eqno(3.3)$$
and
$$
\Gamma^\perp\subset {\cal G}_{>-1}\ ,
\eqno(3.4{\rm a})$$
 allow for establishing {\it a differential polynomial gauge fixing
algorithm} whereby one can construct
out of  $J(x)$  in (3.1) a complete set of gauge invariant  differential
polynomials.

Before proving this result,
we discuss some consequences of
the conditions, which we shall need later.
In the present situation $\Gamma$,
$\Gamma^\perp$ and ${\cal G}$ are graded by the eigenvalues of
${\rm ad}_H$, and first we note that
(3.4a) is {\it equivalent} to
$$
{\cal G}_{\geq 1}\subset \Gamma \ .
\eqno(3.4{\rm b})$$
Indeed,  this follows from the fact that the spaces ${\cal G}_h$ and
${\cal G}_{-h}$ are dual to each other with respect to
the Cartan-Killing form, which is a consequence of
its non-degeneracy and invariance under ${\rm ad}_H$.
Of course, here and below the grading is the one defined by $H$,
and we note that ${\cal G}_{\pm 1}$ are non-trivial because of (3.2).
The condition given by (3.4a) plays a technical role in our considerations,
but perhaps it can be argued
for also physically, on the basis that it ensures
that the conformal weights of the primary field components of
$j(x)$ in (3.1) are {\it non-negative} with respect to $L_H$ (2.10).
Second, let us observe that in our situation $M$ satisfying
(3.2) is {\it uniquely determined}, that is,
there is no possibility of shifting it by elements from $\Gamma^\perp$,
simply because there are no grade $-1$ elements in $\Gamma^\perp$,
on account of (3.4a).
Equation (3.3) means that the operator ${\rm ad}_M$ maps
$\Gamma$ into $\Gamma^\perp$ in an {\it injective} manner,
and for this reason we  refer to (3.3)  as the {\it non-degeneracy
condition}.
Combining the non-degeneracy condition with (3.2), (3.4a) and (2.7)
we see
that our gauge algebra $\Gamma$ can contain only {\it positive} grades:
$$
\Gamma \subset {\cal G}_{>0}\ .
\eqno(3.5)$$
This implies that
every $\gamma\in \Gamma$ is represented by a nilpotent operator
in any finite dimensional representation of ${\cal G}$, and that
$$
{\cal G}_{\geq 0}\subset \Gamma^\perp\ .
\eqno(3.6)$$
It follows from (3.2) that $[H,{\cal K}_M ]\subset {\cal K}_M$,
which is telling us that ${\cal K}_M$ is also graded, and we see
from (3.3) and (3.4b) that
$$
{\cal K}_M \subset {\cal G}_{<1}\ .
\eqno(3.7)$$
Finally, we wish to establish a certain relationship between
the dimensions of ${\cal G}$ and ${\cal K}_M$.
For this purpose we consider an arbitrary
complementary space ${\cal T}_M$ to ${\cal K}_M$,
defining a linear direct sum decomposition
$$
{\cal G}={\cal K}_M +{\cal T}_M\ .
\eqno(3.8)$$
It is easy to see that for the 2-form $\omega_M$ we have
$\omega_M({\cal K}_M,{\cal G})=0$, and
the restriction of $\omega_M$
 to ${\cal T}_M$ is a {\it symplectic} form, in
other words:
$$
\omega_M({\cal T}_M ,{\cal T}_M ) \qquad {\rm is}
\quad {\rm non\!-\!degenerate}\ .
\eqno(3.9)$$
(We note in passing that
 ${\cal T}_M$ can be identified with the tangent space at $M$
to the coadjoint orbit of $G$ through $M$, and in this picture
$\omega_M$ becomes the Kirillov-Kostant
symplectic form of the orbit [34].)
The non-degeneracy condition (3.3) says
that one can choose the space ${\cal T}_M$
in (3.8) in such a way that $\Gamma \subset {\cal T}_M$.
One then obtains the inequality
$$
{\rm dim}(\Gamma )\leq {1\over 2}
{\rm dim}({\cal T}_M)={1\over 2}
\bigl( {\rm dim}({\cal G})-{\rm dim}({\cal K}_M )\bigr) \ ,
\eqno(3.10)$$
where the factor ${1\over 2}$ arises since $\omega_M$ is a symplectic form on
 ${\cal T}_M$,
which vanishes, by (2.6), on the subspace $\Gamma\subset {\cal T}_M$.

After the above clarification of the meaning
of conditions (3.3) and (3.4),  we now wish to show that
they indeed allow for exhibiting a
 complete set of gauge invariant differential polynomials
among the gauge invariant functions.
Generalizing the arguments of [5,13,15],
this will be achieved by demonstrating that
an arbitrary current $J(x)$ subject to
(3.1) {\it can be brought to a certain normal form by
a unique gauge transformation which depends
on} $J(x)$ {\it in a differential polynomial way.}

A normal form suitable for this purpose can be associated
to any {\it graded} subspace $\Theta\subset {\cal G}$ which is
{\it dual to} $\Gamma$ {\it with respect
to the} 2-{\it form} $\omega_M$.
Given such a space $\Theta$, it is possible
 to choose bases
$\gamma^i_h$ and $\theta^j_k$ in
$\Gamma$ and $\Theta$ respectively such that
$$
\omega_M(\gamma_h^l,\theta^i_k) = \delta_{il} \delta _{hk},
\eqno(3.11)$$
where the subscript $h$ on $\gamma_h^l$ denotes the grade,
 and the indices
$i$ and $l$ denote the additional labels which are
necessary to specify the base vectors at fixed grade.
It is to be noted that, by definition, the
 subsript $k$ on elements $\theta_k^j\in \Theta$
 does not denote the grade, which is $(1-k)$.
The normal (or {\it reduced}) form corresponding to $\Theta$
is given by the following equation:
$$
J_{\rm red}(x)=M+j_{\rm red}(x)
{\quad}{\rm where}{\quad}
j_{\rm red}(x)\in \Gamma^\perp\cap\Theta^\perp\ .
\eqno(3.12)$$
In other words, the set of reduced
currents is obtained by supplementing the first class constraints
of equation (2.3)
by the gauge fixing condition
$$
{\chi}_{\theta} (x)=\langle J(x),\theta\rangle -\langle M,\theta\rangle
=0\ ,
\qquad \forall \theta\in \Theta\ .
\eqno(3.13)$$
We call a gauge
which can be  obtained  in the above manner a Drinfeld-Sokolov (DS)
gauge.
It is not hard to see that the space
${\cal V}=\Gamma^\perp\cap \Theta^\perp$ is a graded subspace of
$\Gamma^\perp$ which is disjoint from the image of
$\Gamma$ under the operator ${\rm ad}_M$ and is
in fact complementary to the image, i.e., one has
$$
\Gamma^\perp =[M,\Gamma ]+{{\cal V}}\ .
\eqno(3.14)$$
It also follows from the non-degeneracy condition (3.3)
that any graded complement ${\cal V}$ in (3.14) can be
obtained
in the above manner, by means of using some $\Theta$.
Thus it is possible to define the DS normal form of the current
directly in terms of a complementary space ${\cal V}$ as well,
as has been done in special cases in [5,13,18].

As the first step in proving that any current in (3.1) is gauge
equivalent to one in the DS gauge,
let us consider the gauge transformation by
$g_h(x^+)=\exp [\sum_l a_h^l(x^+)\gamma_h^l]$ for some fixed grade $h$.
Suppressing the summation over $l$, it can be written
as\footnote*{Throughout the chapter, all equations involving
 gauge transformations,
Poisson brackets, etc., are to be evaluated by using a
fixed time, since
they are all consequences of equation (2.1).
By this convention, they are valid both on the canonical phase space
and on the chiral KM phase space belonging to space of solutions
of the theory.}
$$
j(x)\rightarrow j^{g_h}(x)=
e^{a_h\cdot\gamma_h}( j(x) + M ) e^{-a_h\cdot\gamma_h}
+ ( e^{a_h\cdot\gamma_h} )^\prime e^{-a_h\cdot\gamma_h} - M\ .
\eqno(3.15)$$
Taking the inner product of this equation with
the basis vectors $\theta^i_k$ in (3.11)
for all $k\leq h$, we see that
there is no contribution from the derivative term.
We also see that the only contribution from
$$
e^{a_h\cdot\gamma_h}j(x)e^{-a_h\cdot\gamma_h}=
j(x)+[a_h(x^+)\cdot \gamma_h , \, j(x)]+\ldots
\eqno(3.16)$$
is the one coming from the first term, since
all commutators containing the elements $\gamma_h^l$
drop out from the inner product in question as a consequence
of the following crucial relation:
$$
[\gamma_h^l, \theta_k^i]\  \in \  \Gamma,
\mtxt{for} k\leq h,  \eqno(3.17)$$
which
follows from (3.4b) by
noting that the grade
of this commutator, $(1+h-k)$, is
at least $1$
for $k \leq h$.
Taking these into account, and computing
the contribution
from those two terms in $j^{g_h}(x)$ which contain $M$
by using (3.11),
 we obtain
$$
\langle\theta_k^i,j^{g_h}(x)\rangle =
\langle \theta_k^i, j(x)\rangle -
 a_h^i(x^+)\delta_{hk},
\mtxt{for all} k\leq h.
\eqno(3.18)
$$
We see from this equation that
$$
\langle\theta_k^i, j(x)\rangle =0
\quad \Longleftrightarrow \quad
\langle\theta_k^i, j^{g_h}(x)\rangle =0\ ,
\mtxt{for} k<h\ ,
\eqno(3.19)$$
and
$$
a^i_h(x^+)= \langle\theta_h^i,j(x)\rangle
\quad  \Rightarrow  \quad
\langle\theta_h^i , j^{g_h}(x)\rangle =0\ ,
\mtxt{for} k=h .
\eqno(3.20)
$$
These last two equations tell us that
if the gauge-fixing
condition $\langle\theta_k^i,j(x)\rangle =0$
is satisfied  for all $k<h$  then we can ensure that the same
condition holds for
$j^{g_h}(x)$ for the {\it extended range of indices} $k\leq h$,
by choosing $a^i_h(x^+)$ to be
$\langle \theta^i_h,j(x)\rangle$. From this
 it is easy to see that the DS gauge (3.13)
can be reached
by an iterative process of gauge transformations,
and the gauge-parameters
$a^i_h(x^+)$ are
unique polynomials in the current at each
stage of the iteration.

In more detail,
let us write the general element $g(a(x^+))\in e^\Gamma$ of the gauge group
as a product in order of descending grades, i.e.,  as
$$
g(a(x^+)) =g_{h_n}\cdot g_{h_{n-1}}\cdots g_{h_1}, \mtxt{with}
g_{h_i}(x^+)=e^{a_{h_i}(x^+)\cdot \gamma_{h_i}}\ ,
\eqno(3.21{\rm a})$$
where
$$
h_n>h_{n-1}>\ldots >h_1
\eqno(3.21{\rm b})$$
is the list of grades occurring in $\Gamma$.
Let us then insert this expression into
$$
j\rightarrow j^{g} =
g (j + M )  g^{-1} + g^\prime g^{-1} -M\ ,
\eqno(3.22{\rm a})$$
and consider the condition
$$
j^g(x)=j_{\rm red}(x)\ ,
\eqno(3.22{\rm b})$$
with $j_{\rm red}(x)$ in (3.12),
 as an equation for the gauge-parameters
$a_h(x^+)$.
One sees from the  above considerations that this equation is
uniquely soluble for the components of the $a_h(x^+)$ and
the solution is a  differential polynomial in $j(x)$.
 This implies that the components of
$j_{\rm red}(x)$ can also be uniquely computed from (3.22),
and {\it the solution  yields a complete set of gauge invariant
differential polynomials of} $j(x)$, which
establishes the required result.
The above iterative procedure
is in fact a convenient tool for computing the gauge invariant
differential polynomials in practice [15].
We remark that, of course, any unique gauge fixing can be used to define
gauge invariant quantities, but they are in general not polynomial,
not even local in $j(x)$.

We also wish to note that an arbitrary linear subspace
of ${\cal G}$ which
is  dual to ${\cal V}$  in (3.14)
with respect to the Cartan-Killing form
can be used in a natural way
as the space of parameters for describing those current
dependent KM transformations which preserve the DS gauge.
In fact, it is possible
to give an algorithm which
computes the ${\cal W}$-algebra and its action on
the other fields of the corresponding constrained WZNW theory
by finding the gauge preserving KM transformations implementing
the ${\cal W}$-transformations.
This algorithm presupposes the existence of such gauge invariant
differential polynomials which reduce to the current components
in the DS gauge, which is ensured by the above gauge fixing algorithm,
but it works without actually computing them.
This issue is treated in detail in [13,18] in special cases,
but the results given there apply also to the general situation
investigated in the above.

\vskip 1.2truecm

\def \mtxt#1{\quad\hbox{{#1}}\quad}
\def\half{{1 \over 2}}
\def \lwd#1{\lower2pt\hbox{$\scriptstyle #1$}}

\noindent
{\bf 3.2. The polynomiality of the Dirac bracket}

\vskip 0.8truecm

It follows from the polynomiality of the gauge fixing
that the components of the gauge fixed current
$j_{\rm red}$ in (3.12) generate a differential polynomial algebra
{\it under Dirac bracket}.
In our proof of the polynomiality we actually only used
that the graded subspace $\Theta$ of ${\cal G}$
is dual to the graded gauge algebra $\Gamma$ with respect to
$\omega_M$ and satisfies the condition
$$
([\Theta\,,\,\Gamma])_{\geq 1}\subset \Gamma\ ,
\eqno(3.23)$$
which is equivalent to the existence of the bases
$\gamma_h^l$ and $\theta_k^i$ satisfying (3.11) and (3.17).
We have seen that this condition follows from (3.3) and (3.4),
but it should be noted that it is a
{\it more general condition}, since
the converse is not true, as is shown by an example at the end of
this section.

Below we wish to give a direct proof
for the polynomiality of the Dirac bracket
algebra belonging to the {\it second class} constraints:
$$
c_\tau (x)=\langle \tau\,,\,J(x)-M\rangle =0
\quad {\rm where}\quad
 \tau\in \{\gamma_h^l\}\cup \{\theta_k^i\}\ .
\eqno(3.24)$$
The proof will shed a new light on the polynomiality condition.
We note that for certain purposes second class constraints
 might be more natural to use
than first class ones since
in the second class formalism one directly deals with the physical
fields.
For example,
 the  ${\cal W}_{\cal S}^{\cal G}$-algebra  mentioned
in the Introduction is very natural from the second class point of
view and can be realized by starting with a number of different first
class systems of constraints,
as we shall see in the next section.

We first
recall that, by definition, the Dirac bracket algebra of the reduced currents
is
$$\eqalign{
\{ j_{\rm red}^u(x), & j_{\rm red}^v(y)\}^* =
\{ j_{\rm red}^u(x),j_{\rm red}^v(y)\}   \cr
-&\sum_{\mu\nu}\int dz^1dw^1
\{ j_{\rm red}^u(x),c_{\mu}(z)\}
\Delta_{\mu\nu} (z,w)\{ c_{\nu}(w),j_{\rm red}^v(y)\}\cr},
 \eqno(3.25)$$
where, for any $u\in {\cal G}$,
$j_{\rm red}^u(x)=\langle u,j_{\rm red}(x)\rangle$ is to be
substituted by $\langle u,J(x)-M\rangle$ under the KM Poisson
bracket, and $\Delta_{\mu\nu}(z,w)$ is the inverse of the kernel
$$
D_{\mu\nu}(z,w)=\{c_\mu( z),c_\nu (w)\}\ ,
\eqno(3.26)$$
in the sense that (on the constraint surface)
$$
\sum_\nu\int dx^1
\Delta_{\mu\nu}(z,x) D_{\nu\sigma}(x,w) =
\delta_{\mu\sigma}\delta(z^1-w^1).
\eqno(3.27)$$

To establish the polynomiality of the Dirac bracket,
it is useful to consider the
{\it matrix differential operator}
$D_{\mu\nu}(z)$
defined by the kernel $D_{\mu\nu}(z,w)$ in the usual way, i.e.,
$$
\sum_\nu D_{\mu\nu}(z) f_\nu (z)=
\sum_\nu \int dw^1 D_{\mu\nu}(z,w)f_\nu (w)\ ,
\eqno(3.28)$$
for a vector of smooth functions $f_\nu (z)$,
which are periodic in $z^1$.  From the structure of
the constraints in (3.24), $c_\tau =(\phi_\gamma , \chi_\theta )$,
one sees that $D_{\mu\nu}(z)$ is a first order
differential operator possessing the following block structure
$$
D_{\mu\nu} = \pmatrix{
D_{\gamma\tilde \gamma} &  D_{\gamma \theta}   \cr
D_{\tilde \theta\tilde \gamma} & D_{\tilde \theta \theta }  \cr
} = \pmatrix{
0 & E \cr
-E^\dagger & F \cr
}, \eqno (3.29)$$
where $E^\dagger$ is the formal Hermitian conjugate of
the matrix $E$,
$(E^\dagger )_{\theta\gamma}=(E_{\gamma\theta})^\dagger$.
It is clear that the Dirac bracket in (3.25) is
a differential polynomial in $j_{\rm red}(x)$ and $\delta (x^1-y^1)$
whenever the inverse operator $D^{-1}(z)$, whose kernel is
$\Delta_{\mu\nu} (z,w)$,
is a differential operator whose  coefficients
are differential polynomials in $j_{\rm red}(z)$.
On the other hand, we see from (3.29) that the operator $D$
is invertible if and only if its block $E$ is invertible, and in
that case
the inverse takes the form
$$
 (D^{-1})_{\mu\nu}= \pmatrix{
(E^\dagger)^{-1}FE^{-1} & -(E^\dagger)^{-1}  \cr
E^{-1} & 0 \cr
}. \eqno(3.30)$$
Since $E(z)$ and
$F(z)$ are polynomial (even linear) in $j_{\rm red}(z)$
and in $\partial_z$ and the
inverse of $F(z)$ does not occur in $D^{-1}(z)$, it follows
that $D^{-1}(z)$ is a polynomial differential operator if and only if
$E^{-1}(z)$
is a polynomial differential operator.

To show that $E^{-1}$ exists and is a polynomial
differential operator we note that
in terms of the basis of $(\Gamma +\Theta)$ in (3.24) the matrix $E$ is
given explicitly by the following formula:
$$
E_{\gamma_h^m,\theta_k^n}(z)=
\delta_{hk}\delta_{mn}+
\langle [\gamma_h^m,\theta_k^n],j_{\rm red}(z)\rangle +
\langle \gamma_h^m,\theta_k^n\rangle \partial_z\ .
\eqno(3.31)$$
The crucial point is that,
by the grading and the property in (3.17),
we have
$$
E_{\gamma_h^m,\theta_k^n}(z)=
\delta_{hk}\delta_{nm}\ ,
\qquad {\rm for}\qquad k\leq h\ .
\eqno(3.32)$$
The matrix $E$ has a block structure labelled by the (block) row and
(block) column
indices $h$ and $k$, respectively, and (3.32) means that the blocks
in the diagonal of $E$ are unit matrices and the blocks
below the diagonal vanish.
In other words, $E$ is of the form
$E=1+\varepsilon$,
where $\varepsilon$
is a strictly upper triangular matrix.
It is clear that such a matrix differential operator is
polynomially invertible, namely by a
{\it finite series} of the form
$$
E^{-1}=
1-\varepsilon +\varepsilon^2 +\ldots +(-1)^N \varepsilon^N\ ,
\qquad (\varepsilon^{N+1}=0),
\eqno(3.33)$$
which finishes our proof of the polynomiality
of the Dirac bracket in (3.25).
One can use the arguments in the above proof to set up
an algorithm
for actually  computing the Dirac bracket.
The proof also shows
that the polynomiality of the Dirac bracket is
 guaranteed whenever $E$ is of the form
$(1+\varepsilon )$ with $\varepsilon$
being {\it nilpotent as a matrix}.
In our case this was ensured by a special grading assumption,
and it appears an interesting question
whether polynomial reductions
can be obtained at all without using some grading structure.

The zero block occurs in $D^{-1}$ in
(3.30) because the second class constraints originate
from the gauge fixing of first class ones.
We note that the presence
of this zero block implies
that the Dirac brackets of the gauge
invariant quantities coincide with their original Poisson
brackets, namely one sees this from
the formula of the Dirac bracket by keeping in mind that
the gauge invariant quantities weakly commute with
the first class constraints.

Finally, we want to show that condition (3.23) is weaker
than (3.3-4). This is best seen by considering an example.
To this let now ${\cal G}$ be the  maximally non-compact real
form of a complex simple Lie algebra.
If
$\{M_-,M_0,M_+\}$ is the principal $sl(2)$ embedding in ${\cal G}$,
with commutation rules as in (3.34) below,
we simply choose the one-dimensional gauge algebra $\Gamma\equiv\{M_+\}$
and take $M\equiv M_-$.
The $\omega_M$-dual to $M_+$ can be taken to be $\theta=M_0$, and
then  (3.23) holds.
To show that conditions (3.3-4) cannot be satisfied, we prove that
a grading operator $H$ for which $[H,M_-]=-M_-$ and
${\cal G}_{\geq 1}^H\subset \Gamma$,
does not exist. First of all,
$[H , M_-]=-M_-$ and $\langle M_-,M_+ \rangle \neq 0$ imply $[H,M_+]=M_+$,
and thus $\Gamma^H_{\geq 1}=\{ M_+ \}$.
Furthermore, writing $H
= (M_0 + \Delta )$, we  find from $[H , M_\pm ]=\pm M_\pm$  that $\Delta$ must
b
   e an
$sl(2)$ singlet in the adjoint of ${\cal G}$. However, in the case
of the principal $sl(2)$ embedding, there is no such singlet in the adjoint,
and hence $H=M_0$.
But then the condition
${\cal G}_{\geq 1}^{M_0}\subset \Gamma$ is not fulfilled.

\vskip 1.2truecm

\def \mtxt#1{\quad\hbox{{#1}}\quad}

\noindent
{\bf 3.3. First class constraints for the
 ${\cal W}_{\cal S}^{\cal G}$-algebras}

\vskip 0.8truecm

Let ${\cal S}=\{ M_-\,,\,M_0\,,\,M_+\}$ be
an $sl(2)$ subalgebra of the simple Lie algebra ${\cal G}$:
$$
[M_0,M_\pm]=\pm M_\pm\ ,\qquad [M_+,M_-]=2M_0\ .
\eqno(3.34)$$
We argued in the Introduction that it is natural to associate
an extended conformal algebra, denoted as ${\cal W}_{\cal S}^{\cal G}$,
to any such $sl(2)$ embedding [16,18].
Namely, we defined the ${\cal W}_{\cal S}^{\cal G}$-algebra to be the Dirac
bracket algebra generated by the components of the constrained KM current
of the the following special form:
$$
J_{\rm red}(x)=M_-+j_{\rm red}(x)\ ,
{\quad\quad}{\rm with}{\quad\quad}
j_{\rm red}(x)\in {\rm Ker}({\rm ad}_{M_+})\ ,
\eqno(3.35)$$
which means that $j_{\rm red}(x)$ is a linear combination of
the $sl(2)$ highest weight states in the adjoint of ${\cal G}$.
This definition is indeed natural in the sense
that the conformal properties are manifest, since,
as we shall see below,
with the exception of the $M_+$-component
the spin $s$ component
of $j_{\rm red}(x)$ turns out to be
a primary field of conformal weight $(s+1)$
with respect to $L_{M_0}$.
Before showing this, we shall construct here
{\it first class} KM constraints underlying the
${\cal W}_{\cal S}^{\cal G}$-algebra,
which will be used in Chapter 4
to construct generalized Toda theories
which realize the ${\cal W}_{\cal S}^{\cal G}$-algebras as
their chiral algebras.
We expect the ${\cal W}_{\cal S}^{\cal G}$-algebras
to play an important organizing role in describing
the (primary field  content of) conformally invariant KM
reductions in general, and shall give arguments in favour of this idea
later.

We wish to find a gauge algebra $\Gamma$ for which
the triple $(\Gamma , H=M_0, M=M_-)$
satisfies our sufficient conditions for polynomiality
and (3.35) represents a DS gauge for the
corresponding conformally invariant first class constraints.
We start by noticing that the dimension of such a $\Gamma$
has to satisfy the  relation
$$
{\rm dim\,}{\rm Ker}({\rm ad}_{M_+})=
{\rm dim\,}{\cal W}_{\cal S}^{\cal G}
={\rm dim\,}{\cal G}-2{\rm dim\,}{\Gamma }\ .
\eqno(3.36)$$
      From this,
since the kernels of ${\rm ad}_{M_\pm}$ are of equal dimension,
we obtain that
$$
{\rm dim\,}\Gamma =
{1\over 2}{\rm dim\,}{\cal G}-
{1\over 2}{\rm dim\,}{\rm Ker}({\rm ad}_{M_-} )\ ,
\eqno(3.37)$$
which means by (3.10) that we are looking for a $\Gamma$ of {\it maximal}
dimension.
By the representation theory of $sl(2)$,
the above equality is equivalent to
$$
{\rm dim\,}\Gamma
={\rm dim\,}{\cal G}_{\geq 1}+{1\over 2}{\rm dim\,}{\cal G}_{1\over 2}\ ,
\eqno(3.38)$$
where the grading is by the, in general half-integral,
eigenvalues of ${\rm ad}_{M_0}$.
We also know, (3.4b) and (3.5), that for our purpose we have to
choose the graded Lie subalgebra $\Gamma$ of ${\cal G}$ in such a way
that ${\cal G}_{\geq 1}\subset \Gamma\subset {\cal G}_{>0}$.
Observe that the non-degeneracy condition (3.3)  is automatically
satisfied for any such  $\Gamma$ since in the present case
${{\rm Ker}({\rm ad}_{M_-})}\subset {{\cal G}_{\leq 0}}$,
and $M_0\in \Gamma^\perp$ is also ensured,
which guarantees the conformal invariance, see (2.13).

It is obvious from the above that in the special case of an {\it integral}
$sl(2)$ subalgebra, for which ${\cal G}_{1\over 2}$ is empty,
one can simply take
$$
\Gamma ={\cal G}_{\geq 1}\ .
\eqno(3.39)$$
For grading reasons,
$$
\omega_{M_-}({\cal G}_{\geq 1},{\cal G}_{\geq 1})=0
\eqno(3.40)$$
holds, and thus one indeed
obtains first class constraints in this way.

One sees from (3.38) that for finding the
gauge algebra in the non-trivial case  of a {\it half-integral}
$sl(2)$ subalgebra,
one should somehow add  half of ${\cal G}_{1\over 2}$
to ${\cal G}_{\geq 1}$,
in order to have the correct dimension.
The key observation for defining the required
 {\it halving} of ${\cal G}_{1\over 2}$ consists in noticing that
the restriction of the 2-form $\omega_{M_-}$ to ${\cal G}_{1\over 2}$
is non-degenerate.
This can be seen as a consequence of (3.9),
 but is also easy to verify directly.
By the well known Darboux normal form  of symplectic forms [34],
there exists a (non-unique) direct sum decomposition
$$
{\cal G}_{1\over 2}={\cal P}_{1\over 2}+{\cal Q}_{1\over 2}
\eqno(3.41)$$
such that $\omega_{M_-}$ vanishes on
the subspaces ${\cal P}_{1\over 2}$ and ${\cal Q}_{1\over 2}$
separately.
The spaces ${\cal P}_{1\over 2}$ and
 ${\cal Q}_{1\over 2}$, which are
the analogues of the usual  momentum and
coordinate subspaces of the  phase space in
analytic mechanics, are
of equal dimension and  dual to each other with respect to
$\omega_{M_-}$.
The point is that the first-classness conditions in (2.6)
are satisfied if we define the gauge algebra to be
$$
\Gamma={\cal G}_{\geq 1}+{\cal P}_{1\over 2}\ ,
\eqno(3.42)$$
by using {\it any symplectic halving} of the above kind.
It is obvious from the construction that
the first class constraints,
$$
J(x)= M_- + j(x)\,\qquad
{\rm with}\qquad j(x)\in \Gamma^\perp\ ,
\eqno(3.43)$$
obtained by using $\Gamma$ in (3.42) satisfy the sufficient
conditions for polynomiality given in Section 3.1.
With this $\Gamma$  we have
$$
\Gamma^\perp ={\cal G}_{\geq 0}+{\cal Q}_{-{1\over 2}}\ ,
\eqno(3.44{\rm a})$$
where ${\cal Q}_{-{1\over 2}}$ is the subspace
of ${\cal G}_{-{1\over 2}}$ given by
$$
{\cal Q}_{-{1\over 2}}=[M_-,{\cal P}_{1\over 2}]\ .
\eqno(3.44{\rm b})$$
By combining (3.42) and (3.44) one also easily verifies
the following direct sum decomposition:
$$
\Gamma^\perp=[M_-,\Gamma]+ {\rm Ker}({\rm ad}_{M_+})\ ,
\eqno(3.45)$$
which  is just (3.14) with ${\cal V}={\rm Ker}({\rm ad}_{M_+})$.
This means that (3.35) is indeed nothing but the equation of a
particular DS gauge for the first class constraints in (3.43), as required.
This special DS gauge is called the {\it highest weight gauge} [13].
Similarly as for any DS gauge,
there exists therefore a basis of gauge invariant differential
polynomials of the current in (3.43)
such that the base elements reduce to the components of
$j_{\rm red}(x)$ in (3.35) by the gauge fixing.
The KM Poisson bracket algebra of these
gauge invariant differential polynomials is
clearly identical to the Dirac bracket algebra of the
corresponding current components,
and we can thus realize the
${\cal W}_{\cal S}^{\cal G}$-algebra
as a KM Poisson bracket algebra of gauge
invariant differential polynomials.

The second class constraints defining the highest weight gauge (3.35)
are natural in the sense that
in this case $\tau$ in (3.24) runs over the
basis of the space
${\cal T}_{M_-}=[M_+\,,\,{\cal G}]$ which is a natural complement
of ${\cal K}_{M_-}={\rm Ker}({\rm ad}_{M_-})$ in ${\cal G}$, eq. (3.8).

In the second class formalism, the conformal action generated by
$L_{M_0}$ on the ${\cal W}_{\cal S}^{\cal G}$-algebra
is given by the  following formula:
$$
\delta^*_{f,M_0}\,
j_{\rm red}(x)\equiv  -\int dy^1\, f(y^+)\,\{L_{M_0}(y) \,,\,
j_{\rm red}(x)\}^* \ ,
\eqno(3.46)$$
where the parameter function $f(x^+)$  refers to the
conformal coordinate transformation $\delta_f\, x^+=-f(x^+)$,
cf. (2.11), and $j_{\rm red}(x)$ is to be substituted by
$J(x)-M_-$
when evaluating the KM Poisson brackets entering into (3.46),
like in (3.25).
To actually evaluate (3.46), we
first replace $L_{M_0}$ by the object
$$
L_{\rm mod}(x)=L_{M_0}(x)-{1\over 2} \langle
M_+\,,\, J^{\prime\prime}(x)\rangle\ ,
\eqno(3.47)$$
which is allowed  under the Dirac bracket
since the difference (the second term) vanishes upon imposing
the constraints.
The crucial point to notice is
that $L_{\rm mod}$ weakly commutes with  {\it all} the
constraints defining (3.35) (not only with the first class ones)
under the KM Poisson bracket.
This implies that with $L_{\rm mod}$ the Dirac bracket in (3.46)
is in fact
identical to the original KM Poisson bracket and by this observation
we easily obtain
$$
\delta^*_{f,M_0}\, j_{\rm red}(x)=f(x^+)\, j_{\rm red}^{\prime}(x)+
f^{\prime}(x^+)\bigl( j_{\rm red}(x)+[M_0,j_{\rm red}(x)])
-{1\over 2}f^{\prime\prime\prime}(x^+)M_+.
\eqno(3.48)$$
This proves that, with the exception of the $M_+$-component,
the $sl(2)$ highest weight components of $j_{\rm red}(x)$ in
(3.35) transform as conformal primary fields, whereby
the conformal content of
${\cal W}_{\cal S}^{\cal G}$ is determined by the decomposition of
the adjoint of ${\cal G}$ under ${\cal S}$ in the
aforementioned manner.
We end this discussion by noting
that in the highest weight gauge $L_{M_0}(x)$ becomes a
linear combination of the $M_+$-component
of $j_{\rm red}(x)$ and
a quadratic expression in the components corresponding to the
singlets of ${\cal S}$ in ${\cal G}$. From
 this we see that $L_{M_0}(x)$ and the
primary fields corresponding to the $sl(2)$ highest weight states
give a basis for the differential polynomials
contained in ${\cal W}_{\cal S}^{\cal G}$,
which is thus indeed a (classical) ${\cal W}$-algebra
in the sense of the general idea in [20].

In the above we proposed a \lq halving procedure'
for finding {\it purely first  class} constraints for which
${\cal W}_{\cal S}^{\cal G}$ appears as the algebra of
the corresponding gauge invariant differential polynomials.
We now wish to clarify the relationship between our method
and the construction in a recent paper by Bais {\it et al} [16], where
the  ${\cal W}_{\cal S}^{\cal G}$-algebra has been described,
in the special case of ${\cal G}=sl(n)$, by using a different
method.
We recall that the ${\cal W}_{\cal S}^{\cal G}$-algebra has been
constructed in [16]
by adding to the first class constraints
defined by the pair
$({\cal G}_{\geq 1}, M_-)$ the
{\it second class} constraints
$$
\langle u\,,\,J(x)\rangle =0\ ,
\qquad {\rm for} \quad \forall\  u\in {\cal G}_{1\over 2}\ .
\eqno(3.49)$$
Clearly, we recover these constraints
by first imposing our complete set of first class constraint
belonging to $(\Gamma , M_-)$
with $\Gamma$ in (3.42), and then partially fixing the gauge by
imposing the condition
$$
\langle u\,,\,J(x)\rangle =0\ ,
\qquad {\rm for} \quad \forall\  u\in {\cal Q}_{1\over 2}\ .
\eqno(3.50)$$
One of the advantages of our construction is that
 by using only first class KM constraints it is easy
to construct generalized Toda theories
which possess ${\cal W}_{\cal S}^{\cal G}$ as their
chiral algebra, for any $sl(2)$ subalgebra, namely by using our
general method  of  WZNW reductions.
This will be elaborated in the next chapter.
We  note that in [16] the authors
were actually also led to  replacing  the
original constraints by a first class
system of constraints, in order to be able to consider the
BRST quantization of the theory.
For this purpose they
introduced
unphysical \lq auxiliary fields' and thus constructed
first class constraints
in an extended phase space.
However, in that construction
one  has to check that the auxiliary fields
finally disappear from the physical quantities.
Another important advantage of our halving procedure
is that it renders the use of any
such auxiliary fields completely
unnecessary, since one can start by imposing
a complete system of first class constraints
on the KM phase space from the very beginning.
We study some aspects of the BRST quantization in Chapter 5,
and we shall see that the Virasoro central charge given in
[16] agrees with the one  computed by
taking our first class constraints as the starting point.

The first class constraints leading to ${\cal W}_{\cal S}^{\cal G}$
are not unique, for example one can consider an arbitrary halving
in (3.41) to define $\Gamma$.
We conjecture that these ${\cal W}$-algebras always occur
under certain natural  assumptions on the constraints.
To be more exact, let us suppose that we have
conformally invariant first class constraints
determined by the pair $(\Gamma ,M_-)$ where $M_-$ is a
{\it nilpotent} matrix and the {\it non-degeneracy} condition (3.3)
holds together with equation (3.37).
By the Jacobson-Morozov theorem, it is possible to extend the
nilpotent generator $M_-$  to an  $sl(2)$ subalgebra
${\cal S}=\{ M_-,M_0,M_+\}$.
It is also worth noting that the conjugacy class of ${\cal S}$ under
the automorphism group of
${\cal G}$ is uniquely determined by the conjugacy class of the
nilpotent element $M_-$.
For this and other questions concerning the theory of $sl(2)$ embeddings
into semi-simple Lie algebras the reader may consult refs.
[32,33,38,39].
We expect that the above assumptions on $(\Gamma ,M_-)$
are sufficient for the existence of a complete set of gauge
invariant differential polynomials and their algebra is isomorphic
to ${\cal W}_{\cal S}^{\cal G}$, where $M_-\in{\cal S}$.
We are not yet able to prove this conjecture in general,
but below we wish to sketch the proof in an
important special case which illustrates
the idea.

Let us assume that we have conformally invariant
first class constraints described by $(\Gamma, M_-,H)$
subject to the sufficient conditions for polynomiality given in
Section 3.1, such that $H$ is an {\it integral grading operator} of ${\cal G}$.
We note that
these are exactly  the assumptions satisfied by the constraints
in the non-degenerate case of the generalized Toda theories
associated to integral gradings [18].
In this case equation (3.37) is actually
automatically satisfied as a consequence of the non-degeneracy
condition (3.3).
One can also show that it is possible to find an $sl(2)$ algebra
${\cal S}=\{M_-,M_0,M_+\}$ for which
in addition to
$[H,M_-]=-M_-$ one  has
$$
[H,M_0]=0
\qquad {\rm and}\qquad
[H,M_+]=M_+\ ,
\eqno(3.51)$$
and that for this $sl(2)$ algebra the relation
$$
{\rm Ker}({\rm ad}_{M_+})\subset {\cal G}_{\geq 0}^H
\eqno(3.52)$$
holds, where the superscript indicates that the grading is defined by $H$.
For the $sl(2)$
subject to (3.51) the latter property is in fact equivalent to
${\rm Ker}({\rm ad}_{M_-})\subset {\cal G}^H_{\leq 0}$,
which is just the non-degeneracy condition as in our case
$\Gamma ={\cal G}^H_{>0}$.
The proof of these statements
is given in Appendix B.

We introduce a definition at this point, which will be used in the rest of
the paper.
Namely, we call an $sl(2)$ subalgebra ${\cal S}=\{M_-, M_0, M_+ \}$ an
$H${\it-compatible} $sl(2)$ from now on
if there exists an integral grading operator $H$ such that
$[H, M_\pm ]=\pm M_\pm$
is satisfied together with the  non-degeneracy condition.
The non-degeneracy condition  can be expressed  in various equivalent
forms, it can be given for example as the relation in (3.52),
and its (equivalent) analogue for $M_-$.

Turning back to the problem at hand,
we now point out that by using the $H$-compatible
$sl(2)$ we have the following direct sum decomposition
of $\Gamma^\perp ={\cal G}^H_{\geq 0}$:
$$
{\cal G}_{\geq 0}^H=[M_-,{\cal G}^H_{>0}]+{\rm Ker}({\rm ad}_{M_+}).
\eqno(3.53)$$
This means that the set of currents of the form (3.35)
represents  a DS gauge for the present first class constraints.
This implies the required result,
that is that the $\cal W$-algebra belonging to
the constraints defined by $\Gamma ={\cal G}^H_{>0}$
together with a non-degenerate $M_-$
is isomorphic to ${\cal W}_{\cal S}^{\cal G}$
with $M_-\in {\cal S}$.
In this example both $L_H(x)$ and $L_{M_0}(x)$
are gauge invariant differential polynomials.
Although the spectrum of ${\rm ad}_H$ is
{\it integral} by assumption,
in some cases  the $H$-compatible $sl(2)$ is
embedded into ${\cal G}$ in a {\it half-integral} manner,
i.e., the  spectrum of ${\rm ad}_{M_0}$ can be half-integral in certain cases.
We shall return to this point later.
We further note that in general it is clearly impossible
to build such an $sl(2)$ out of $M_-$ for which $H$
would play the role of $M_0$.
It is possible to prove that in those
cases there is no full set of primary fields
with repect to $L_H$ which would
complete this Virasoro density to a generating set
of the corresponding differential polynomial
${\cal W}$-algebra.
We have seen that such a conformal basis is manifest
for ${\cal W}_{\cal S}^{\cal G}$, which seems to indicate
that in the present situation the conformal structure defined by
the $sl(2)$, $L_{M_0}$, is preferred in comparison to the
one defined by $L_H$.

We also would like to mention an interesting general fact about
the ${\cal W}_{\cal S}^{\cal G}$-algebras, which will be used
in the next section.
Let us consider the decomposition of ${\cal G}$ under the $sl(2)$ subalgebra
${\cal S}$.
In general, we shall find singlet states and they span a Lie subalgebra
in the Lie subalgebra ${\rm Ker}({\rm ad}_{M_+})$ of ${\cal G}$.
Let us denote this zero spin subalgebra as ${\cal Z}$.
It is easy to see that we  have the semi-direct sum decomposition
$$
{\rm Ker}({\rm ad}_{M_+}) ={\cal Z} + {\cal R} ,\qquad
[{\cal Z} ,{\cal R} ]\subset {\cal R} ,\qquad
[{\cal Z} ,{\cal  Z}]\subset {\cal Z} ,
\eqno(3.54)$$
where ${\cal R}$ is the linear space spanned by the rest of the highest weight
states, which have non-zero spin.
It is not hard to prove that the subalgebra
of the original KM algebra  which belongs to ${\cal Z}$, survives the
reduction to ${\cal W}_{\cal S}^{\cal G}$.
In other words, the Dirac brackets of the ${\cal Z}$-components
of the highest weight gauge current, $j_{\rm red}$ in (3.35), coincide
with their original KM Poisson brackets, given by (2.1).
Furthermore, this ${\cal Z}$ KM subalgebra acts on
the ${\cal W}_{\cal S}^{\cal G}$-algebra
by the  corresponding original
KM transformations, which preserve the highest weight gauge:
$$
J_{\rm red}(x)\rightarrow
e^{a^i(x^+)\zeta_i}\,J_{\rm red}(x)\,e^{-a^i(x^+)\zeta_i}
+ (e^{a^i(x^+)\zeta_i})^{\prime}\,e^{-a^i(x^+)\zeta_i},
\eqno(3.55)$$
where the $\zeta_i$ form a basis of ${\cal Z}$.
In particular, one sees that the ${\cal W}_{\cal S}^{\cal G}$-algebra
inherites
the semi-direct sum structure given by (3.54) [16].
The point we wish to make is that it is possible
to {\it further reduce} the ${\cal W}_{\cal S}^{\cal G}$-algebra by
applying the general
method of conformally invariant
KM reductions to the present ${\cal Z}$ KM symmetry.
In principle, one can generate a huge number of
new conformally invariant systems out of the
${\cal W}_{\cal S}^{\cal G}$-algebras in this way, i.e.,
by applying conformally invariant constraints
to their singlet KM subalgebras.
For example, if one can find a subalgebra of ${\cal Z}$ on which the
Cartan-Killing
form of ${\cal G}$ vanishes, then one can
consider the  obviously conformally invariant reduction
obtained by  constraining the corresponding components of $j_{\rm red}$
in (3.35) to zero.
We do not explore these \lq secondary'
reductions of the ${\cal W}_{\cal S}^{\cal G}$-algebras
in this paper.
However, their potential importance will be highlighted by
the example of the next  section.

Finally, we note that, for a half-integral
$sl(2)$, one can consider (instead of using $\Gamma$ in (3.42))
also those conformally
invariant first class constraints which are defined by the triple
$(\Gamma , M_0 , M_- )$ with any graded $\Gamma$ for which
${\cal G}_{\geq 1}\subset \Gamma\subset ({\cal G}_{\geq 1}+{\cal P}_{1\over
2})$
   .
The polynomiality conditions of Section 3.1 are  clearly satisfied with
any such non-maximal $\Gamma$, and
the corresponding extended conformal algebras
are in a sense between the KM and ${\cal W}_{\cal S}^{\cal G}$-algebras.

\vskip 1.2truecm

\def \mtxt#1{\quad\hbox{{#1}}\quad}
\def\half{{1 \over 2}}

\def\d{\delta}

\def\pa{\partial}
\def\i{\int\!d^2x\,}

\def\t{{\rm Tr}\,}

\def\la{\langle}
\def\ra{\rangle}
\def\di{D\!\!\!\!/}


\noindent
{{\bf 3.4. The }${\cal W}_{\cal S}^{\cal G}$
 {\bf interpretation of the} $W^l_n${\bf-algebras}}

\vskip 0.8truecm

The $W_n^l$-algebras  are certain conformally invariant
reductions of the $sl(n,R)$ KM algebra introduced by Bershadsky [26]
using a mixed set of {\it first class and second class}
constraints.
It is known [16] that the simplest non-trivial case $W_3^2$,
originally proposed by Polyakov [27],
coincides with the ${\cal W}_{\cal S}^{\cal G}$-algebra belonging to the
highest root $sl(2)$ of $sl(3,R)$.
The purpose of this section is to understand whether or not
these reduced KM systems fit into our framework,
which is based on using {\it purely first class} constraints,
and to uncover  their possible connection with the
${\cal W}_{\cal S}^{\cal G}$-algebras in the general case.
(In  this section,  ${\cal G}=sl(n,R)$.)
In fact, we shall construct here purely first class KM
constraints leading to the $W_n^l$-algebras.
The construction will demonstrate
that the $W_n^l$-algebras can in general be identified as
{\it further reductions of particular}
${\cal W}^{\cal G}_{\cal S}${\it-algebras}.
The secondary reduction process is obtained by means of the singlet KM
subalgebras of the relevant ${\cal W}_{\cal S}^{\cal G}$-algebras,
in the manner mentioned in the previous section.

By definition [26],
the KM reduction yielding the $W_n^l$-algebra is obtained by
constraining the current to take the following form:
$$
J_{\rm B}(x)=M_-+j_{\rm B}(x), \qquad j_{\rm B}(x) \in \Delta^\perp ,
\eqno(3.56)$$
where $\Delta$ denotes the set of all strictly upper triangular
$n \times n$ matrices and
$$M_-=e_{l+1,1}+e_{l+2,2}+...+e_{n,n-l},
\eqno(3.57)$$
the $e$'s being the standard $sl(n,R)$ generators ($l \leq
n-1$), i.e., $M_-$ has 1's all along the $l$-th slanted line below
the diagonal.
The current in (3.56) corresponds to imposing the
constraints $\phi_{\delta} (x)=0$ for all $\delta\in \Delta$,
like in (2.3).
Generally, these constraints comprise first and second class
parts, where
the first class part
is the one belonging
to the subalgebra ${\cal D}$ of $\Delta$ defined by
the relation $\omega_{M_-}({\cal D},\Delta )=0$, (see (2.4)).
The second class part belongs to the complementary
space, ${\cal C}$, of ${\cal D}$ in $\Delta$.
In fact, for $l=1$ the constraints are  the usual first
class ones which yield the standard ${\cal W}$-algebras,
but the second class part is non-empty for $l>1$.
The above KM reduction is so constructed that it is conformally
invariant, since the constraints weakly commute with
the Virasoro density $L_{H_l}(x)$, see (2.10),
where $H_l = {1\over l} H_1$ and $H_1$ is the standard grading
operator of $sl(n,R)$, for which $[H_1\,,\,e_{ik}]=(k-i)e_{ik}$.

We start  our construction by extending the
nilpotent generator $M_-$ in (3.57)
to an $sl(2)$ subalgebra ${\cal S}=\{ M_- , M_0 , M_+\}$.
In fact,
parametrizing $n = ml + r$ with
$m = \bigl[{n\over l}\bigr]$ and
$0 \le r < l$,
we can take
$$
M_0 = {\rm diag}\Bigl(
      \overbrace{{m\over2}, \cdots}^{r \rm\;times},\,
      \overbrace{{{m-1}\over2}, \cdots}^{(l-r) \rm\;times},\,
      \cdots,
      \overbrace{-{m\over2}, \cdots}^{r \rm\;times} \Bigr),
\eqno(3.58)
$$
where the mutiplicities, $r$ and $(l - r)$, occur alternately and
end with $r$.
The meaning of
this formula is that
the fundamental of $sl(n,R)$ branches into $l$ irreducible
representations under ${\cal S}$, $r$ of spin $m\over 2$ and $l-r$
of spin ${m-1}\over 2$.
The explicit form of $M_+$
is a certain linear combination of the
$e_{ik}$'s with
$(k-i)=l$, which is straightforward to compute.

We  describe next the first and the second
 class parts
of the constraints in (3.56) in more detail
by using the grading defined by $M_0$.
We observe first that in terms of this grading
the space $\Delta$ admits the decomposition
$$
\Delta =\Delta_0 +{\cal G}_{1\over 2}+{\cal G}_1 + {\cal G}_{> 1}\ .
\eqno(3.59)$$
       From this and the definition of $\omega_{M_-}$,
the subalgebra ${\cal D}$ comprising the first class part
can also be decomposed into
$$
{\cal D}={\cal D}_0 + {\cal D}_1 + {\cal G}_{>1}\ ,
\eqno(3.60)$$
where
$$
{\cal D}_0 = {\rm Ker\,}({\rm ad}_{M_-})\cap \Delta_0
\eqno(3.61)$$
 is the set of the $sl(2)$ singlets
in $\Delta$, and ${\cal D}_1$ is a
subspace of ${\cal G}_1$ which we do not need to
specify.
By combining (3.59) and (3.60), we see that
the complementary space ${\cal C}$, to which the
second class part belongs, has the structure
$$
{\cal C}={\cal Q}_0 + {\cal G}_{1\over 2} + {\cal P}_1 \ ,
\eqno(3.62)$$
where the subspace ${\cal Q}_0$ is complementary to ${\cal D}_0$
in ${\Delta }_0$,
 and ${\cal P}_1$ is  complementary to
${\cal D}_1$ in ${\cal G}_1$.
The 2-form $\omega_{M_-}$ is non-degenerate on ${\cal C}$ by construction,
and this implies by the grading that the spaces ${\cal Q}_0$  and ${\cal P}_1$
are symplectically conjugate to each other, which is reflected by the
notation.

We shall construct a gauge algebra, $\Gamma$, so that
Bershadsky's constraints will be recovered
by a partial gauge fixing from the first class ones belonging
to $\Gamma$.
As a generalization of the halving procedure of the previous
section, we take the following ansatz:
$$
\Gamma = {\cal D} + {\cal P}_{{1\over 2}}+{\cal P}_1 \ ,
\eqno(3.63)$$
where  ${\cal P}_{1\over 2}$ is defined
by means of some symplectic halving
${\cal G}_{{1\over 2}}={\cal P}_{{1\over 2}}+{\cal Q}_{{1\over 2}}$,
like in (3.41).
It is important to notice that this equation can be
recasted into
$$
\Gamma = {\cal D}_0 + {\cal P}_{{1\over 2}}+{\cal G}_{\geq 1}\ ,
\eqno(3.64)$$
which  would be just the familiar formula (3.42)
if ${\cal D}_0$ was not here.
By using (3.57) and (3.58),
${\cal D}_0$ can be identified
as the set of $n\times n$ block-diagonal
matrices, $\sigma$, of the following form:
$$\sigma=\hbox{block-diag} \{\Sigma_0,\sigma_0,\Sigma_0,
.....,\Sigma_0,\sigma_0,\Sigma_0 \},
\eqno(3.65)$$
where the $\Sigma_0$'s and the $\sigma_0$'s
are identical
copies of strictly upper triangular
$r \times r$ and $(l-r) \times (l-r)$
matrices respectively.
This implies that
$$
{\rm dim\,}{\cal D}_0= {1 \over 4}[l(l-2)+(l-2r)^2]\ ,
\eqno(3.66)$$
which shows that ${\cal D}_0$ is  non-empty except
when $l=2,\,r=1$,
which is the case of $W_n^2$ with  $n={\rm odd}$.
The fact that ${\cal D}_0$ is
in general non-empty gives us
a trouble at this stage, namely, we have now no guarantee that
the above $\Gamma$
is actually a {\it subalgebra} of ${\cal G}$.
By using the grading and
the fact that ${\cal D}_0$ is a subalgebra,
we see that $\Gamma$ in (3.64) becomes a
subalgebra if and only if
$$
[ {\cal D}_0\,,\,{\cal P}_{{1\over 2}}]
\subset {\cal P}_{{1\over 2}} .
\eqno(3.67)$$
We next show that it
is possible
to find such a \lq good halving' of ${\cal G}_{1\over 2}$ for which
${\cal P}_{{1\over 2}}$
satisfies (3.67).

For this purpose,
we use yet another grading here.
This grading is provided by using the
particular diagonal matrix, $H\in {\cal G}$,
which we construct out of $M_0$ in (3.58) by
first adding ${1\over 2}$ to its half-integral eigenvalues,
and then substracting a multiple of the unit
matrix so as to make the result traceless.
In the adjoint representation,
we then have
${\rm ad}_H = {\rm ad}_{M_0}$ on the tensors, and
${\rm ad}_H = {\rm ad}_{M_0} \pm {1/ 2}$ on
the spinors.
We notice from this that the $H$-grading is
an integral grading.
In fact, the relationship between the two gradings
allows us to  define a  good halving of ${\cal G}_{1\over 2}$
as follows:
$$
{\cal P}_{{1\over 2}}\equiv  {\cal G}_{{1\over 2}}\cap {\cal G}^H_1\,,
\qquad {\rm and}\qquad
{\cal Q}_{{1\over 2}}\equiv  {\cal G}_{{1\over 2}}\cap {\cal G}^H_0\,.
\eqno(3.68)$$
Since $M_-$ is of grade $-1$
with respect to both gradings,
the spaces given by (3.68)
clearly yield a sympectic halving of ${\cal G}_{{1\over 2}}$
with respect to $\omega_{M_-}$.
That this is a good halving,  i.e.,
it
ensures the condition
(3.67), can also be seen easily by observing that ${\cal D}_0$ has
grade $0$ in the $H$-grading, too.
Thus we obtain the required subalgebra $\Gamma$ of
${\cal G}$ by using this particular ${\cal P}_{{1\over 2}}$
in (3.64).

Let us consider now the
first class constraints corresponding to
the above constructed  gauge algebra $\Gamma$,
$\phi_\gamma(x) = 0$ for $\gamma \in \Gamma$,
which bring the
current into the form
$$
J_{\Gamma}(x)=M_- + j_{\Gamma}(x)\ ,
\qquad  j_{\Gamma}(x)\in \Gamma^\perp\ .
\eqno(3.69)$$
It is easy to verify that the original constraint surface (3.56)
can be recovered from (3.69) by a partial gauge fixing in such a
way that the residual gauge transformations are exactly the ones
belonging to the space ${\cal D}$.
In fact,  this is achieved by fixing the gauge freedom corresponding
to the piece $({\cal P}_{1\over 2}+{\cal P}_1)$ of $\Gamma$, (3.63), by
imposing the partial gauge fixing condition
$$
\phi_{q_i} (x) = 0 \ ,
\qquad  q_i\in ({\cal Q}_0+{\cal Q}_{1\over 2}),
\eqno(3.70)$$
where the $q_i$ form a basis of the space
$({\cal Q}_0+{\cal Q}_{1\over 2})$ and the $\phi_{q}$'s are defined
like in (2.3).
This implies that the  reduced phase space defined by
the constraints in  (3.69) is the same as the one
determined by the original  constraints (3.56).
In conclusion,  our purely first class constraints, (3.69),
have the same physical content as Bershadsky's original
mixed set of constraints, (3.56).

Finally, we give the relationship between Bershadsky's
$W_n^l$-algebras and the $sl(2)$ systems.
Having seen that the reduced KM phase spaces carrying the
$W_n^l$-algebras can be realized by starting from the first class constraints
in (3.69), it follows from (3.64) that the $W_n^l$-algebras coincide
with particular ${\cal W}_{\cal S}^{\cal G}$-algebras
if and only if the space ${\cal D}_0$ is empty, i.e., for
$W_n^2$ with $n={\rm odd}$.
In order to establish the ${\cal W}_{\cal S}^{\cal G}$ interpretation
of $W_n^l$
in the general case, we point out that the reduced  phase space can be reached
from (3.69) by means of the following two step process based on the $sl(2)$
structure.
Namely, one can proceed by first
fixing the gauge freedom corresponding to the piece
$({\cal P}_{1\over 2}+{\cal G}_{\geq 1})$ of $\Gamma$,
and then fixing the rest of the gauge freedom.
Clearly,  the constraint surface resulting  in the first step
is the same as the one obtained
by putting to zero those components of the highest weight gauge
current representing ${\cal W}_{\cal S}^{\cal G}$ which correspond to
${\cal D}_0$.
The final reduced phase space is obtained in the second
step by fixing the gauge freedom
generated by the constraints belonging to ${\cal D}_0$,
which we have seen to be the space
of the upper triangular singlets of ${\cal S}$.
Thus we can conclude that $W_n^l$ can be regarded as a further
reduction of the corresponding ${\cal W}_{\cal S}^{\cal G}$,
where the \lq secondary reduction' is of the type mentioned
at the end of Section 3.3.
One can exhibit primary field bases for the $W_n^l$-algebras
and describe their structure in detail in terms of the underlying
${\cal W}_{\cal S}^{\cal G}$-algebras by further analysing
the secondary reduction, but this is outside the scope of
the present paper, see [37].

\vfill\eject

\centerline{\bf 4. Generalized Toda theories}

\vskip 0.8truecm

Let us remind ourselves that, as has been detailed in the Introduction,
the standard conformal Toda field theories can be naturally regarded as
reduced WZNW theories, and as a consequence these theories
possess the chiral algebras
${{\cal W}_{\cal S}^{\cal G}}\times {\tilde {\cal W}_{\cal S}^{\cal G}}$
as their canonical symmetries,
where $\cal S$ is the principal $sl(2)$ subalgebra
of the maximally non-compact real Lie algebra $\cal G$.
It is  natural to seek for WZNW reductions leading to
effective field theories which would realize
${{\cal W}_{\cal S}^{\cal G}}\times {\tilde {\cal W}_{\cal S}^{\cal G}}$
as their chiral algebras {\it for any}
$sl(2)$ {\it subalgebra} ${\cal S}$ {\it of any simple real Lie algebra}.
The main purpose of this chapter is to obtain, by combining the results
of sections 2.3 and 3.3, {\it generalized Toda theories}
meeting the above requirement in the non-trivial case of the
{\it half-integral} $sl(2)$ subalgebras of the simple Lie algebras.
Before turning to describing these new theories, next we
briefly recall the main features of those generalized Toda theories,
associated to the {\it integral gradings} of the simple Lie algebras,
which have been studied before [3,4,14-18].
The simplicity of the latter theories
will motivate some subsequent developments.

\vskip 1.2 truecm
\noindent
{\bf 4.1. Generalized Toda theories associated with
integral gradings}

\vskip 0.8truecm

The WZNW reduction leading to the generalized Toda theories
in question is
set up by considering an integral grading operator $H$ of ${\cal G}$,
and taking the special case
$$
\Gamma ={\cal G}^H_{\geq 1}
\qquad {\rm and}\qquad
\tilde \Gamma = {\cal G}^H_{\leq -1}\ ,
\eqno(4.1)$$
and any non-zero
$$
M\in {\cal G}^H_{-1}
\qquad {\rm and}\qquad
\tilde M\in {\cal G}^H_{1}\ ,
\eqno(4.2)$$
in the general construction given in Section 2.3.
We note that
by an integral grading operator  $H\in {\cal G}$
we mean a diagonalizable element
whose spectrum
in the adjoint of ${\cal G}$ consists
of integers and contains $\pm 1$,
and that ${\cal G}_n^H$ denotes the grade $n$ subspace defined by $H$.
In the present case ${\cal B}$ in (2.25b)
is the subalgebra ${\cal G}_0^H$ of ${\cal G}$,
and, because of the grading structure, the properties
expressed by equation (2.34) hold.
Thus the effective field equation reads as
(2.37) and the corresponding action is given by
the simple formula
$$
I_{\rm eff}^H(b)=S_{\rm WZ}(b)-
 \int\,d^2x\,\langle b\tilde M b^{-1},M\rangle \ ,
\eqno(4.3)$$
where the field $b$ varies in the little group $G^H_0$ of $H$
in $G$.

Generalized, or non-Abelian, Toda theories
of this type have been
first investigated by Leznov and Saveliev [1,3], who defined these theories
by postulating their Lax potential,
$$
{\cal A}^H_+=\partial_+b \cdot b^{-1} +M\ ,
\qquad\qquad
{\cal A}^H_-=-b\tilde M b^{-1}\ ,
\eqno(4.4)$$
which they obtained by  considering the problem that if one
requires a ${\cal G}$-valued pure-gauge  Lax potential
to take some special form, then the consistency
of the system of equations coming from the
zero curvature condition becomes a non-trivial problem.
In comparison, we have seen in Section 2.3
that in the WZNW framework the Lax potential
originates from the chiral zero curvature equation (1.9), and the
consistency and the integrability of the effective theory
arising from the reduction is automatic.

It was shown in [3,4,16] in the special case when
$H$, $M$ and $\tilde M$ are taken to be the standard
generators of an integral  $sl(2)$ subalgebra of ${\cal G}$,
that the non-Abelian Toda equation allows for conserved chiral
currents underlying its exact integrability.
These currents then generate chiral ${\cal W}$-algebras
of the type ${\cal W}_{\cal S}^{\cal G}$, for
integrally embedded $sl(2)$'s.

By means of the argument given in Section 3.3,
we can establish the structure of the chiral algebras of a wider class
of non-Abelian Toda systems [18].
Namely, we see that if $M$ and $\tilde M$ in (4.2) satisfy the
non-degeneracy conditions
$$
{\rm Ker}({\rm ad}_M)\cap {\cal G}^H_{\geq 1} =\{ 0 \}
\qquad{\rm and}\qquad
{\rm Ker}({\rm ad}_{\tilde M})\cap {\cal G}^H_{\leq -1} =\{ 0 \}\ ,
\eqno(4.5)$$
then the
${\rm left}\times {\rm right}$ chiral algebra of the
corresponding generalized
Toda theory  is isomorphic to
${{\cal W}_{{\cal S}_-}^{\cal G}}\times {\tilde {\cal W}_{{\cal S}_+}^{\cal
G}}$, where ${\cal S}_-$ (${\cal S}_+$) is an $sl(2)$ subalgebra of
${\cal G}$ containing the nilpotent generator  $M$ ($\tilde M$),
respectively.
The $H$-compatible $sl(2)$ algebras ${\cal S}_\pm$
occurring here are {\it not always integrally embedded} ones.
Thus for certain {\it half-integral} $sl(2)$ algebras
${\cal W}_{\cal S}^{\cal G}$ can be realized in a generalized Toda theory
of the type (4.3).
As we would like to have generalized Toda theories
which possess  ${\cal W}_{\cal S}^{\cal G}$ as their  symmetry algebra
for an arbitrary $sl(2)$ subalgebra,
we have to ask whether the theories given above are
already enough for this purpose or not.
This leads to the technical  question as to
whether for  every half-integral $sl(2)$ subalgebra
${\cal S}=\{M_-,M_0,M_+\}$
 of ${\cal G}$ there exists an integral grading operator $H$
such that ${\cal S}$ is an $H$-compatible $sl(2)$,
in the sense introduced in Section 3.3.
The answer to this question is negative, as
proven in Appendix C, where the relationship
between integral gradings and $sl(2)$ subalgebras is  studied in detail.
Thus we have to find new integrable conformal
field theories for our purpose.

\vskip 1.2truecm

\noindent
{\bf 4.2. Generalized Toda theories for half-integral
$sl(2)$ embeddings}

\vskip 0.8truecm

In the following we exhibit a generalized Toda theory possessing
the ${\rm left}\times {\rm right}$ chiral algebra
${{\cal W}_{\cal S}^{\cal G}}\times {\tilde {\cal W}_{\cal S}^{\cal G}}$
for an arbitrarily chosen half-integral $sl(2)$ subalgebra
${\cal S}=\{M_-\,,\,M_0\,,\,M_+\}$ of the arbitrary but non-compact
simple real Lie algebra ${\cal G}$.
Clearly, if one imposes first class constraints
of the type described in Section 3.3 on the
currents of the WZNW theory then the resulting
effective field theory will have the
required chiral algebra.
We shall choose the left and right gauge algebras in such a way to be
dual to each other with respect to the Cartan-Killing form.

Turning to the details,
first we choose a direct sum decomposition
of ${\cal G}_{1\over 2}$
of the type in (3.41), and then define the {\it induced decomposition}
${\cal G}_{-{1\over 2}}={\cal P}_{-{1\over 2}}+{\cal Q}_{-{1\over 2}}$
to be given by the subspaces
$$
{\cal Q}_{-{1\over 2}}\equiv {\cal P}_{1\over 2}^\perp
\cap {\cal G}_{-{1\over 2}}
=[M_-\,,\,{\cal P}_{1\over 2}]
\quad {\rm and}\quad
{\cal P}_{-{1\over 2}}\equiv {\cal Q}_{1\over 2}^\perp
\cap {\cal G}_{-{1\over 2}}
=[M_-\,,\,{\cal Q}_{1\over 2}]\ .
\eqno(4.6)$$
It is easy to see that the 2-form ${\omega}_{M_+}$ vanishes on the
above subspaces of ${\cal G}_{-{1\over 2}}$  as a consequence
of the vanishing of  ${\omega}_{M_-}$  on the
corresponding  subspaces of ${\cal G}_{1\over 2}$.
Thus we can take the left and right gauge algebras to be
$$
\Gamma =({\cal G}_{\geq 1}+{\cal P}_{1\over 2})
\quad{\rm and}\quad
\tilde\Gamma =({\cal G}_{\leq -1}+{\cal P}_{-{1\over 2}})\ ,
\eqno(4.7)$$
with the constant matrices $M$ and $\tilde M$
entering the constraints given by $M_-$ and $M_+$, respectively.
The duality hypothesis of Section 2.3 is
obviously satisfied by this construction.

In principle, the action and the Lax potential of the effective
 theory can be obtained by
specializing the general formulas of Section 2.3 to the
present particular case.
In our case
$$
{\cal B}={\cal Q}_{{1\over 2}}+{\cal G}_0+{\cal Q}_{-{1\over 2}}\ ,
\eqno(4.8)$$
and the physical modes, which are given by the entries of $b$
in the generalized Gauss decomposition $g=a b c$
with $a\in e^\Gamma$ and $c\in e^{\tilde\Gamma}$,
 are now  conveniently
parametrized as
$$
b(x)=\exp [q_{1\over 2}(x)]
\cdot g_0(x)\cdot \exp [q_{-{1\over 2}}(x)]\ ,
\eqno(4.9)$$
where $q_{\pm {1\over 2}}(x)\in {\cal Q}_{\pm {1\over 2}}$ and
$g_0(x)\in G_0$, the little group of $M_0$ in $G$.
Next we introduce some notation which will be useful for describing
the effective theory.

The operator ${\rm Ad}_{g_0}$ maps
${\cal G}_{-{1\over 2}}$ to itself and, by writing the general
element $u$ of ${\cal G}_{-{1\over 2}}$ as a two-component
column vector
$u=(u_1\ u_2)^t$
with $u_1\in {\cal P}_{-{1\over 2}}$ and $u_2\in {\cal Q}_{-{1\over
2}}$,
we can designate this operator
as a $2\times 2$ matrix:
$$
{{\rm Ad}_{g_0}}
_{\vert {\cal G}_{-{1\over 2}}}
=\pmatrix{X_{11}(g_0)&X_{12}(g_0)\cr
          X_{21}(g_0)&X_{22}(g_0)\cr}\ ,
\eqno(4.10)$$
where, for example, $X_{11}(g_0)$ and $X_{12}(g_0)$ are linear
operators mapping ${\cal P}_{-{1\over 2}}$ and
${\cal Q}_{-{1\over 2}}$ to ${\cal P}_{-{1\over 2}}$, respectively.
Analogously, we introduce the notation
$$
{{\rm Ad}_{g_0^{-1}}}
_{\vert {{\cal G}_{{1\over 2}}}}
=
\pmatrix{ Y_{11}(g_0) & Y_{12}(g_0)\cr
          Y_{21}(g_0) & Y_{22}(g_0)\cr}\ ,
\eqno(4.11)$$
which corresponds to writing the general element of
${\cal G}_{1\over 2}$ as a column vector, whose  upper and lower
components belong to ${\cal P}_{1\over 2}$ and
${\cal Q}_{1\over 2}$, respectively.

The action functional of the effective field theory resulting
from the  WZNW  reduction at hand reads as follows:
$$\eqalign{
I^{\cal S}_{\rm eff}(g_0,q_{1\over 2},&q_{-{1\over 2}})=S_{\rm WZ}(g_0)
-\int\,d^2x\,\langle g_0M_+g_0^{-1}\,,\,M_-\rangle\cr
&+\int\,d^2x\,\bigl(
\langle\partial_- q_{1\over 2}\,,\,
g_0 \partial_+q_{-{1\over 2}} g_0^{-1}\rangle
+\langle\eta_{1\over2}\,,\,
X^{-1}_{11}\cdot\eta_{-{1\over 2}}\rangle\bigr)\,,\cr}
\eqno(4.12{\rm a})$$
where the objects $\eta_{\pm {1\over 2}}\in {\cal P}_{\pm {1\over 2}}$
are given by the formulae
$$
\eta_{1\over 2}=
[M_+,q_{-{1\over 2}} ]+Y_{12}\cdot\partial_- q_{1\over 2}
\quad{\rm and}\quad
\eta_{-{1\over 2}}=
[M_-, q_{1\over 2}]-X_{12}\cdot\partial_+q_{-{1\over 2}}.
\eqno(4.12{\rm b})
$$
The Euler-Lagrange equation of this action is the zero curvature
condition of the following Lax potential:
$$\eqalign{
{\cal A}^{\cal S}_+=&M_-+\partial_+g_0\cdot g_0^{-1}+
g_0 (\partial_+ q_{-{1\over 2}}+X_{11}^{-1}\cdot
\eta_{-{1\over 2}})g_0^{-1}\,,\cr
{\cal A}_-^{\cal S}=&-g_0M_+g_0^{-1}
-\partial_-q_{1\over 2}+Y_{11}^{-1}\cdot\eta_{1\over 2}\,.\cr}
\eqno(4.13)$$

The above new (conformally invariant) effective action and
Lax potential are among the main results of the present paper.
Clearly, for an integrally embedded $sl(2)$  this action
and Lax potential simplify to the ones given by equation (4.3) and (4.4).

The derivation of the above formulae is not completely
straightforward, and next we wish to sketch the main steps.
First, let us remember that, by (2.29a), to specialize the general
effective action given by (2.40) and the Lax potential given by (2.32)
to our situation, we should express the objects
$\partial_+c c^{-1}$ and $a^{-1}\partial_-a$ in terms of $b$ by using
the constraints on $J$ and $\tilde J$, respectively.
(In the present case it would be tedious to compute the inverse matrix
of $V_{ij}$ in (2.27), which would be needed for using directly (2.29b).)
For this purpose it turns out to be convenient to parametrize
the WZNW field $g$ by using the grading defined by the $sl(2)$, i.e., as
$$
g=g_+\cdot g_0\cdot g_-
\quad{\rm where}\quad
g_+=a\cdot\exp [q_{1\over 2}],
\quad
g_-=\exp [q_{-{1\over 2}}]\cdot c\,.
\eqno(4.14)$$
We recall that the fields $a$, $c$, $g_0$ and $q_{\pm {1\over 2}}$ have
been introduced  previously
by means of  the parametrization
 $g=a b c$, with $b$
in (4.9).
Also for later convenience, we  write $g_\pm$ as
$$
g_+=\exp[r_{\geq 1}+p_{1\over 2}+q_{1\over 2}]
\quad{\rm and}\quad
g_-=\exp[r_{\leq -1}+p_{-{1\over 2}}+q_{-{1\over 2}}]\,.
\eqno(4.15)$$
Note that here and below the subscript denotes the grade of the
variables, and $p_{\pm {1\over 2}}\in {\cal P}_{\pm {1\over 2}}$.
In our case this parametrization of $g$ is advantageous,
since, as shown below, the use of the grading structure facilitates solving
the constraints.

For example, the left constraint are restrictions on
$J_{<0}$, for which  we have
$$
J_{<0}=(g_+ g_0 N g_0^{-1} g_+^{-1})_{<0}
\quad {\rm with} \quad N=\partial_+ g_-\cdot g_-^{-1}\,.
\eqno(4.16)$$
By considering this equation grade by grade,
starting from the lowest grade, it is easy to see that the constraints
corresponding to ${\cal G}_{\geq 1}\subset \Gamma$ are equivalent to the
relation
$$
N_{\leq -1}=g_0^{-1} M_- g_0\ .
\eqno(4.17)$$
The remaining left constraints set the ${\cal P}_{-{1\over 2}}$
part of $J_{-{1\over 2}}$ to zero, and to unfold
these constraints first we note that
$$
J_{-{1\over 2}}=[p_{1\over 2}+q_{1\over 2}\,,\,M_-]
+g_0\cdot N_{-{1\over 2}}\cdot g_0^{-1}\,,
\quad{\rm with}\quad
N_{-{1\over 2}}=\partial_+ p_{-{1\over 2}}+\partial_+ q_{-{1\over 2}}\,.
\eqno(4.18)$$
By using the notation introduced in (4.10), the
vanishing of the projection of $J$ to
${\cal P}_{-{1\over 2}}$
is written as
$$
[q_{1\over 2}\,,\,M_-]
+X_{11}\cdot \partial_+ p_{-{1\over 2}}
+X_{12}\cdot \partial_+ q_{-{1\over 2}}=0\ ,
\eqno(4.19)$$
and from this we obtain
$$
\partial_+ p_{-{1\over 2}}=
X_{11}^{-1}\cdot
\bigl\{ [M_-\,,\,q_{1\over 2}]-X_{12}\cdot \partial_+ q_{-{1\over
2}}\bigr\}\ .
\eqno(4.20)$$
Combining our previous formulae, finally we obtain that on the constraint
surface of the WZNW theory
$$
N =g_0^{-1} M_- g_0 +
\partial_+ q_{-{1\over 2}}+
X_{11}^{-1}(g_0)\cdot
\bigl\{ [M_-\,,\,q_{1\over 2}]-X_{12}(g_0)\cdot \partial_+ q_{-{1\over
2}}\bigr\}\ .
\eqno(4.21)$$
A similar analysis applied to the right constraints
yields that they are  equivalent to the
following equation:
$$
-g_+^{-1}\cdot \partial_- g_+=
-g_0 M_+ g_0^{-1}
-\partial_- q_{1\over 2}+Y_{11}^{-1}(g_0)\cdot
\bigl\{ [M_+\,,\,q_{-{1\over 2}}]+Y_{12}(g_0)\cdot \partial_- q_{{1\over
2}}\bigr\}\, .
\eqno(4.22)$$

By using the relations established above, we can at this stage
easily compute $b^{-1}Tb=\partial_+c c^{-1}$ and
$b\tilde Tb^{-1}=a^{-1}\partial_-a$ as well, and substituting these
into (2.40), and using the Polyakov-Wiegmann identity to
rewrite $S_{\rm WZ}(b)$ for $b$ in (4.9),  results in
the action in (4.12) indeed.
The Lax potential in (4.13) is obtained from the general expression
in (2.32) by an additional \lq gauge transformation' by the field
$\exp [-q_{1\over 2}]$, which made the final result simpler.
Of course, for the above analysis we have to restrict ourselves to
a neighbourhood of the identity
where the operators $X_{11}(g_0)$ and $Y_{11}(g_0)$ are invertible.

The choice of the constraints leading to the  effective theory
(4.12) guarantees that the chiral algebra of this theory is
the required one,
${\cal W}_{\cal S}^{\cal G}\times \tilde {\cal W}_{\cal S}^{\cal G}$,
and thus one should be able to express the
${\cal W}$-currents in terms of the local fields in the action.
To this first we recall that in Section 3.1 we have given
an algorithm for constructing the gauge invariant
differential polynomials $W(J)$.
The point we wish to make is that the expression of
the gauge invariant object $W(J)$
in terms of the local fields in (4.12)
is simply $W(\partial_+b\, b^{-1} +T(b))$, where $b$ is given by (4.9).
Applying the reasoning of [40,18] to the present case,
this follows since the function $W$ is form-invariant
under any gauge transformation of its argument, and
the quantity $(\partial_+b\, b^{-1}+T(b))$
is obtained by a (non-chiral) gauge transformation from $J$,
namely by the gauge transformation defined by the field
$a^{-1}\in e^\Gamma$, see equations (2.31-2).
(In analogy, when considering a right moving
${\cal W}$-current one gauge transforms
the argument $\tilde J$ by the field $c\in e^{\tilde \Gamma}$.)
We can in principle compute the object $T(b)$,
as explained in the above, and thus we have an algorithm for finding
the formulae of the $W$'s in terms of
the local fields $g_0$ and $q_{\pm {1\over 2}}$.

The conformal symmetry of the effective theory (4.12)
is determined by the left and right Virasoro densities
$L_{M_0}(J)$ and $L_{-M_0}(\tilde J)$, which survive the reduction.
To see this conformal symmetry  explicitly,
it is useful to extract the {\it Liouville
field} $\phi$ by means of the decomposition
$g_0=e^{\phi M_0}\cdot \hat g_0$, where $\hat g_0$ contains the
generators from ${\cal G}_0$ orthogonal to $M_0$.
One can easily rewrite the action in terms of the new variables
and then its conformal symmetry  becomes manifest since
$e^{\phi}$ is of conformal weight $(1,1)$,
$\hat g_0$ is conformal scalar, and the fields
$q_{\pm {1\over 2}}$ have conformal weights $({1\over 2},0)$
and $(0,{1\over 2})$, respectively.
This assignment of the conformal weights can be established
in a number of ways, one can for example derive it from the
corresponding conformal symmetry transformation of the WZNW
field $g$ in the gauged WZNW theory, see eq. (5.30).
We also note that the action (4.12) can be made generally covariant
and thereby our generalized Toda theory can be re-interpreted
as a theory of two-dimensional gravity
since $\phi$ becomes the gravitational Liouville mode [14].

We would like to point out the relationship between the generalized
Toda theory  given by (4.12) and certain non-linear integrable equations
which have been associated to the half-integral $sl(2)$ subalgebras
of the simple Lie algebras by Leznov and Saveliev, by using a different
method. (See, e.g., equation (1.24) in the review paper
in {\it J. Sov. Math.} referred to in [3].)
To this we note that, in the half-integral case, one can
also consider that WZNW reduction
which is defined by imposing
the left and right constraints corresponding
to the subalgebras ${\cal G}_{\geq 1}$ and ${\cal G}_{\leq -1}$
of $\Gamma$ and $\tilde \Gamma$ in (4.7).
In fact, the Lax potential of the effective field theory corresponding to this
WZNW reduction coincides with the Lax potential postulated by Leznov and
Saveliev to set up their theory.
Thus, in a sense, their theory lies between the WZNW
theory and our generalized Toda theory which has been obtained by
imposing a larger set of first class KM constraints.
This means that the theory  given by (4.12)
can also be regarded as a reduction of their theory.

There is a certain freedom in constructing a field theory
possessing the required chiral algebra ${\cal W}^{\cal G}_{\cal S}$,
for example, one has a freedom of choice in the
halving procedure used here to set up the gauge algebra.
The theories in (4.12) obtained
by using different halvings in
equation (3.41) have their chiral algebras in common, but
it is not quite obvious if these theories  are always
completely equivalent local Lagrangean field theories or not.
We have not investigated this \lq equivalence problem'
in general.

A special case of this problem arises from the fact that
one can expect that in some cases the theory in
(4.12) is equivalent to one of the form (4.3).
This is certainly so in those cases
when for the half-integral $sl(2)$
of $M_0$ and $M_\pm$ one can find an integral grading operator $H$ such that:
$\rm (i)$
$[H\,,\,M_\pm ]=\pm M_\pm$,
$\rm (ii)$
${\cal P}_{1\over 2}+{\cal G}_{\geq 1}={\cal G}^H_{\geq 1}$,
$\rm (iii)$
${\cal P}_{-{1\over 2}}+{\cal G}_{\leq -1}={\cal G}^H_{\leq -1}$,
$\rm (iv)$
${\cal Q}_{-{1\over 2}}+{\cal G}_0
+{\cal Q}_{1\over 2}={\cal G}^H_0$,
where one uses  the $M_0$ grading and
the $H$-grading on the left- and on the right hand sides of these
conditions, respectively.
By definition,
we call the halving
${\cal G}_{1\over 2}={\cal P}_{1\over 2}+{\cal Q}_{1\over 2}$
an $H${\it -compatible halving}
if these conditions are met.
(We note in passing that an $sl(2)$ which allows for an
$H$-compatible halving is automatically an $H$-compatible $sl(2)$
in the sense defined in Section 3.3, but, as  shown in Appendix C,
 not every $H$-compatible $sl(2)$ allows for an
$H$-compatible halving.)
Those generalized Toda theories in (4.12) which have been
obtained by using $H$-compatible halvings in the
WZNW reduction
can be rewritten in the simpler
form (4.3) by means of a renaming of the variables, since in this case
the relevant first class constraints are in the overlap of the ones
which have been considered for the integral gradings and for the
half-integral $sl(2)$'s to derive the respective theories.
Since the form of the action in (4.3) is much simpler than the one in
(4.12), it appears
important to know the list of those $sl(2)$ embeddings which allow for
an $H$-compatible halving,
i.e., for which conditions ${\rm (i)}\ldots {\rm (iv)}$
can be satisfied with some
integral grading operator $H$ and  halving.
We study this group theoretic question
for the $sl(2)$ subalgebras of the maximally non-compact
real forms of the classical Lie algebras in Appendix C.
We show that the existence of an $H$-compatible halving is
 a very restrictive condition
on the half-integral $sl(2)$ subalgebras of
the symplectic and
orthogonal Lie algebras, where such a halving  exists only
for the special $sl(2)$
embeddings listed at the end of Appendix C.
In contrast, it turns out that for
${\cal G}=sl(n,R)$
an $H$-compatible halving can be found for every $sl(2)$ subalgebra,
since in this case one can construct  such a halving
by proceeding similarly as we did in Section 3.4 (see (3.68)).
This means that
in the case of ${\cal G}=sl(n,R)$  any chiral algebra
${\cal W}_{\cal S}^{\cal G}$ can be realized
in a generalized Toda theory associated to an integral grading.

It is interesting to observe that those theories which can
be alternatively written in both forms (4.3) and (4.12) allow for
several conformal structures.
This is so since in this case at least two different Virasoro densities,
namely $L_H$ and $L_{M_0}$, survive the WZNW reduction.

\vskip 1.2truecm

\noindent
{\bf 4.3. Two examples of generalized Toda theories}
\vskip 0.8truecm

We wish to illustrate here the general construction of the previous section
by  working out two examples.
First we shall describe a
generalized Toda theory associated to the
highest root $sl(2)$ of $sl(n+2,R)$.
This is a half-integral $sl(2)$ embedding, but, as  we shall see
explicitly, the theory (4.12) can in this case be
recasted in the form (4.3), since the corresponding halving  is $H$-compatible.
We note that the ${\cal W}$-algebras defined by these
$sl(2)$ embeddings have been investigated before
by using auxiliary fields in [29].
It is perhaps worth stressing that our method does not require
the use of auxiliary fields when reducing the WZNW theory
to the generalized Toda theories which possess these ${\cal W}$-algebras
as their symmetry algebras, see also Section 5.3.
According to the group theoretic analysis  in  Appendix C,
the simplest case when  a ${\cal W}_{\cal S}^{\cal G}$-algebra defined
by a half-integral $sl(2)$ embedding
cannot be realized
in a theory of the type (4.3) is
the case of ${\cal G}=sp(4,R)$.
As our second example, we shall
elaborate on the generalized Toda theory in (4.12)
which realizes
the ${\cal W}$-algebra belonging to the highest root  $sl(2)$ of $sp(4,R)$.

\vskip 1.2truecm

\par\noindent
{\bf i)  Highest root} $sl(2)$ {\bf of} $sl(n+2,R)$\par\noindent
\vskip 0.8truecm
In the usual basis where the Cartan subalgebra consists of diagonal
matrices, the
$sl(2)$ subalgebra ${\cal S}$ is generated by the elements
$$
M_0={1\over 2}\pmatrix{1&\cdots&0\cr 0&0_n&0\cr0&\cdots&-1\cr}
\qquad\hbox{and}\qquad
M_+=M_-^t=\pmatrix{0&\cdots&1\cr 0&0_n&0\cr 0&\cdots&0\cr}.
\eqno(4.23)
$$
Note that here and below dots mean $0$'s in the entries of
the various matrices.
The adjoint of $sl(n+2)$ decomposes into one triplet, $2n$ doublets
and $n^2$ singlets under this ${\cal S}$. It is convenient to parametrize
the general element, $g_0$, of the little group of $M_0$
as
$$
g_0=e^{\phi M_0}\cdot e^{\psi T}\cdot \pmatrix{1&\dots&0\cr 0&\tilde g_0&0\cr
0&\cdots&1\cr},\quad\hbox{where}\quad
T={1\over 2+n}\pmatrix{n&\cdots&0\cr 0&-2I_n&0\cr 0&\cdots&n\cr}
\eqno(4.24)
$$
is trace orthogonal to $M_0$ and $\tilde g_0$ is from $sl(n)$.
We note that  $T$ and $M_0$ generate the centre of the corresponding
subalgebra, ${\cal G}_0$.
We consider the halving of ${\cal G}_{\pm{1\over 2}}$ which is defined by
the subspaces
${\cal P}_{\pm {1\over 2}}$  and ${\cal Q}_{\pm {1\over 2}}$
consisting
of matrices of the following form:
$$\eqalign{
p_{{1\over 2}}&=\pmatrix{0& p^t&0\cr 0&0_n&0\cr 0&\cdots&0\cr} ,\qquad
q_{{1\over 2}}=\pmatrix{0&\cdots&0\cr 0&0_n& q\cr 0&\cdots&0\cr} , \cr
p_{-{1\over 2}}
   &=\pmatrix{0&\cdots&0\cr \tilde p&0_n&0\cr 0&\cdots&0\cr} ,\qquad
q_{-{1\over 2}}=\pmatrix{0&\cdots&0\cr 0&0_n&0\cr 0&\tilde q^{\,t}&0\cr},\cr}
\eqno(4.25)$$
where $q$ and $\tilde p$ are $n$-dimensional column vectors
and $p^t$ and $\tilde q{\,^t}$ are $n$-dimensional row vectors, respectively.
One sees
that the ${\cal P}$ and ${\cal Q}$ subspaces of
${\cal G}_{\pm{1\over 2}}$ are invariant under the adjoint
action of $g_0$, which means that the block-matrices in (4.10) and
(4.11) are diagonal, and thus
$\eta_{\pm {1\over 2}}=[M_\pm,q_{\mp {1\over 2}}]$.
One can also verify that
$X_{11}=e^{-{1\over 2} \phi -\psi} \tilde g_0$, and that using this
the effective action (4.12) can  be written as follows:
$$\eqalign{
I_{\rm eff}(g_0,q_{{1\over 2}},q_{-{1\over 2}})=S_{\rm
WZ}(g_0)-\int d^2x\,\Big[e^{\phi}
-&e^{-{1\over 2}\phi+\psi}\;
(\partial_+ \tilde q)^t\cdot \tilde g_0^{-1} \cdot (\partial_-q)\cr
+&e^{{1\over 2} \phi+\psi}\;
\tilde q^t\cdot \tilde g_0^{-1} \cdot q \Big],\cr}
\eqno(4.26)$$
where dot means usual matrix multiplication.
With respect to the conformal structure defined by $M_0$,
$e^\phi$ has weights $(1,1)$,
the fields
$q$ and $\tilde q$ have half-integer weights $({1\over 2},0)$ and
$(0,{1\over 2})$, respectively,
$\psi$ and  $\tilde g_0$ are conformal
scalars.
In particular, we see that $\phi$
is the Liouville mode with respect to this conformal structure.
\par

In fact, the halving considered
in (4.25)  can be written like the one in (3.68),
by using the integral grading operator $H$ given explicitly as
$$
H=M_0+{1\over 2}T={1\over n+2}\;\pmatrix{n+1&0\cr 0&-I_{n+1}\cr}.
\eqno(4.27)$$
It is an $H${\it-compatible halving} as one
can  verify that it satisfies  the conditions ${\rm (i)}\ldots {\rm (iv)}$
mentioned at the end of Section 4.2, see also Appendix C.
It follows that our reduced WZNW theory can also be regarded
as a generalized Toda theory associated with the integral grading
$H$.  In other words,
it is possible to identify the effective action
(4.26) as a special case of the one in (4.3).
To see this in concrete terms,
it is convenient to
 parametrize the little group of $H$ as
$$
b=\exp(q_{1\over 2}) \cdot g_0 \cdot \exp(q_{-{1\over 2}}),\qquad
\hbox{where}\qquad g_0= e^{\Phi H} \cdot e^{\xi S} \cdot
\pmatrix{1&\cdots&0\cr 0& \tilde g_0&0\cr 0&\cdots&1\cr},
\eqno(4.28)
$$
and $S=M_0-({{n+2}\over{2n}})T$ is trace orthogonal to $H$.
It is easy to check that by inserting this decomposition
 into the effective action (4.3)
and using the Polyakov-Wiegmann identity one recovers indeed the effective
action (4.26), with
$$
\phi=\Phi+\xi\qquad\hbox{and}\qquad \psi={1\over 2}\Phi-{2+n\over 2n}\xi.
\eqno(4.29)$$
The conformal structure
defined by $H$ is different from the one defined by $M_0$.
In fact, with respect to the former conformal
structure $\Phi$ is the Liouville mode
and  all other fields, including $q$ and $\tilde q$,
are conformal scalars. \par

\vskip 1.2truecm\par

\noindent
{\bf ii) Highest root} $sl(2)$ {\bf of} $sp(4,R)$\par\noindent
\vskip 0.8truecm

We use the convention when
the symplectic
matrices have the form
$$
g=\pmatrix{A&B\cr C&-A^t\cr},\qquad
{\rm where}\qquad  B=B^t\,,\ C=C^t\,,
\eqno(4.30)$$
and the Cartan subalgebra is diagonal.
The $sl(2)$ subalgebra ${\cal S}$  corresponding to the highest root of
$sp(4,R)$ is generated by the matrices
$$
M_0={1\over 2} (e_{11} - e_{33}),\quad
M_+=e_{13}\, ,\quad {\rm and}\quad M_-=e_{31}\ ,
\eqno(4.31)$$
where $e_{ij}$ denotes  the elementary $4\times 4$ matrix containing a
single $1$ in the $ij$-position.
The adjoint of $sp(4)$ branches into $\underline{3}+2\cdot
\underline {2}+3\cdot \underline{1}$ under ${\cal S}$. The three singlets
generate an $sl(2)$ subalgebra different from ${\cal S}$, so that the little
group of $M_0$ is $GL(1)\times SL(2)$. $GL(1)$ is generated by
$M_0$ itself and the corresponding field is the Liouville mode.
Using
usual Gauss-parameters for the $SL(2)$, we can parametrize the
little group of $M_0$ as
$$
g_0=e^{\phi M_0}
\pmatrix{1&0&0&0\cr 0&e^\psi+\alpha\beta e^{-\psi}&0&\alpha e^{-\psi}\cr
0&0&1&0\cr 0&\beta e^{-\psi}&0&e^{-\psi}\cr}.
\eqno(4.32)$$
We decompose the ${\cal G}_{\pm{1\over 2}}$ subspaces (spanned
by the two doublets) into their ${\cal P}$ and ${\cal Q}$
parts as follows
$$
p_{{1\over 2}}+q_{{1\over 2}}=
\pmatrix{0&p&0&q\cr 0&0&q&0\cr0&0&0&0\cr 0&0&-p&0\cr},\qquad
p_{-{1\over 2}}+q_{-{1\over 2}}=
\pmatrix{0&0&0&0\cr \tilde p&0&0&0\cr0&\tilde q&0&-\tilde p\cr
\tilde q&0&0&0\cr}.
\eqno(4.33)$$
Now the little group, or more precisely the $SL(2)$ generated
by the three singlets, mixes the ${\cal P}$ and ${\cal Q}$ subspaces
of ${\cal G}_{-{1\over 2}}$ so that the matrices $X_{ij}$ and
$Y_{ij}$ in (4.10) and (4.11) possess off-diagonal elements:
$$
X_{ij}=e^{-{1\over 2}\phi}\pmatrix{e^\psi+\alpha\beta e^{-\psi}&
\alpha e^{-\psi}\cr \beta e^{-\psi}&e^{-\psi}\cr},\qquad
Y_{ij}=X_{ji}.
\eqno(4.34)$$
Inserting this into (4.12) yields the following effective action:
$$\eqalign{
I_{\rm eff}^{\cal S}(g_0,q,\tilde q)=&S_{\rm WZ}(g_0)-\int d^2x\,
 \Biggl[e^\phi -2e^{-{1\over 2} \phi-\psi}
(\partial_-q)\cdot (\partial_+\tilde q)\cr
&+2e^{{1\over 2}\phi}\,{\big(\tilde q+e^{-{1\over 2}\phi-\psi}
\beta\partial_-q\big) \cdot\big(q+e^{-{1\over 2}\phi-\psi}\alpha\partial_+
\tilde q\big)\over
 e^\psi+\alpha\beta e^{-\psi}}\Biggl] ,
\cr}
\eqno(4.35)$$
for the Liouville mode $\phi$, the conformal scalars $\psi,\;\alpha,\;
\beta$ and the fields $q$, $\tilde q$ with weights $({1\over 2},0)$
and $(0,{1\over 2})$, respectively.\par
It is easy to see directly from its formula that it is impossible
to obtain  the above action
 as a special case of (4.3).
Indeed, if the expression in (4.35) was obtained from (4.3) then
the non-derivative term $\sim \tilde q\,q(e^\psi+\alpha\beta e^{-\psi})^{-1}$
could only be gotten from the second term in (4.3), but, since $g_0$
and $b$ are matrices of unit determinant, this
 term could never produce the
denominator in the non-derivative term in (4.35).

\vfill\eject

\def\d{\delta}

\def\db{\delta_{\rm B}}
\def\bc{\bar c_+}

\def\pa{\partial}
\def\i{\int\!d^2x\,}

\def\t{{\rm Tr}\,}

\def\la{\langle}
\def\ra{\rangle}
\def\di{{\rm \dim}\,}

\centerline {\bf 5. Quantum framework for WZNW reductions}
\vskip 0.8truecm

In this chapter we study
the quantum version of the WZNW reduction
by using the path-integral formalism and also re-examine
some of the classical aspects discussed in the previous chapters.
We first show that the
configuration space path-integral
of the constrained WZNW theory can be realized
by the gauged WZNW theory of Section 2.2.
We then point out that the
effective action of the reduced theory, (2.40),
can be derived by integrating out
the gauge fields in
a convenient gauge, the physical gauge,
in which the gauge degrees of freedom are
frozen.
A nontrivial feature of the
quantum theory may appear in the
path-integral measure.  We shall find that
for
the generalized Toda theories associated with
integral gradings
the effective measure takes the form determined from the symplectic
structure of the reduced theory.
This means that
in this case the quantum Hamiltonian reduction results in the
quantization of the reduced classical theory; in other words,
the two procedures, the reduction and the quantization, commute.
We shall also exhibit the
${\cal W}$-symmetry of the effective
action for this example.
By using the gauged WZNW theory, we can
construct the BRST formalism for the WZNW reduction
in the general case.
For conformally invariant reductions, this allows for computing the
corresponding Virasoro centre explicitly.
In particular,
we derive here a nice formula for the Virasoro centre of
${\cal W}_{\cal S}^{\cal G}$
for an arbitrary $sl(2)$ embedding.
We shall verify that our result agrees with
the one obtained in [16], in spite of the apparent
difference in the structure of the constraints.

\vskip 1.2truecm

\noindent{\bf 5.1. Path-integral for constrained WZNW theory}
\vskip 0.8truecm

In this section we wish to set up the path-integral formalism for
the constrained WZNW theory.  For this, we recall that
classically the reduced theory has been obtained
by imposing a set of
first-class constraints in the {\it Hamiltonian
formalism}.  Thus what we should do is to write down the
path-integral of the WZNW theory first in phase space
with the constraints implemented and then find the corresponding
configuration
space expression.
The phase space path-integral
can formally be defined once the canonical
variables
of the theory are specified.
A practical way to find the canonical variables is the
following [41].
Let us start from the WZNW action $S_{\rm WZ}(g)$ in
(1.2) and parametrize
the group element $g \in G$
in some arbitrary way,
$g = g(\xi)$.
We shall regard the parameters $\xi^a$,
$a=1,...,{\rm dim}\,G$, as
the canonical coordinates in the theory.
To find the canonical momenta, we introduce
the 2-form
${\cal A} = {1\over 2}  {\cal A}_{ab}(\xi)\, d\xi^a d\xi^b $
to rewrite the Wess-Zumino term as
$$
       {1\over 3} \t (dg\, g^{-1})^3 = d {\cal A}.
\eqno(5.1)
$$
The 2-form ${\cal A}$
is well-defined only locally on $G$, since the Wess-Zumino 3-form is closed
but not exact.
Fortunately we
do not need to specify ${\cal A}$ explicitly below.
We next define $N_{ab}(\xi)$ by
$$
\Bigl( {{\pa g}\over{\pa\xi^a}} \Bigr)\, g^{-1}  =  N_{ab}(\xi) T^b,
\eqno(5.2)
$$
where $T^b$ are the generators of ${\cal G}$.
The matrix $N$ is easily shown to be non-singular, det$N \ne 0$.
Upon writing $S_{\rm WZ}(g) = \i {\cal L}(g)$,
the canonical momentum conjugate to $\xi^a$ is found to be
$$
\Pi_a = {{\pa {\cal L}}\over{\pa \pa_0 \xi^a}}
      = \kappa \Bigl[ N_{ab}(\xi)(\pa_0 g\,g^{-1})^b -
        {\cal A}_{ab}(\xi)\pa_1 \xi^b \Bigr].
\eqno(5.3)
$$
The Hamiltonian of the WZNW theory is then given by
$H = \int dx^1 {\cal H}$ with
$$
{\cal H} = \Pi_a \pa_0 \xi^a - {\cal L}
         = {1 \over 2\kappa} \t \bigl[ P^2
           + (\kappa \pa_1 g\,g^{-1})^2 \bigr],
\eqno(5.4)
$$
where
$$
P^a = (N^{-1})^{ab}(\Pi_b + \kappa {\cal A}_{bc}\pa_1\xi^c).
\eqno(5.5)
$$
Since $P = \kappa \pa_0 g\,g^{-1}$ in the original variables,
the Hamiltonian density takes the usual
Sugawara form as expected.

Classically, the constrained WZNW theory has been defined
as the usual WZNW theory with its KM phase space
reduced by the set of
constraints given by (2.16), which in the canonical variables read
$$
\eqalign
{
\phi_i &= \la \gamma_i, P + \kappa (\pa_1 g\,g^{-1} -  M) \ra = 0, \cr
\tilde\phi_i &= \la \tilde \gamma_i,
     g^{-1} P g - \kappa (g^{-1} \pa_1 g +  \tilde M) \ra = 0,
}
\eqno(5.6)
$$
with the bases $\gamma_i \in \Gamma$,
$\tilde\gamma_i \in \tilde\Gamma$.
As in Section 2.2, no relationship is assumed here
between the two subalgebras, $\Gamma$ and $\tilde\Gamma$.
Now we write down the phase space path-integral for the constrained
WZNW theory.  According to Faddeev's prescription [42]
it is defined as
$$
\eqalignno{
Z = \int d\Pi d\xi \,\delta (\phi)\delta(\tilde\phi)
     \delta(\chi) &\delta(\tilde\chi) \det \vert \{\phi,\chi\} \vert
     \det \vert \{\tilde\phi,\tilde\chi\} \vert               \cr
     &\times \exp \Bigl(\,i \i ( \Pi_a \pa_0 \xi^a - {\cal H}) \Bigr),
&(5.7)
}
$$
where we implement the first class constraints
by inserting $\delta(\phi)$ and $\delta(\tilde\phi)$
in the path-integral.
The $\delta$-functions of $\chi$ and $\tilde\chi$ refer to
gauge fixing conditions corresponding to
the constraints, $\phi$ and $\tilde\phi$,
which act as generators of gauge
symmetries.  By introducing
Lagrange-multiplier fields, $A_- = A_-^i\gamma_i$ and
$A_+ = A_+^i \tilde\gamma_i$,
(5.7) can be written as
$$
\eqalignno{
Z = \int d\Pi & d\xi dA_+ dA_- \delta(\chi) \delta(\tilde\chi)
                            \det \vert \{\phi,\chi\} \vert
                       \det \vert \{\tilde\phi,\tilde\chi\} \vert      \cr
              &\times \exp \Bigl(\,i\i
          \bigl[ \t (\Pi \pa_0 \xi +  A_-\phi +
                     A_+\tilde\phi) - {\cal H} \bigr] \Bigr).
&(5.8)
}
$$
By changing the momentum variable
from $\Pi_a$ to $P^a$ in (5.5), the
measure acquires a determinant factor,
$d\Pi = dP \det N$, and
the integrand of the exponent in (5.8) becomes
$$
\eqalignno{
  \t &(\Pi \pa_0 \xi +  A_-\phi +
                    A_+\tilde\phi) - {\cal H}        \cr
       &=  \kappa \t \Bigl[ -{1\over 2} \bigl( {1\over\kappa}
             P \bigr)^2
            + {1\over\kappa} P (A_- + gA_+g^{-1} +  \pa_0 g\,g^{-1})
            -  N^{-1} {\cal A}\, \pa_1 \xi
              (\pa_0 g\,g^{-1})                             \cr
          &  \qquad -{1\over 2} (\pa_1 g\,g^{-1})^2
           + A_- (\pa_1 g\,g^{-1} - M)
           - A_+ (g^{-1} \pa_1 g + \tilde M) \Bigr].
&(5.9)
}
$$
Since the matrix
$N(\xi)$ is independent of $P$, we can easily perform the
integration over $P$
provided that the remaining $\delta$-functions and the determinant factors
are also $P$-independent.  We can choose the gauge fixing
conditions, $\chi$ and $\tilde\chi$, so that this is true.
(For example, the physical gauge which we will
choose in the next
section fulfills this demand.)
Then we end up with the following formula of the
configuration space path-integral:
$$
Z = \int d\xi \det N \, dA_+ dA_- \delta (\chi) \delta(\tilde\chi)
         \det \vert \{\phi,\chi\} \vert
         \det \vert \{\tilde\phi,\tilde\chi\} \vert \,
               e^{i I(g,A_-,A_+)},
\eqno(5.10)
$$
where $I(g,A_-,A_+)$ is
the gauged WZNW action (2.18).
We note that the measure for the coordinates
in this path-integral is the
invariant Haar measure,
$$
d\mu(g) = \prod_a d\xi^a \det N = \prod_a (dg\,g^{-1})^a.
\eqno(5.11)
$$
This is a consequence of the fact that
the phase space measure in (5.7) is invariant
under canonical transformations to which the group transformations
belong.

The above formula for the configuration space path-integral
means that the gauged WZNW theory provides the Lagrangian realization
of the Hamiltonian reduction, which we have already seen on the
basis of a classical argument in Section 2.2.

\vskip 1.2truecm

\noindent{\bf 5.2. Effective theory in the physical gauge}
\vskip 0.8truecm

Having seen how the constrained WZNW theory is realized as
the gauged WZNW theory, we next
discuss the effective theory which arises
when we eliminate
all the unphysical degrees of freedom
in a particularly convenient gauge, the physical gauge.
We shall rederive,
in the path-integral
formalism,
the effective action which
appeared in the classical context earlier in this paper.
For this purpose,
within this section we restrict our attention to the
left-right dual reductions considered in Section 2.3.
It, however, should be noted that
this restriction is not absolutely necessary
to get an effective action by the method given below.
In this respect, it is also worth noting that
Polyakov's 2-dimensional gravity action in the
light-cone gauge can be regarded
as an effective action in a non-dual reduction, which is obtained
by imposing a constraint only on the left-current
for $G = SL(2)$ [43,12].  We will not pursue the non-dual cases here.

To eliminate
all the unphysical gauge
degrees of freedom, we simply gauge them away
from $g$, i.e., we gauge fix the Gauss decomposed
$g$ in (2.25) into the form
$$
g = abc \rightarrow b.
\eqno(5.12)
$$
More specifically, with the
parametrization
$a(x) = \exp{[\sigma_i(x)\gamma_i]}$, $c(x) =
\exp{[\tilde\sigma_i(x)\tilde\gamma_i]}$ we define the
{\it physical gauge} by
$$
\chi_i = \sigma_i = 0, \qquad \tilde\chi_i = \tilde\sigma_i = 0.
\eqno(5.13)
$$
We here note that for this gauge the
determinant factors in (5.8) are actually constants.
Now the effective action is obtained by
performing the $A_\pm$ integrations in (5.10).
The integration of $A_-$ gives rise to the delta-function,
$$
\prod_i \delta \Bigl(\la \gamma_i, bA_+ b^{-1} +
        \pa_+b\,b^{-1} - M\ra\Bigr),
\eqno(5.14)
$$
with $\gamma_i \in \Gamma$ normalized by the duality condition
(2.22).
One then notices that the delta-function (5.14)
implies exactly condition (2.29) with $\pa_+c\,c^{-1}$ replaced
by $A_+$.  Hence, with the help of the matrix $V_{ij}(b)$ in (2.27)
and $T(b)$ in (2.29), it can be rewritten as
$$
(\det V)^{-1} \,\delta \Bigl(A_+ - b^{-1} T(b) b\Bigr).
\eqno(5.15)
$$
Finally, the integration of $A_+$ yields
$$
Z = \int d\mu_{\rm eff}(b)\, e^{I_{\rm eff}(b)},
\eqno(5.16)
$$
where $I_{\rm eff}(b)$ is the effective action
(2.40)\footnote*
{
Actually, the effective action always takes the form (2.40) if one
restricts the WZNW field to be of the form $g = abc$ with
$a \in e^\Gamma$, $c\in e^{\tilde\Gamma}$ and $b$ such that
$V_{ij}(b)$ is invertible.  The duality between $\Gamma$
and $\tilde\Gamma$ is not necessary but
can be used to ensure this technical assumption.
},
and $d\mu_{\rm eff}(b)$ is the effective measure given by
$$
d\mu_{\rm eff}(b) = (\det V)^{-1}\,
                      d\mu(g)\delta(\sigma)\delta(\tilde\sigma)
                  = (\det V)^{-1}\,{{d\mu(g)}\over{d\sigma d\tilde\sigma }}
                    \biggr\vert_{\sigma = \tilde\sigma = 0}.
\eqno(5.17)
$$

Of course, as far as the effective action is concerned, the
path-integral approach should give the same result as the classical
one, because the integration
of the gauge fields is Gaussian and hence
equivalent to the classical elimination
of the gauge fields by their field equations.
However, a non-trivial feature
may arise at the quantum level when the
effective path-integral measure (5.17)
is taken into account.
Let us examine the effective measure in the simple case where
the space
${\cal B} = (\Gamma + \tilde\Gamma)^{\bot}$,  with  which
$b \in e^{\cal B}$,
forms a subalgebra of ${\cal G}$ satisfying (2.34),
and thus the effective action in (5.16) simplifies to
$$
I_{\rm eff}(b) = S_{\rm WZ}(b) - \kappa \i \la b \tilde M  b^{-1},
M\ra.
\eqno(5.18)
$$
In this case,
the 1-form appearing
in the measure $d\mu(g)$ of (5.11),
$$
dg\,g^{-1} = da\, a^{-1} + a (db\, b^{-1}) a^{-1}
 + a b (dc\, c^{-1}) b^{-1} a^{-1},
\eqno (5.19)
$$
turns out, in the physical gauge, to be
$$
dg\,g^{-1} \bigr
\vert_{\sigma = \tilde\sigma = 0} =  \gamma_i d\sigma_i + db\,b^{-1}
+ V_{ij}(b)\tilde \gamma_i d\tilde\sigma_j. \eqno(5.20)
$$
As a result, the determinant factor in (5.17) is cancelled by
the one coming from (5.20), and
the effective measure
admits a simple form:
$$
d\mu_{\rm eff}(b) = db\, b^{-1}.
\eqno(5.21)
$$
The point is that this is exactly the measure
which is determined from the symplectic structure
of the effective theory (5.18) obtained by
the {\it classical} Hamiltonian reduction.
This tells us that in
this case the {\it quantum}
Hamiltonian reduction results in the
quantization of the reduced classical theory.
In particular, since the above assumption for ${\cal B}$ is satisfied
for the generalized Toda theories associated with integral gradings,
we conclude that
these generalized Toda theories
are equivalent to
the corresponding
constrained (gauged) WZNW theories even at the quantum level,
 i.e., including the measure.
This result has been established before in the special case of the
standard Toda theory (1.1) in [44], where
the measure
$d\mu_{\rm eff}(b)$ is simply given by $\prod_i d\varphi^i$.

We end this section by noting that it is
not clear whether the
measure determined from the symplectic structure of
the reduced classical theory
is identical to the effective measure (5.17) in general.
In the general case both measures in question
could become quite involved and thus one would need some geometric
argument to see if they are identical or not.

\vskip 1.5truecm

\def\d{\delta}

\def\pa{\partial}
\def\i{\int\!d^2x\,}

\def\t{{\rm Tr}\,}

\def\la{\langle}
\def\ra{\rangle}

\noindent
{\bf 5.3.} {\bf The} ${\cal W}${\bf -symmetry of the generalized Toda
action} $I_{\rm eff}^H(b)$
\vskip 0.8truecm

In the previous section we have seen the quantum equivalence of
the  generalized Toda theories given by (4.3)
and the corresponding constrained WZNW theories.
It follows from their WZNW origin that the generalized Toda theories
possess conserved ${\cal W}$-currents.
It is thus natural to expect that their effective actions,
$I_{\rm eff}^H$ in (4.3) and $I^{\cal S}_{\rm eff}$ in (4.12),
allow for {\it symmetry transformations
yielding the ${\cal W}$-currents as the corresponding
Noether currents}.
We  demonstrate below that this is indeed the case
on the example of the theories associated with integral gradings,
when the action takes a simple form.
We however believe that there are symmetries
of the effective action
 corresponding to the conserved chiral currents inherited
from the KM algebra for any reduced WZNW theory.

Let us consider a gauge invariant differential polynomial
$W(J)$ in the constrained WZNW theory giving rise to the effective
theory  described by the action  in (4.3).
In terms of the generalized Toda field $b(x)$,
this conserved ${\cal W}$-current is given by the differential polynomial
$$
W_{\rm eff}(\beta )=W(M+\beta ),
\qquad {\rm where}\qquad
\beta  \equiv \partial_+b\, b^{-1}.
\eqno(5.22)$$
This equality [34,15] holds because the constrained current
 $J$ and $(M+\beta )$ (which is, incidentally, just the Lax potential
${\cal A}_+^H$ in (4.4)) are related by a gauge transformation,
as we have seen.
By choosing some test function $f(x^+)$,
we now associate to $W_{\rm eff}(\beta )$ the following
transformation of the field $b(x)$:
$$
\d_W b(y)=\Bigl[ \i f(x^+) {{\d W_{\rm eff}(x)}\over{\d \beta
(y)}}\Bigr]
\cdot b(y)\ ,
\eqno(5.23)
$$
and we wish to show that $\delta_W b$ is a symmetry of the action
$I_{\rm eff}^H(b)$.
Before proving this,
we notice, by combining the definition in (5.23) with (5.22),
 that
 $(\delta_W b) b^{-1}$ is  a polynomial expression in $f$,
$\beta$ and their $\partial_+$-derivatives up to some finite order.

We start the proof by noting that the change of the action under an
arbitrary variation $\delta b$ is given by the formula
$$\eqalign{
\delta I_{\rm eff}^H(b)& =
- \int d^2y\,
\langle \delta b\,b^{-1}(y)\,,\,
b(y) {{\delta I^H_{\rm eff}}\over {\delta b(y)}} \rangle \cr
&= - \int d^2y\,
\langle \delta b\,b^{-1}(y)\,,\,  \partial_- \beta (y) +[b(y) \tilde M
b^{-1}(y)
    , M]
\rangle\,. \cr}
\eqno(5.24)$$
In the next step,
we use the field equation to replace $\partial_-\beta$ by $-[b\tilde M b^{-1} ,
   M]$
in the obvious equality
$$
\pa_- W_{\rm eff}(x) = \int d^2y\,
\la {{\d W_{\rm eff}(x)}\over{\d \beta (y)}},\pa_- \beta (y) \ra ,
\eqno(5.25)
$$
and then, from the fact that $\partial_-W_{\rm eff}=0$ on-shell,
we obtain the following identity:
$$
\int d^2y \,
\la {{\d W_{\rm eff}(x)}\over{\d \beta (y)}},[b(y)\tilde M b^{-1}(y), M] \ra
=0\,, \eqno(5.26)$$
Of course, the previous argument only implies that (5.26) holds  on-shell.
However,  we now make the crucial observation that
 (5.26) is  an {\it
off-shell identity}, i.e., it is valid for any field $b(x)$
not only for the solutions of the field equation.
This follows by noticing that the object in (5.26) is
a local expression in $b(x)$ containing only $x^+$-derivatives.
In fact,  any such object which vanishes on-shell has to vanish also off-shell,
because one can find
solutions of the field equation for which
the $x^+$-dependence of
the field $b$ is prescribed
in an arbitrary way at an arbitrarily chosen fixed value of $x^-$.

By using the above observation, it is easy to  show that
$\delta_W b$ in (5.23) is indeed a symmetry of the action.
First, simply inserting (5.23) into (5.24), we have
$$
\delta_W I_{\rm eff}^H(b) =
- \int d^2x\, f(x^+) \int d^2y\,
\langle {{\d W_{\rm eff}(x)}\over{\d \beta (y)}}\,,\,
 \partial_- \beta (y) +[b(y) \tilde M b^{-1}(y) , M]\rangle .
\eqno(5.27)$$
We then rewrite this equation as
$$
\delta_W I_{\rm eff}^H(b) =- \i f(x^+) \pa_- W_{\rm eff}(x) ,
\eqno(5.28)$$
with the aid of the identities (5.26) and (5.25).
This then proves that
$$
\delta_W I_{\rm eff}^H(b) =0\,,
\eqno(5.29)$$
since the integrand in (5.28) is a total derivative, thanks
to $\partial_- f =0$.
One can also see,
from equation (5.23), that $W_{\rm eff}$ is the Noether charge density
corresponding to the symmetry transformation  $\delta_W b$
of $I_{\rm eff}^H(b)$.

\vskip 1.2truecm

\noindent{\bf 5.4. BRST formalism for WZNW reductions}
\vskip 0.8truecm

Since the constrained WZNW theory can be regarded as the gauged
WZNW theory (2.18), one is naturally led to construct the BRST formalism
for the theory as
a basis for quantization.  Below we discuss the BRST formalism
based on the gauge symmetry (2.19) and thus
return to the general situation of Section 5.1 where no
relationship between the two subalgebras,
$\Gamma$ and $\tilde\Gamma$, is supposed.

Prior to the construction
we here note how the conformal symmetry
is realized in the gauged WZNW theory when there is an
operator $H$ satisfying the condition (2.13).
(For simplicity, in what follows we discuss the symmetry
associated to the left-moving sector.)
In fact, with such $H$ and a chiral test function $f^+(x^+)$
one can define the following transformation,
$$
\eqalign{
\d g &= f^+\pa_+g + \pa_+ f^+ Hg,    \cr
\d A_- &= f^+\pa_+A_- + \pa_+ f^+ [H, A_-],   \cr
\d A_+ &= f^+\pa_+A_+ +\pa_+f^+ A_+,
}
\eqno(5.30)
$$
which leaves the gauged WZNW action
$I(g,A_-,A_+)$ invariant.
This corresponds exactly to
the conformal transformation
in the constrained WZNW theory
generated by the
Virasoro density $L_{H}$ in (2.10), as can be confirmed by observing
that (5.30) implies the conformal action (2.11) for the current
with $f(x^+)=f^+(x^+)$.
We shall derive
later the Virasoro density as the Noether
charge density in the BRST system.

Turning to the construction of the BRST formalism,
we first choose the space $\Gamma^* \subset {\cal G}$
which is dual to $\Gamma$ with
respect to the Cartan-Killing form (and similarly $\tilde\Gamma^*$
dual to $\tilde\Gamma$).
Following the standard procedure [45] we
introduce two sets of ghost, anti-ghost and
Nakanishi-Lautrup fields,
$\{c \in \Gamma,\,\,\bc, B_+ \in \Gamma^*\}$ and
$\{b \in \tilde\Gamma,\,\, \bar b_-, B_- \in \tilde\Gamma^*\}$.
The BRST transformation corresponding to
the (left-sector of the) local gauge transformation
(2.19) is given by
$$
\eqalign{
\db g &= -cg,                                 \cr
\db A_- &= D_- c,                                 \cr
\db c &= -c^2,                                      \cr
}\qquad
\eqalign{
\db \bc &= i B_+,                                       \cr
\db  B_+ &= 0,                                            \cr
\db ({\rm others}) &= 0,
}
\eqno(5.31)
$$
with
$D_\pm = \pa_\pm \mp [A_\pm,\quad]$.
After defining the BRST transformation
$\bar\delta_{\rm B}$ for
the right-sector in an analogous way, we
write the BRST action by adding a
gauge fixing term
and a ghost term to the gauged action,
$$
I_{\rm BRST} = I(g, A_-, A_+) + I_{\rm gf} + I_{\rm ghost}.
\eqno(5.32)
$$
The additional terms can be constructed by
the manifestly BRST invariant expression,
$$
\eqalignno{
I_{\rm gf} + &I_{\rm ghost}
    = -i\kappa (\d_{\rm B} + \bar\d_{\rm B})
        \i \bigl( \la \bc, A_-\ra + \la\bar b_-, A_+\ra\bigr) \cr
    &= \kappa \i
        \bigr( \la B_+, A_-\ra + \la B_-, A_+\ra +
    i \la \bc, D_- c\ra + i\la \bar b_-, D_+ b\ra\bigr),
&(5.33)
}
$$
where we have chosen the gauge fixing conditions as $A_\pm = 0$.
Then the path-integral for the BRST system is given by
$$
Z = \int d\mu(g)\, dA_+ dA_- dc\, d\bc
  db\, d\bar b_- dB_+ dB_-
         \, e^{i I_{\rm BRST}},
\eqno(5.34)
$$
which, upon integration of the ghosts and the
Nakanishi-Lautrup fields, reduces to (5.10).
(Strictly speaking, for this we have to generalize the gauge fixing
conditions in (5.10) to be dependent on the gauge fields.)
By this construction
the nilpotency, $\db^2 = 0$,
and the BRST invariance of the action, $\db I_{\rm BRST} = 0$, are
easily checked.

It is, however, convenient to deal with
the  simplified BRST theory obtained by
performing the
trivial integrations of $A_\pm$ and $B_\pm$ in (5.34),
$$
I_{\rm BRST} (g,c,\bc,b,\bar b_-)
 = S_{\rm WZ}(g) + i \kappa \i \bigl( \la \bc, \pa_- c\ra +
    \la \bar b_-, \pa_+ b\ra\bigr).
\eqno(5.35)
$$
We note that
this effective BRST theory is not merely a sum of a free WZNW sector
and free ghost sector as it appears, but rather it consists of
the two
interrelated sectors in the
physical space specified by the BRST charge defined below.
At this stage the BRST transformation
which leaves the simplified BRST action (5.35) invariant reads
$$
\eqalign{
\db g &= -cg,                                 \cr
\db c &= -c^2,                                      \cr
}\qquad
\eqalign{
\db \bc &=
- \pi_{\Gamma^*}\Bigl[ i(\pa_+g\,g^{-1} - M_-) +
                           (c\bc + \bc c) \Bigr], \cr
\db ({\rm others}) &= 0, \cr
}
\eqno(5.36)
$$
where $\pi_{\Gamma^*} = \sum_i \vert \gamma^*_i \ra \la \gamma_i
\vert$ is the projection operator onto the dual space $\Gamma^*$
with the normalized bases, $\la \gamma_i, \gamma^*_j \ra = \d_{ij}$.
     From the associated
conserved Noether current, $\pa_- j^{\rm B}_+ = 0$,
the BRST charge $Q_{\rm B}$
is defined to be
$$
Q_{\rm B} = \int dx^+ j^{\rm B}_+(x)
= \int dx^+ \la c, \pa_+ g\,g^{-1} -M - c \bc\ra.
\eqno(5.37)
$$
The physical space is then specified by the condition,
$$
Q_{\rm B} \vert {\rm phys} \rangle = 0.
\eqno(5.38)
$$
In the simple case of the WZNW reduction which leads
to the standard Toda theory,
the BRST charge (5.37) agrees with the one discussed earlier [46].

In the case where there is an $H$ operator
which guarantees the conformal invariance,
the BRST system also has the corresponding
conformal symmetry,
$$
\eqalign{
\d g &= f^+\pa_+g + \pa_+ f^+ Hg,    \cr
\d c &= f^+\pa_+c + \pa_+ f^+ [H, c],   \cr
\d \bc &= f^+\pa_+ \bc + \pa_+ f^+ (\bc + [H, \bc]),   \cr
}\qquad
\eqalign{
\d b &= f^+\pa_+b, \cr
\d \bar b_- &= f^+\pa_+ \bar b_-, \cr
            &  \cr
}
\eqno(5.39)
$$
inherited from the one (5.30) in the gauged WZNW theory.
If the $H$ operator further provides a grading, one
finds from (5.39) that the currents of grade
$-h$ have the (left-) conformal weight
$1 - h$, except the $H$-component,
which is not a primary field.
Similarly, the ghosts $c$, $\bc$ of grade
$h$, $-h$ have the conformal weight
$h$, $1 - h$, respectively, whereas the ghosts $b$, $\bar b$ are
conformal scalars.
Now we define the total Virasoro density operator $L_{\rm tot}$
from the associated Noether current,
$\pa_- j^{\rm C}_+ = 0$, by
$$
\int dx^+ j^{\rm C}_+(x) = {1\over\kappa} \int dx^+ f^+(x^+) L_{\rm tot}(x).
\eqno(5.40)
$$
The (on-shell) expression is found to be the sum of the two parts,
$L_{\rm tot} = L_H + L_{\rm ghost}$, where $L_H$ is indeed the
Virasoro operator (2.10) for the WZNW part, and
$$
L_{\rm ghost} = i\kappa \bigl(
\la \bc, \pa_+ c\ra + \pa_+ \la H, c\bc + \bc c \ra \bigr),
\eqno(5.41)
$$
is the part for the ghosts.
The conformal invariance of the BRST charge, $\d Q_{\rm B} = 0$, or
equivalently, the BRST invariance of
the total conformal charge, $\db L_{\rm tot} = 0$, are readily
confirmed.

Let us find the Virasoro centre of our BRST system.  The total
Virasoro centre $c_{\rm tot}$ is given by the sum of the two contributions,
$c$ from the WZNW part and $c_{\rm ghost}$ from the ghost one.  The Viraso
centre from $L_H$ is given by
$$
c = {{k\, {\rm dim}\,{\cal G}}\over{k + g}} - 12 k \la H, H \ra,
\eqno(5.42)
$$
where $k$ is the level of the KM algebra and
$g$ is the dual Coxeter number.
On the other hand, the ghosts contribute to the Virasoro centre by
the usual formula,
$$
c_{\rm ghost} = -2 \sum_{\Gamma}\bigl[ 1 + 6 h (h - 1) \bigr],
\eqno(5.43)
$$
where the summation is performed over the
eigenvectors of ${\rm ad}_H$ in the subalgebra $\Gamma$.
(One can confirm (5.43) by performing the
operator product expansion with $L_{\rm ghost}$ in (5.41).)

\vskip 1.2truecm

\def\d{\delta}

\def\db{\delta_{\rm B}}
\def\bc{\bar c_+}

\def\pa{\partial}
\def\i{\int\!d^2x\,}

\def\t{{\rm Tr}\,}

\def\la{\langle}
\def\ra{\rangle}
\def\di{{\rm \dim}\,}

\noindent
{\bf 5.5. The Virasoro centre in two examples}
\vskip 0.8truecm

By elaborating on the general result of the previous section,
we here derive explicit formulas for
the total Virasoro  centre in
two important special cases of the WZNW reduction.
\bigskip
\noindent
{\bf i) The generalized Toda theory} $I_{\rm eff}^H(b)$
\smallskip
In this case
the summation
in (5.43) is  over the eigenstates of ${\rm ad}_H$ with eigenvalues
$h>0$, since $\Gamma ={\cal G}^H_{>0}$.
We can establish a concise formula for $c_{\rm tot}$, (5.46) below,
by using the following group theoretic facts.

First, we  can assume that the grading operator $H\in {\cal G}$
is from the Cartan subalgebra of the complex simple Lie
algebra ${\cal G}_c$ containing ${\cal G}$.
Second, the scalar product
$\langle \ ,\ \rangle$ defines a natural isomorphism
between the Cartan subalgebra and the space of roots,
and we introduce the notation $\vec \delta$ for the vector
in root space corresponding to $H$ under this isomorphism.
More concretely, this means that we  set $H=\sum_i \delta_i H_i$
by using  an orthonormal Cartan basis,
$\langle H_i , H_j\rangle =\delta_{ij}$.
Third, we recall the {\it strange formula} of Freudenthal-${\rm de\, Vries}$
[47],
which (by taking into account the normalization of
$\langle \ ,\ \rangle$
and the duality between the root space and the Cartan subalgebra)
reads
$$
\di {\cal G} = {12\over g}  \vert \vec \rho \vert^2 \ ,
\eqno(5.44)
$$
where $\vec \rho$ is the Weyl vector, given by half the sum of
the positive roots.
Fourth, we choose the simple positive
roots in such a way that the corresponding step operators,
which are in general in ${\cal G}_c$ and not in ${\cal G}$,
have non-negative grades with respect to $H$.

By using the above conventions,
it is straightforward to obtain the following expressions
$$
\eqalign{
\sum_{h > 0} 1 & = {\rm dim}\, \Gamma =
              {1\over 2}({\rm dim}\,{\cal G} - \di {\cal G}_0^H), \qquad
\sum_{h > 0} h  = 2 (\vec\rho \cdot \vec \d), \cr
\sum_{h > 0} h^2 & = {1\over 2} {\rm tr}\,
        ({\rm ad}_H)^2 = g \la H, H \ra = g \vert \vec \d \vert^2 ,
}
\eqno(5.45)
$$
for the corresponding terms in (5.43).
Substituting these into (5.43) and also (5.44) into (5.42),

one can finally establish
the following nice formula of the total Virasoro
centre [14]:
$$
c_{\rm tot} = c + c_{\rm ghost}
            = \di {\cal G}_0^H - 12 \Bigl\vert \sqrt{k + g}\,\vec \d
               - { 1 \over {\sqrt{k + g}}}\,\vec \rho \Bigr\vert^2.
\eqno(5.46)
$$
In particular, in the case of the reduction leading to
the standard Toda theory (1.1)
the result (5.46) is consistent  with the one directly obtained in
the reduced theory [8,10].

\bigskip
\noindent
{\bf ii) The ${\cal W}_{\cal S}^{\cal G}$-algebra for half-integral
$sl(2)$ embeddings}
\smallskip

For $sl(2)$ embeddings the role of the $H$
is played
by $M_0$ and in the half-integral case we have
$\Gamma = {\cal G}_{\ge 1} + {\cal P}_{1\over2}
= {\cal G}_{> 0} - {\cal Q}_{1\over2}$.
It follows that the value of the  total Virasoro centre
can now be obtained by substracting the contribution of the
\lq missing ghosts'  corresponding to ${\cal Q}_{1\over 2}$,
which is ${1\over 2} \di {\cal G}_{1\over 2}$, from the
expression in (5.46).
We thus obtain that in this case
$$
c_{\rm tot} =
              N_t - {1\over 2} N_s - 12 \Bigl\vert \sqrt{k + g}\,\vec \d
               - { 1 \over {\sqrt{k + g}}}\,\vec \rho \Bigr\vert^2 ,
\eqno(5.47a)
$$
where
$$
N_t =\di {\cal G}_0\ ,
\qquad {\rm and}\qquad
N_s = \di {\cal G}_{1\over 2}\ ,
\eqno(5.47b)$$
are the number of tensor and spinor multiplets in the decomposition
of the adjoint of ${\cal G}$ under the $sl(2)$ subalgebra ${\cal S}$,
respectively.
We note that, as proven by Dynkin [39],
it is possible to choose  a system of positive simple roots
so that the grade of the corresponding step operators is from the set
$\{ 0, {1\over 2}, 1\}$,  and that
$\vec \delta$ is
(${1\over 2} \times$) the so called
{\it defining vector} of the $sl(2)$ embedding in  Dynkin's terminology.

As has been mentioned in Section 3.3,
Bais {\it et al} [16] (see also [29])
studied a similar reduction
of the KM algebra for half-integral $sl(2)$ embeddings where
all the current components corresponding to  ${\cal G}_{>0}$
are constrained from the very beginning.
In their system,
the constraints (3.49) of ${\cal G}_{1\over 2}$, being inevitably
second-class, are
modified into
first-class by
introducing an auxiliary field to each
constraint of ${\cal G}_{1\over2}$.  Accordingly, the auxiliary fields
give rise to the extra contribution
$-{1\over2}\di {\cal G}_{1\over2}$ in the total Virasoro centre.
It is clear that adding this  to the sum of the WZNW and ghost parts
(which is of the form (5.46) with $M_0$ substituted for $H$),
renders the total Virasoro centre of their system
identical to that of our system, given by  (5.47).
This result is natural if we recall
the fact that their reduced phase space (after complete
gauge fixing) is actually identical to ours.
It is obvious that our method,
which is based on purely first-class KM constraints
and does not require auxiliary fields, provides  a simpler
way to reach the identical reduced theory.

\vfill\eject

\centerline{\bf 6. Discussion}

\vskip 0.8truecm

The main purpose of this paper has been to study the general structure
of the Hamiltonian reductions of the WZNW theory.
Considering the number of interesting examples resulting from the reduction,
this problem appears important for the theory of two-dimensional
integrable systems and in particular for conformal field theory.

Our most important result perhaps is that we established the
gauged WZNW setting of the Hamiltonian reduction by first class constraints
in full generality.
It was then used here to set up the BRST formalism in the
general case, and for obtaining the effective actions for the left-right
dual reductions.
We hope that the general framework we set up will be
useful for further studies of this very rich problem.

The other major concern of the paper has been to investigate the
${\cal W}$-algebras and their field theoretic realizations arising
from the WZNW reduction.
We found first class KM constraints leading to the
${\cal W}_{\cal S}^{\cal G}$-algebras which allowed us to construct
generalized Toda theories realizing these interesting extended
conformal algebras.
We believe that the $sl(2)$-embeddings underlying the
${\cal W}_{\cal S}^{\cal G}$-algebras
are to play an important
organizing role in general for
understanding the structure, especially the
primary field content, of the conformally invariant reduced KM
systems.  We illustrated this idea
by showing that the $W_n^l$-algebras
are nothing but further
reductions of ${\cal W}_{\cal S}^{\cal G}$-algebras
belonging to particular $sl(2)$-embeddings (see also [37]).
In our study of ${\cal W}$-algebras we employed two (apparently) new
methods, which are likely to have a wider range of applicability than
what we exploited here.
The first is {\it the method of symplectic halving} whereby  we constructed
purely first class KM constraint for the ${\cal W}_{\cal S}^{\cal G}$
as well as for the $W_n^l$-algebras.
The second is what we call {\it the} $sl(2)${\it-method},
which can be summarized by saying that  if one has conformally invariant
first class constraints given by some $(\Gamma , M_-)$  with $M_-$
nilpotent, then one should build the $sl(2)$  containing $M_-$ and
try to analyse the system in terms of this $sl(2)$.
We used this method to investigate, in the non-degenerate case,
the generalized Toda sytems belonging to integral gradings, and also
to provide
the ${\cal W}_{\cal S}^{\cal G}$-interpretation of the
$W_n^l$-algebras.

We wish to remark here that,
as far as we know,
the technical problem concerning the inequivalence of
those ${\cal W}_{\cal S}^{\cal G}$-algebras which belong to
group theoretically inequivalent $sl(2)$ embeddings
has not been tackled yet.

It is well known [22] that the standard ${\cal W}$-algebras
can be identified as the second Poisson bracket structure of
the generalized KdV hierarchies of Drinfeld-Sokolov [5].
A  similar relationship between
${\cal W}$-algebras and  KdV type hierarchies
has been established very recently in more general cases
[28,48,49].
In particular, the  $W_n^l$-algebras have been
related to the so called fractional KdV hierarchies.
It would be clearly worthwhile to study in general the relationship
between the generalized Drinfeld-Sokolov hierachies of [48]
and the ${\cal W}_{\cal S}^{\cal G}$-algebras together with their
further reductions, see also [16,17].

We gave a general local analysis of the effective
theories arising in the left-right dual case of
the reduction, and investigated in particular
the generalized Toda theories obtained by the reduction in some detail.
In the case of the generalized Toda
theories associated with the integral gradings
we exhibited the way in which the ${\cal W}$-symmetry operates
as an ordinary symmety of the action, and demonstrated that
the quantum Hamiltonian reduction is consistent
with the canonical quantization of the reduced classical theory.
It would be nice to have the analogous problems
under control also in more general cases.
In our analysis we restricted
the considerations to Gauss-decomposable fields.
The fact that the Gauss decomposition may break down can
introduce apparent singularities in the local description
of the effective theories, but
the WZNW description is inherently global and remains
valid for non Gauss-decomposable fields as well [12,13].
It is hence an interesting problem to further analyze the  global
(topological) aspects of the phase space of the reduced WZNW
theories.

We should also note that it is possible to remove the
technical assumption of left-right duality.
In particular, the study of purely chiral WZNW reductions
could be of importance, as they are likely to give
natural generalizations of Polyakov's $2 {\rm d}$
gravity action [43,12].

In this paper we assumed the existence of a gauge invariant
Virasoro density $L_H$, of the form given by (2.10),
for obtaining conformally invariant reductions.
Based on this assumption, we came to realize that,
when $H$ provides a grading of $\Gamma$ and $M$,
the $sl(2)$ built out of $M=M_-$ plays an important role.
However, the example of
Appendix A
indicates that there is
another class of conformally
invariant reductions where the form of the surviving Virasoro density
is different from that of an $L_H$.
The study of this novel way of preserving
the conformal invariance may open up a
new perspective on conformal reductions of the WZNW
theory
as well as on ${\cal W}$-algebras.

There are many further interesting questions
related to the Hamiltonian reductions of the WZNW theory,
which we could not mention in this paper.
We hope to be able to present those
in future publications.

\bigskip
\medskip
\noindent
{\bf Acknowledgement.}
We wish to thank B. Spence for a suggestion which
has been crucial for us for understanding the ${\cal W}$-symmetry
of the Toda action.

\bigskip
\medskip
\noindent
{\bf Note added.}
After finishing this paper, there appeared a
preprint [50] also advocating the importance of $sl(2)$ structures
in classifying ${\cal W}$-algebras.

\vfill\eject

\def\d{\delta}
\def\a{\alpha}

\def\pa{\partial}
\def\i{\int\!d^2x\,}
\def\\{{\hfil\break }}

\def\gi{g^{-1}}

\def\rr{\rangle}
\def\ll{\langle}
\def\half{{1 \over 2}}

\centerline{\bf Appendix A: A solvable but not nilpotent gauge algebra}

\vskip 0.8truecm

In all the cases of the reduction we considered in Chapters 3 and 4,
the gauge algebra $\Gamma$ was a graded nilpotent subalgebra of
${\cal G}$.
On the other hand, we have seen in Section 2.1 that the
first-classness of the constraints imply that $\Gamma$ is solvable.
We want here to discuss a constrained WZNW model for which the gauge
algebra is {\it solvable but not nilpotent}.
Interestingly enough, it turns out that in this example no $H$
satisfying (2.13)
exists which would render the constraints conformally invariant.
However, conformal invariance can still be maintained, showing
clearly that the existence of such an $H$ is
only a sufficient but not a
necessary condition.

We choose the Lie algebra ${\cal G}$ to be $sl(3,R)$ and the gauge
algebra $\Gamma$ as generated by the following three generators
$$\gamma_1=E_{\a_1}=\pmatrix{0&1&0 \cr 0&0&0 \cr 0&0&0 \cr},
\quad \gamma_2=E_{\a_1+\a_2}=\pmatrix{0&0&1 \cr 0&0&0 \cr 0&0&0},
\eqno(A.1a) $$
$$\gamma_3={1 \over \sqrt{3}} (2H_1+H_2)+{1 \over
2} (E_{\a_2}-E_{-\a_2})=
\pmatrix{{1 \over \sqrt{3}}&0&0 \cr
0&-{1 \over {2\sqrt{3}}}&{1 \over 2} \cr
0&-{1 \over 2}&-{1 \over {2\sqrt{3}}} \cr},
\eqno(A.1b)
$$
\noindent where the Cartan-Weyl generators are normalized by
$[H_i,E_{\pm\a_i}]=\pm E_{\pm\a_i}$ and
$[E_{\a_i},E_{-\a_i}]=2H_i$, for the simple positive roots
$\alpha_i$.
Note that, being diagonalizable over the complex numbers,
$\gamma_3$ is not a nilpotent operator.
The algebra of $\Gamma$ is
$$
[\gamma_1,\gamma_2]=0, \quad
[\gamma_1,\gamma_3]=-{\sqrt{3} \over 2}\gamma_1 + {1 \over
2}\gamma_2, \quad
[\gamma_2,\gamma_3]=-{1 \over 2}\gamma_1 -{\sqrt{3} \over 2}\gamma_2
.  \eqno(A.2) $$
\noindent
It is easy to verify that $\Gamma$ is
a solvable, not-nilpotent Lie algebra.
It qualifies as a gauge algebra since
${\rm Tr\,}(\gamma_i\, \gamma_j)=0$.

It is readily checked that the spaces $\Gamma^\perp$ and
$[\Gamma , \Gamma]^\perp$
are given by
$$\eqalign{
&\Gamma^{\perp} =
{\rm span}\{H_2,E_{\a_1},E_{\a_1+\a_2},2H_1+\sqrt{3}E_{\a_2},
2H_1-\sqrt{3}E_{-\a_2}\}, \cr
&[\Gamma,\Gamma]^{\perp} = {\rm span}\{H_1,H_2,E_{\a_1},
E_{\a_1+\a_2},E_{\a_2},E_{-\a_2}\}. \cr}
\eqno(A.3) $$
\noindent Thus
$[\Gamma,\Gamma]^{\perp}/\Gamma^{\perp}$,
which is the space of the $M$'s leading to first class constraints, is
one-dimensional, and we can take
$$ M =\mu Y \equiv
{\mu \over \sqrt{3}} (4H_1+2H_2) =
{\mu \over \sqrt{3}}\pmatrix{2 & 0 & 0 \cr 0 & -1 & 0 \cr 0 & 0 & -1 \cr}\,
\eqno(A.4) $$
without loss of generality.

The next question is the conformal invariance. As discussed in Section 2.1,
a sufficient condition for conformal invariance
is provided by the existence of a
(modified) Virasoro density
$L_{H}=L_{KM}-\pa_x\langle H,J(x) \rangle$
weakly commuting with the constraints.
For this to work, the generator
$H$ must satisfy the three conditions
in (2.13).
However,  it is an easy matter to show that those
conditions are contradictory in the present case,
and therefore no such $H$ exists.

The above analysis can also be carried out for the simpler gauge
algebra spanned by $\gamma_3$ only.
This gauge algebra is obviously nilpotent, since it is Abelian.
Nevertheless, the previous conclusions remain: There exists no
$H$ which would render the first class constraints
conformally invariant, for any $M\neq 0$ from
$[\Gamma,\Gamma]^{\perp}/\Gamma^{\perp}$.
This shows the importance of the
gauge generators being nilpotent operators, rather than the gauge
algebra being nilpotent.
It would be interesting to know whether there is always an $H$
satisfying (2.13) for gauge algebras consisting of
nilpotent operators.

\medskip
Although there is no $H$ such that the constraints are preserved by
$L_H$, we can nevertheless  construct another Virasoro density
$\Lambda$ which does preserve the constraints. It is given by
$$ \Lambda(x) = L_{KM}(x) - {\mu} \ll \gamma_3^t,J(x) \rr.
\eqno(A.5) $$
For $M$ given in (A.4), the constraints read
$$ \ll \gamma_1,J(x) \rr = \ll \gamma_2,J(x) \rr = 0 \,,\qquad
\ll \gamma_3,J(x) \rr = \mu\,,
\eqno(A.6) $$
and are checked to weakly commute with $\Lambda$:
$\{\Lambda(x),\ll \gamma_i,J(y)\rr \} \approx 0$ on the constraint
surface (A.6). (Note that, when going from $L_{KM}$ to $\Lambda$, we
have not changed the conformal central charge, which is classically
zero.) Therefore we expect the reduced theory to be
invariant under the conformal transformation
generated by $\Lambda$ being its Noether charge density.
We now proceed to show that it is indeed the case.
Before doing this, we display the form of $\Lambda$ on the
constraint surface:
$$ \Lambda(x) = T_1^2(x) + T_2^2(x)\,, \eqno(A.7a) $$
$$ T_1=\half \ll E_{\a_2}+E_{-\a_2},J \rr\;, \qquad
   T_2=\ll H_2,J \rr. \eqno(A.7b) $$

\medskip
Following the analysis of Section 2.3, we take the left and right gauge
algebras to be dual to each other ($\langle \gamma_i,\tilde\gamma_j
\rangle = \delta_{ij}$)
$$ \Gamma = {\rm span}\{\gamma_1,\gamma_2,\gamma_3\}, \quad
\tilde \Gamma = {\rm span}\{\tilde\gamma_1,\tilde\gamma_2,\tilde
\gamma_3\} = {\rm span}\{\gamma_1^t,\gamma_2^t,\gamma_3^t\},
\eqno(A.8) $$
\noindent and consider $M=\mu Y$ and $\tilde M=\nu Y^t=\nu Y$.
We write the $SL(3,R)$ group elements as $g=a \cdot b \cdot c$, with
$a \in \exp\Gamma$, $b \in \exp{{\cal H}}$ and $c \in \exp{\tilde
\Gamma}$, with ${\cal H}={\rm span}\{Y,H_2\}$ the Cartan subalgebra.
We did not conform to the general prescription
given in Section 2.3, which required to write
$g = abc$ with $b\in \exp {\cal B}$ for
a space ${\cal B}$ complementary to
$\Gamma +\tilde \Gamma $ in ${\cal G}$, eqs.(2.25-26).
Had we done
that, the resulting effective action would have looked much more
complicated. Here, we simply take a set of coordinates in which the
action looks simple.

The reduction yields
an effective theory for the group-valued field $b$, of
which the effective action is given by (2.40) with (2.29b).
Using the parametrization $b=\exp{(\a Y)} \cdot
\exp{(2\beta H_2)}$, the explicit form of the effective action is
$$ I_{{\rm eff}}(\a,\beta) =  \i \Bigl\{
\partial_+\a \partial_-\a + \partial_+\beta \partial_-\beta
-{(\pa_+\a - \mu)(\pa_- \a - \nu) \over {\cosh ^2 {\beta}}}
\Bigr\}.
\eqno(A.9) $$
By inspection, we see that this effective action is going to be
conformally invariant if the field $\beta$ is a scalar, and if the
transformation of $\a$ is such that $\mu - \partial_+\a$ and $\nu -
\partial_-\a$ are (1,0) and (0,1) vectors respectively. It implies
that, under a conformal transformation $x^{\pm} \longrightarrow
x^{\pm}-f^{\pm}(x^{\pm})$, the fields $\a$ and $\beta$ transform as
$$\eqalign{
& \d\a = f^+\,(\partial_+\a - \mu) +
         f^-\,(\partial_-\a - \nu), \cr
& \d\beta =  f^+\,\partial_+\beta +
             f^-\,\partial_-\beta. \cr }
\eqno(A.10)
$$

We now want to show our previous claim: the action (A.9) is
conformally invariant under the conserved Virasoro density
$\Lambda(x)$, which reproduces the $f^+$-transformations (A.10)
by Poisson brackets. (The $f^-$-transformations could also be
realized by constructing the corresponding Virasoro density
$\tilde\Lambda$ in the right-handed sector in
a similar way.)  For this, we first note that
in terms of the reduced variables $\a$ and $\beta$ the two current
components $T_1$ and $T_2$ of (A.7b) read
$$ T_1=- (\mu - \partial_+\a)\,\tanh {\beta}\,,\quad {\rm and}
\quad T_2= \partial_+\beta.  \eqno(A.11) $$
These expressions can be obtained as follows. Writing $g=a \cdot b
\cdot c$ and using the constraints (2.29b), the constrained current reads
$$ J = a[T(b) + \partial_+b \cdot b^{-1}]
a^{-1} + \partial_+a \cdot a^{-1}, \eqno(A.12) $$
with $T(b)$ given by (2.29).
Although neither $T_1$ nor $T_2$ is gauge invariant, the quantity we
want to compute, $\Lambda(x)$, is gauge invariant. As a result, it
cannot depend on the gauge variables contained in $a$. Hence we can
just as well put $a=1$ in (A.12). Doing that, the definitions (A.7b)
yield (A.11). We thus find the following expression for $\Lambda$:
$$ \Lambda =  (\mu - \partial_+\a)^2 \, \tanh ^2 {\beta} +
 (\partial_+\beta)^2.   \eqno(A.13)  $$
It is an easy matter to show, by using the field equations obtained
from the action (A.9),
$$
\eqalign{
   &\sinh^2 \beta\,\pa_+\pa_-\alpha + \tanh \beta\,
    \bigl[ \pa_+\beta (\pa_-\a - \nu) + \pa_-\beta (\pa_+\a - \mu) \bigr]
     = 0\, , \cr
   &\cosh^2 \beta\,\pa_+\pa_-\beta - \tanh \beta\,
      (\pa_-\a - \nu)(\pa_+\a - \mu)
     = 0\, ,
}
\eqno(A.14)
$$
that $\Lambda$ is indeed chiral,
satisfying
$$
\partial_-\Lambda = 0\,.  \eqno(A.15)
$$
Moreover one also checks the following Poisson brackets
$$
\eqalign{
& \{\Lambda(x),\a(y)\} = -(\partial_+\a - \mu) \, \d
(x^1-y^1)\,, \cr
& \{\Lambda(x),\beta(y)\} = -(\partial_+\beta) \, \d
(x^1-y^1),\cr}  \eqno(A.16)
$$
which reproduce the transformations (A.10).
Thus the density $\Lambda$ features all what
is expected from the Noether charge density associated with
the conformal symmetry.

Finally, we present here for completeness the general solution of the
equations of motion (A.14).
Along the lines of Section 2.3, it can be obtained as follows:
$$ \eqalign{
  \a  &= (\eta_L + \eta_R)
         + \tan^{-1} \Bigl[{{\sinh (\theta_L - \theta_R)} \over {
 \sinh (\theta_L + \theta_R)}} \tan (\lambda_L - \rho_R) \Bigr]
        + \mu x^+ + \nu x^-,
\cr
\cosh (2\beta) &=  \cosh (2\theta_L) \cosh (2\theta_R)
    + \sinh (2\theta_L) \sinh (2\theta_R)
      \cos (2(\lambda_L - \rho_R)),
}
\eqno(A.17)
$$
where $\{\eta_L, \lambda_L, \theta_L \}$ and
$\{ \eta_R, \rho_R, \theta_R \}$ are arbitrary functions of $x^+$ and
$x^-$ only, respectively,
and the three functions of each chirality are related by the equations,
$$\pa_+ \eta_L + \pa_+ \lambda_L\, \cosh (2\theta_L) = 0\, ,  \qquad
  \pa_- \eta_R + \pa_- \rho_R\, \cosh (2\theta_R) = 0\, .
\eqno(A.18)
$$

\vfill\eject

\centerline{{\bf Appendix B}: $H${\bf-compatible} $sl(2)$
{\bf and the non-degeneracy condition}}

\vskip 0.8truecm

   Our purpose in this technical appendix is to
analyse the notion of
the $H$-compatible $sl(2)$ subalgebra, which has been
introduced in Section 3.3.
We recall that the $sl(2)$ subalgebra ${\cal S}=\{ M_-,M_0,M_+\}$ of the simple
Lie algebra ${\cal G}$ is called $H$-compatible if $H$ is an integral grading
operator, $[H\,,\,M_\pm ] = \pm M_\pm $,
and $M_\pm$ satisfy  the
non-degeneracy conditions
$$
{\rm Ker}({\rm ad}_{M_\pm})\cap {\cal G}_\mp^H=\{ 0\}.
\eqno(B.1)$$
Note that the second property in this definition
is equivalent to the fact that ${\cal S}$ commutes with
$(H-M_0)$.
We prove here the results stated in Section 3.3, and also
establish an  alternative form  of the non-degeneracy condition,
which will be used in Appendix C.

Let us first consider an arbitrary (not necessarily integral) grading operator
$
   H$
of ${\cal G}$ and  some  non-zero element $M_-$ from ${\cal G}^H_{-1}$.
We wish to show  that to each such pair $(H,M_-)$ there exists
an  $sl(2)$ subalgebra ${\cal S}=\{ M_-,M_0,M_+\}$ for which $M_+ \in {\cal
G}_{
   +1}^H$.
   ommutes
To exhibit the ${\cal S}$-triple in question, we  need the
Jacobson-Morozov theorem, which has already been mentioned in Section 3.3.
In addition, we shall also  use  the following lemma, which can be
found in [33] (Lemma 7 on page 98, attributed to Morozov).

\medskip
\noindent
{\bf Lemma:} Let ${\cal L}$ be a finite-dimensional Lie algebra over
a field of characteristic $0$ and suppose ${\cal L}$ contains
elements $h$ and $e$ such that $[h\,,\,e]=-e$ and
$h\in [{\cal L}\,,\,e]$.
Then there exists an element $f\in {\cal L}$ such that
$$
[h\,,\,f]=f {\quad}{\rm and}{\quad} [f\,,\,e]=2h\ .
\eqno(B.2)$$
\medskip

Turning to the proof,  we first use the Jacobson-Morozov theorem  to find
generators $(m_-,m_0,m_+)$ in ${\cal G}$ completing  $m_-\equiv M_-$
to an $sl(2)$ subalgebra.
We then decompose the elements $m_0$ and $m_+$ into
their components of definite grade, i.e., we write
$$
m_0=\sum_n m_0^n
\quad {\rm and} {\quad}
m_+=\sum_n m_+^n\ ,
\eqno(B.3)$$
where $n$ runs over the spectrum of the grading operator  $H$.
Since $M_-$ is of grade $-1$, it follows from
the $sl(2)$ commutation relations that
$$
[m_0^0\,,\,M_-]=-M_-
\quad {\rm and} \quad
[m_+^1\,,\,M_-]=2m_0^0\ ,
\eqno(B.4)$$
and these relations tell us that  $h=m_0^0$ and $e=M_-$
satisfy the conditions of the above lemma.
Thus there exists an element  $f$ satisfying (B.2), which we can
write as $f=\sum_n f^n$ by using the $H$-grading again.
The proof is finished by verifying that $M_+\equiv f^1$ and
$M_0\equiv m_0^0$ together with $M_-$ span the required
$sl(2)$ subalgebra of ${\cal G}$.

    From now on, let  $H$ be an {\it integral} grading operator.
For an element  $M_\pm$   of grade $\pm 1$, respectively,
the pair $(H,M_\pm)$ is called {\it non-degenerate}
if it satisfies the corresponding condition in (B.1).

We claim that if ${\cal S}=\{ M_-,M_0,M_+\}$ is an $sl(2)$ for which
 the  generators
 $M_\pm$ are from ${\cal G}^H_{\pm 1}$,
then the non-degeracy of the pairs $(H,M_-)$
and $(H,M_+)$ are {\it equivalent statements}.
This will  follow immediately
from the $sl(2)$ structure if we prove that the {\it non-degeneracy
of the pair $(H,M_\pm )$ is equivalent to the following equality}:
$$
{\rm dim\,}{\rm Ker}({\rm ad}_{M_\pm})={\rm dim\,}{\cal G}_0^H\ .
\eqno(B.5)$$

It is enough to prove this latter statement  for a pair
$(H,M_-)$, since then for a pair $(H,M_+)$ it can be obtained
by changing $H$ to $-H$.
To prove this
let us first rearrange the identity
$$
{\rm dim\,}{\cal G}={\rm dim\,}{\rm Ker}({\rm ad}_{M_-})+
{\rm dim\,}[M_-,{\cal G}]
\eqno(B.6)$$
by using the grading as
$$\eqalign{
{\rm dim\,}{\rm Ker}({\rm ad}_{M_-})-{\rm dim\,}{\cal G}_0^H=&
\bigl\{ {\rm dim\,}{\cal G}_+^H-{\rm dim\,}[M_-,{\cal G}^H_+]\bigr\}\cr
&+\bigl\{ {\rm dim\,}{\cal G}_-^H
-{\rm dim\,}[M_-,{\cal G}_0^H +{\cal G}_-^H]\bigr\}\cr}.
\eqno(B.7)$$
Since both terms on the right hand side of this equation are
non-negative, we see that
$$
{\rm dim\,}{\rm Ker}({\rm ad}_{M_-})\geq {\rm dim\,}{\cal G}_0^H\ ,
\eqno(B.8)$$
and equality is achieved here if and only if
$$
{\rm dim\,}{\cal G}^H_+={\rm dim\,}[M_-,{\cal G}_+^H]
\quad {\rm and}\quad
[M_-,{\cal G}^H_0+{\cal G}_-^H] ={\cal G}_-^H\ .
\eqno(B.9)$$
On the other hand, we can show that the two equalities in (B.9) are actually
equivalent to each other.
To see this, let us assume that the second equality in (B.9) is not true.
This is clearly equivalent to
the existence of some non-zero $u\in {\cal G}_+^H$ such that
$\langle u\,,\,[M_-,{\cal G}_0^H+{\cal G}_-^H]\rangle =\{ 0\}$.
By the invariance and the non-degeneracy of the Cartan-Killing form,
this is in turn
equivalent to $[M_-,u]=0$, which means that the first equality
in (B.9) is not true.
By noticing  that the first equality in  (B.9) is just
the non-degeneracy condition for the pair $(H,M_-)$,
we can
conclude that the non-degeneracy condition is indeed equivalent to the
equality in (B.5).

We wish to mention  a consequence of the  results
proven in the above.
To this let us consider a non-degenerate pair $(H,M_-)$.
By our more general result, we know that there exists
such an  $sl(2)$ subalgebra  ${\cal S}=\{ M_-,M_0,M_+\}$
for which  $M_+$ is from ${\cal G}^H_{+ 1}$.
The point to mention is that this ${\cal S}$ is an $H$-compatible
$sl(2)$ subalgebra,
as has already been sated in Section 3.3.
In fact, it is now easy to see that
this follows from the equivalence of (B.1) with (B.5) by taking
into account that the kernels of ${\rm ad}_{M_{\pm}}$ are of equal
dimension by the $sl(2)$ structure.

\vfill\eject

\def\d{\delta}
\def\a{\alpha}

\def\pa{\partial}
\def\i{\int\!d^2x\,}
\def\\{{\hfil\break }}

\def\gi{g^{-1}}

\def\rr{\rangle}
\def\ll{\langle}
\def\half{{1 \over 2}}

\centerline{{\bf Appendix C:} $H${\bf-compatible} $sl(2)$ {\bf embeddings and
halvings}}

\vskip 0.8truecm
In Section 3.3, we showed  that, given a triple
$(\Gamma,M,H)$ satisfying the conditions for first-classness,
conformal invariance and polynomiality
(eqs. (2.6), (2.13) and (3.2-4)), the corresponding ${\cal
W}$-algebra is isomorphic to ${\cal W}_{{\cal S}}^{{\cal G}}$,
provided that $H$ is an integral grading operator. Here ${\cal S}
=\{M_-,M_0,M_+\}$ is some $sl(2)$ subalgebra containing $M_-=M$. A
natural question is what $sl(2)$ subalgebras arise in this
way, or equivalently, given an arbitrary $sl(2)$ subalgebra,
can the resulting ${\cal W}_{{\cal S}}^{{\cal G}}$-algebra be
obtained as the ${\cal W}$-algebra corresponding to the triple
$(\Gamma,M,H)$, for some integral grading operator $H\,$?
Whether this occurs or
not depends only on how the $sl(2)$ is embedded, and it is
therefore a pure group-theoretic question.
According to Section 3.3, the $sl(2)$
subalgebras
having this property are the $H$-compatible ones.
This appendix is
devoted to establishing when a given $sl(2)$ embedding is
$H$-compatible, and if so, what the corresponding $H$ is.

\medskip
The question of an $sl(2)$ being $H$-compatible is very much
related to another one, which was mentioned at the end of Section
4.2. We noted that in some instances, a generalized Toda theory
associated to an $sl(2)$ embedding could as well be regarded as a
Toda theory associated to an integral grading operator $H$.
This means that the effective action
of the theory  is a special case of  both (4.12)
and (4.3) at the same time.
We have seen that this is the  case when the
corresponding  halving is $H$-compatible,
i.e.,  when the Lie algebra decomposition ${\cal G}
= ({\cal G}_{\geq 1} + {\cal P}_{\half}) + ({\cal Q}_{\half}+{\cal
G}_0+{\cal Q}_{-\half}) + ({\cal P}_{-\half}+{\cal G}_{\leq -1})$
(subscripts are $M_0$-grades) can be nicely recasted into ${\cal G}
= {\cal G}_{\geq 1}^H + {\cal G}_0^H + {\cal G}_{\leq -1}^H$.
Our second problem, addressed at the end of the appendix, is to find
the list of those $sl(2)$ subalgebras
which allow for an $H$-compatible halving.
Clearly, an $sl(2)$ subalgebra which possesses an $H$-compatible
halving is also  $H$-compatible in the above sense, but it will turn out
that the converse is not true.

\medskip
Let ${\cal S}=\{M_-,M_0,M_+\}$ be an $sl(2)$ subalgebra embedded
in a maximally non-compact real simple Lie algebra ${\cal G}$. For
the classical algebras $A_l$, $B_l$, $C_l$ and $D_l$,
these real forms are respectively $sl(l+1,R)$, $so(l,l+1,R)$,
$sp(2l,R)$ and $so(l,l,R)$. (We do not consider
the exceptional Lie algebras.)
For ${\cal S}$ to be an $H$-compatible $sl(2)$, one
should find an $H$ in ${\cal G}$ with the following properties:
\medskip \parindent 45pt
\item{1.} ${\rm ad}_H$ is diagonalizable with eigenvalues being integers,
\item{2.} $H-M_0$ must commute with the ${\cal S}$-triple,
\item{3.} ${\rm dim\,} {\rm Ker}({\rm  ad}_H )=
{\rm dim\,} {\rm Ker}({\rm  ad}_{M_\pm})$.
\medskip \parindent 20pt
\noindent
We remark that here the equivalence of relations
(B.1) and (B.5), proven in the previous appendix, has been taken
into account.
Under conditions 1-3, the decomposition

$$\Gamma^{\perp}=[M_-,\Gamma]+{\rm Ker(ad}_{M_+})  \eqno(C.1)$$

\noindent holds, where $\Gamma={\cal G}_{\geq 1}^H$ in the $(\Gamma,M_-,H)$
setting, or $\Gamma={\cal P}_{{1 \over 2}}+{\cal G}_{\geq 1}^{M_0}$
in the $sl(2)$ setting, respectively. (For clarity, note that these
two gauge algebras  are in general not equal.)
As a consequence,  $J_{\rm red}(x)=M_- +j_{\rm red}(x)$ with $j_{\rm red}(x)
\in {\rm Ker(ad}_{M_+})$ is a DS gauge in both settings, and thus
the ${\cal W}$-algebras are the same.

\medskip
In order to answer the question of whether an $sl(2)$ embedding is
$H$-compatible, it is useful to know what these embeddings actually
are. For a
classical {\it complex}  Lie algebra ${\cal G}_c$, this question
has been completely answered by Malcev (and Dynkin for the
exceptional complex Lie algebras) [39]. The result can be
nicely stated in terms of the way the fundamental vector
representation
reduces into irreducible representations of the $sl(2)$:
\medskip
\item{$A_l$ :} the $sl(2)$ reduction of the
 (l+1)-dimensional representation can
be arbitrary,
\item{$B_l$ :} the $(2l+1)$-dimensional representation of
$B_l$ reduces
in such a way that the multiplicity of each $sl(2)$ {\it spinor}
appearing in the reduction is even,
\item{$C_l$ :} the $2l$-dimensional representation of $C_l$ reduces
in such a way that the multiplicity of each $sl(2)$ {\it tensor}
appearing in the reduction is even,
\item{$D_l$ :} same restriction as the $B_l$ series: the spinors
come in pairs.
\medskip
\noindent The above conditions are necessary and sufficient, i.e.,
every possible $sl(2)$  content
satisfying the above requirements
actually occurs for
some $sl(2)$ embedding.
Moreover, for the classical complex Lie algebras,
the way the fundamental reduces
completely specifies the $sl(2)$ subalgebra, up to
automorphisms of the embedding ${\cal G}_c$ [39].

The  above description of the $sl(2)$ embeddings remains valid for
the maximally non-compact classical real Lie algebras,
except the last statement.
First of all, this means that the above restrictions apply
to the possible decompositions of the fundamental under the
$sl(2)$ subalgebras in the real case as well.
It is also obvious that those $sl(2)$ embeddings for
which the content of the fundemantal is different are
inequivalent.
The converse however ceases to be true in the real case
in general:
inequivalent $sl(2)$ subalgebras  can have the same
multiplet content in the fundamental of ${\cal G}$.
The answer to the  problem of
$H$-compatibility  will in fact be provided by looking  more
closely at the decomposition of the fundamental
of ${\cal G}$ under the $sl(2)$ subalgebra in question,
as will be clear below.

\bigskip
As an immediate consequence of condition 2, $H-M_0$ is an $sl(2)$
invariant and can only depend on the value of the Casimir. If,
in the reduction of the fundamental of ${\cal G}$, a
spin $j$ representation occurs with multiplicity $m_j$, the
$sl(2)$ generators $\vec M$ and $H$ can be written
$$ \vec M = \sum_j \vec M^{(j)} \times I_{m_j}, \eqno(C.2{\rm a}) $$
$$ H = M_0 + \sum_j I_{2j+1} \times D(j), \eqno(C.2{\rm b}) $$
\noindent where $I_n$ denotes the unit $n \times n$ matrix, and the
$D(j)$'s are $m_j \times m_j$ diagonal matrices. Hence, within each
irreducible representation of $sl(2)$, $H$ is equal to $M_0$
shifted by a constant. Obviously, this is also true in the adjoint
representation and, in turn, this implies that
ad$_H$ takes the value zero at most once
in each $sl(2)$ multiplet in the adjoint of ${\cal G}$.
    From  condition 3, ad$_H$ must take the value zero exactly once,
i.e., each $sl(2)$ representation must
intersect ${\rm Ker}({\rm  ad}_H)$ exactly once. In
particular, the $sl(2)$ singlets must be ${\rm ad}_H$-eigenvectors with
zero eigenvalue.

\medskip The trivial solution $H = M_0$ exists whenever ad$_{M_0}$
is diagonalizable on the integers, i.e., when the reduction of the
fundamental of ${\cal G}$ is either purely tensorial or purely
spinorial. From now on, we suppose that the reduction involves
both kinds of $sl(2)$ representations.

\bigskip
\medskip
\noindent 1) $A_l$ algebras.

\noindent
The problem for the $A_l$ series is simple to solve since, in this
case, an $H$ {\it always} exists. As a proof, we explicitly give an
$H$ which fulfills all the requirements. In (C.2b), we set

$$D(j) = \cases{\lambda \cdot I_{m_j} & if $j \in N$, \cr
(\lambda + {1 \over 2}) \cdot I_{m_j} & if $j \in N+{1 \over 2}$, \cr}
\eqno(C.3)  $$

\noindent where $\lambda$ is a constant that makes $H$ traceless.
In order to show that the $H$ so defined has the required
properties, we recall that for the $A_l$ algebras, the adjoint
representation is obtained by tensoring the fundamental with its
contragredient. As a result, the roots are the differences of the
weights of the fundamental (up to a singlet) and we have

$$ {\rm ad}_H = {\rm ad}_{M_0} + [D(j_1) - D(j_2)],
\eqno(C.4) $$

\noindent where $j_1$ and $j_2$ are the spins of the states in the
fundamental representation from which a given state in the adjoint
representation is formed. That the conditions 1-3 are satisfied is
obvious from the fact that ad$_H = {\rm ad}_{M_0}$ on tensors and
ad$_H = {\rm ad}_{M_0} \pm {1 \over 2}$ on spinors, with $+{1 \over
2}$ occurring as many times as $-{1 \over 2}$.

It should be pointed out that (C.3) is by no means the only solution.
Since in the product $j_1 \times j_2$, the highest weights
have an $M_0$-eigenvalue at least equal to $\vert j_1-j_2 \vert$,
another solution is given by $D(j)=(\lambda +j) \cdot I_{m_j}$.

\bigskip
\medskip
\noindent 2) $C_l$ algebras.

\noindent
For the symplectic algebras, the adjoint representation is obtained
from the {\it symmetric} product of the fundamental with itself and we
therefore have

$$ {\rm ad}_H = {\rm ad}_{M_0} + [D(j_1)+D(j_2)].
\eqno(C.5)$$

\noindent Since the symmetric product of a tensor with itself produces
a singlet, which must belong to ${\rm Ker}({\rm ad}_H)$, we have $2D(t)=0$ for
every integer $j=t$. Hence in the fundamental representation, $H=M_0$
on tensors. Similarly, the symmetric product of a spinor with itself
always produces a triplet, one member of which must belong to
${\rm Ker}({\rm ad}_H)$. This implies that the diagonal
entries of $2D(s)$ are either $0$
or $\pm 1$, for every half-integer
$j=s$. However $D(s)$ cannot have a zero on the diagonal, because ad$_H$
would not be integral on the representations contained in $s \times t$.
Therefore, in the fundamental, $H=M_0 \pm {1 \over 2}$ on spinors.

Let us now look at the $m_s$ spinor representations of spin $s$, say
$s^1,s^2,\ldots,s^{m_s}$. The product $s^i \times s^j$
of any two of those contains a singlet, and that implies
$D(s^i)+D(s^j)=0$. This equality must hold for any pair of spin $s$
representations, which is impossible unless $m_s \leq 2$.

Let us consider the restriction $g_s$ of the symplectic form
to the spin $s$ representations. The restricted form
is non-degenerate, because the original non-degenerate metric is
block-diagonal with respect to the eigenvalues of the $sl(2)$
Casimir.

If $m_s=1$, then the $H$ given by $M_0 \pm \half \cdot I$ on the
unique spin $s$ representation, should be in the symplectic algebra:
$g_sH + H^tg_s=0$. Since $M_0$ is already symplectic, we require
that the identity be symplectic, which is impossible for a
non-degenerate form. Hence $m_s$ must be 2.

If $m_s=2$, $H-M_0$ and $g_s$ look like (in the basis where $M_0$
and $H$ are diagonal)

$$ H-M_0 = \pm \pmatrix{\half & 0 \cr 0 & -\half \cr}, \qquad
   g_s = \pmatrix{a & b \cr -b^t & c \cr },  \eqno(C.6) $$

\noindent where the blocks $a$ and $c$ are antisymmetric. $H-M_0$ being
symplectic leads to $a=c=0$.

\medskip
To summarize, for an integral $H$ to exist, the $sl(2)$ embedding
must be such that: (i) the multiplicity of any spinor representation
in the fundamental of ${\cal G}$ is 2, (ii) if $(s,s')$ is such
a pair of spinors, they must be the dual of each other with respect to
the symplectic form. If these two conditions are met, then $H$ is
given in the fundamental by

$$ H = \cases{M_0 & on tensors, \cr
              M_0 {\rm +/-} \half & on a pair of spinors $s/s'$. \cr }
\eqno(C.7) $$

\noindent Conditions 1-3 are satisfied since (C.7) implies ${\rm
ad}_H = {\rm ad}_{M_0}$ on singlets, ${\rm ad}_H = {\rm ad}_{M_0}
\pm (1$ or 0) on tensors and ${\rm ad}_H = {\rm ad}_{M_0} \pm \half$
on spinors.

\bigskip
\noindent 3) $B_l$ and $D_l$ algebras.

\noindent
The analysis here is similar to what has been done in 2), and
we can  therefore go through the proof quickly.

For the orthogonal algebras, the adjoint is got from the {\it
antisymmetric} product of the fundamental with itself and we
still have

$$ {\rm ad}_H = {\rm ad}_{M_0} + [D(j_1)+D(j_2)].
\eqno(C.8)$$

\noindent The antisymmetric product of a tensor (spinor) with itself
produces a triplet (singlet), so that with respect to the symplectic
algebras, the situation is reversed in the sense that the tensors
and the spinors have their roles interchanged: $H=M_0 \pm \half$ on
tensors, $H=M_0$ on spinors and $m_t \leq 2$ for any tensor
representation of spin $t$.

If as in 2), we look at the restriction $g_t$ of the orthogonal
metric to the spin $t$ tensors, we have $m_t=2$ on account of the
non-degeneracy of $g_t$. From this, we get at once that there can
be no solution for the $B_l$ algebras. Indeed, the fundamental being
odd-dimensional, at least one tensor representation must come on
its own.

On the $2(2t+1)$-dimensional subspace made up by the two spin $t$
tensors, $H-M_0$ and $g_t$ take the form

$$ H-M_0 = \pm \pmatrix{\half & 0 \cr 0 & -\half \cr}, \qquad
   g_s = \pmatrix{a & b \cr b^t & c \cr },  \eqno(C.9) $$

\noindent where $a$ and $c$ are now symmetric. Requiring that
$H-M_0$ be orthogonal, we again obtain $a=c=0$.

\medskip
Therefore, for the orthogonal algebras, we get the following
conclusions. There is no solution for the $B_l$ series if the
$sl(2)$ embedding is not integral. As to the $D_l$ series, the
$sl(2)$ embedding must be such that: (i) every tensor in the
fundamental of ${\cal G}$ has a multiplicity equal to 2, (ii) if
$(t,t')$ is such a pair of tensors, they must be the dual of each
other with respect to the orthogonal metric. In this case, $H$ is
given in the fundamental by

$$ H = \cases{M_0 {\rm +/-} \half & on a pair of tensors $t/t'$, \cr
              M_0 & on spinors. \cr }
\eqno(C.10) $$

\vskip 0.8truecm
Summarizing the analysis, the
$H$-compatible $sl(2)$ embeddings are the following ones:
\medskip \parindent 36pt
\item{$A_l$ :} any $sl(2)$ subalgebra,
\item{$B_l$ :} only the integral $sl(2)$'s,
\item{$C_l$ :} those for which each {\it spinor} occurs
in the fundamental of $C_l$ with a multiplicity 0 or 2, the
pairs of spinors being symplectically dual,
\item{$D_l$ :} those for which each {\it tensor} occurs
in the fundamental of $D_l$ with a multiplicity 0 or 2, the
pairs of tensors being orthogonally dual.
\medskip \parindent 20pt
\noindent The reader may wish to check that the above results are
consistent with the isomorphisms $B_2 \sim C_2$ and $A_3 \sim D_3$.

\vskip 0.8truecm
We now come to the second question alluded to at the beginning of
this appendix, namely the problem of $H$-compatible halvings.
      From the definition, an $sl(2)$ subalgebra allows for an
$H$-compatible halving if in addition to conditions 1-3 one also has
\medskip \parindent 45pt
\item{4.} ${\cal P}_{\half} + {\cal G}_{\geq 1} = {\cal G}_{\geq
1}^H$, $\ {\rm and}\ $
${\cal P}_{-\half} + {\cal G}_{\leq -1} = {\cal G}_{\leq
-1}^H$.
\medskip \parindent 20pt
\noindent In particular, this fourth condition implies ${\cal
G}_0^{M_0} \subset {\cal G}_0^H$. So we readily obtain that $H$ and
$M_0$ must satisfy

$$ {\rm ad}_H = {\rm ad}_{M_0}, \qquad {\rm on} \quad {\rm tensors,}
\eqno(C.11) $$

\noindent since we know, from the previous analysis, that ad$_H -
{\rm ad}_{M_0}$ is a constant in every representation (condition 2).
Therefore, we can simply look at those solutions of the first
problem which satisfy (C.11) and check if  condition 4 is fully
satisfied or not. We get that the $sl(2)$ embeddings
allowing for an $H$-compatible halving are as follows:
\medskip \parindent 36pt
\item{$A_l$ :} any $sl(2)$ subalgebra. There are only two
solutions for $H$ given by setting in (C.2b): $D(j)=(\lambda \pm
\epsilon(j))\cdot I_{m_j}$ with $\epsilon(j)=0/\half$ for a tensor/spinor,
\item{$B_l$ :} only the integral $sl(2)$'s with $H=M_0$,
\item{$C_l$ :} only the integral $sl(2)$'s,
\item{$D_l$ :} the integral $sl(2)$'s, and those for
which the fundamental of $D_l$ reduces into spinors and two
singlets, with $H$ given by (C.10).
\medskip \parindent 20pt

\vfill\eject

\centerline{\bf References}

\vskip 0.8truecm

\item{[1]}
A. N. Leznov and M. V. Saveliev, {\sl Lett. Math. Phys.} {\bf 3} (1979) 489;
 {\sl Commun. Math. Phys.} {\bf 74} (1980) 111.
\item{[2]}
A. V. Mikhailov, M. A. Olshanetsky and A. M. Perelomov,
{\sl Commun. Math. Phys.} {\bf 79)} (1981) 473.
\item{[3]}
A. N. Leznov and M. V. Saveliev,
{\sl Lett. Math. Phys.} {\bf 6} (1982) 505;
{\sl Commun. Math. Phys.} {\bf 83} (1983) 59;
{\sl J. Sov. Math.} {\bf 36} (1987) 699;
{\sl Acta Appl. Math.} {\bf 16} (1989) 1.
\item{[4]}
M. V. Saveliev,
{\sl Mod. Phys. Lett. A} {\bf 5} (1990) 2223.
\item{[5]}
V. Drinfeld and V. Sokolov, {\sl J. Sov. Math.} {\bf 30}
(1984) 1975.
\item{[6]}
D. Olive and N. Turok,
{\sl Nucl. Phys.} {\bf B257} (1985) 277;
\item{}
L. A. Ferreira and D. I. Olive,
{\sl Commun. Math. Phys.} {\bf 99}  (1985) 365.
\item{[7]}
B. Kostant,
{\sl Adv. Math.} {\bf 34} (1979) 195;
\item{}
A. M. Perelomov,
\lq\lq Integrable Systems of Classical Mechanics and Lie Algebras",
Birkh\"auser Verlag, Basel-Boston-Berlin, 1990.
\item{[8]}
P. Mansfield, {\sl Nucl. Phys.} {\bf B208} (1982) 277; {\bf B222} (1983) 419;
\item{}
T. Hollowood and P. Mansfield, {\sl Nucl. Phys.} {\bf B330} (1990)
720.
\item{[9]}
J.-L. Gervais and A. Neveu, {\sl Nucl. Phys.} {\bf B224} (1983)
329;
\item{}
E. Braaten, T. Curtright, G. Ghandour and C. Thorn ,
{\sl Phys. Lett.} {\bf 125B} (1983) 301.
\item{[10]}
A. Bilal and J.-L. Gervais, {\sl Phys. Lett.} {\bf 206B} (1988) 412;
{\sl Nucl. Phys.} {\bf B314} (1989) 646;
{\sl Nucl. Phys.} {\bf B318} (1989) 579.
\item{[11]}
O. Babelon, {\sl Phys. Lett.} {\bf 215B} (1988) 523.
\item{[12]}
P. Forg\'acs, A. Wipf, J. Balog, L. Feh\'er and L. O'Raifeartaigh,
{\sl Phys. Lett.} {\bf 227B} (1989) 214.
\item{[13]}
 J. Balog, L. Feh\'er, L. O'Raifeartaigh, P. Forg\'acs
and A. Wipf,
{\sl Ann. Phys.} (N. Y.)  {\bf 203} (1990) 76;
{\sl Phys. Lett.} {\bf 244B} (1990) 435.
\item{[14]}
L. O'Raifeartaigh and A. Wipf,
{\sl Phys. Lett.} {\bf 251B} (1990) 361.
\item{[15]}
L. O'Raifeartaigh, P. Ruelle, I. Tsutsui and A. Wipf,
{\it  W-Algebras for Generalized Toda Theories},
Dublin preprint DIAS-STP-91-03,
{\sl Commun. Math.
Phys.}, to appear.
\item{[16]}
F. A. Bais, T. Tjin and P. Van Driel, {\sl Nucl. Phys.} {\bf B357}
(1991) 632.
\item{[17]}
T. Tjin and P. Van Driel,
{\it Coupled WZNW-Toda models and Covariant KdV hierarchies},
Amsterdam preprint IFTA-91-04.
\item{[18]}
L. Feh\'er, L. O'Raifeartaigh, P. Ruelle, I. Tsutsui and A. Wipf,
{\it  Generalized Toda theories
and W-algebras associated with integral gradings},
Dublin preprint DIAS-STP-91-17, {\sl Ann. Phys.
}, to appear.
\item{[19]}
P. Mansfield and B. Spence, {\sl Nucl. Phys.} {\bf B362} (1991) 294.
\item{[20]}
A. B. Zamolodchikov, {\sl Theor. Math. Phys.} {\bf 65} (1986) 1205.
\item{[21]}
V. A. Fateev and S. L. Lukyanov,
{\sl Int. J. Mod. Phys. } {\bf A3} (1988) 507;
\item{}
S. L. Lukyanov and V. A. Fateev,
{\it Additional Symmetries and Exactly Soluble Models in Two
Dimensional Conformal Field Theory},
Kiev preprints ITF-88-74R, ITF-88-75R, ITF-88-76R.
\item{[22]}
K. Yamagishi, {\sl Phys. Lett.} {\bf 205B} (1988) 466;
\item{}
P. Mathieu, {\sl Phys. Lett.} {\bf 208B} (1988) 101;
\item{}
I. Bakas, {\sl Phys. Lett.} {\bf 213B} (1988) 313;
\item{}
D.-J. Smit, {\sl Commun. Math. Phys.} {\bf 128} (1990) 1.
\item{[23]}
B. Feigin and E. Frenkel,
{\sl Phys. Lett.} {\bf 246B} (1990) 75;
\item{}
J. M. Figueroa-O'Farrill, {\sl Nucl. Phys.} {\bf B343} (1990) 450;
\item{}
H. G. Kausch and G. M. T. Watts,
{\sl Nucl. Phys.} {\bf B354} (1991) 740;
\item{}
R. Blumenhagen, M. Flohr, A. Kliem, W. Nahm, A. Recknagel and
R. Varnhagen, {\sl Nucl. Phys.} {\bf B361} (1991) 255.
\item{[24]}
P. Di  Francesco, C. Itzykson and J.-B. Zuber,
{\it Classical W-Algebras},
preprint PUTP-1211 S.Ph.-T/90-149;
\item{}
V. A. Fateev and S. L. Lukyanov,
{\it Poisson-Lie Groups and Classical W-Algebras},
Paris preprint PAR-LPTHE 91-16.
\item{[25]}
G. Sotkov and M. Stanishkov,
{\sl Nucl. Phys. } {\bf B356} (1991) 439;
\item{}
A. Bilal, V. V. Fock and I. I. Kogan, {\sl Nucl. Phys.} {\bf B359}
(1991) 635.
\item{[26]}
M. Bershadsky, {\sl Commun. Math. Phys.} {\bf 139} (1991) 71.
\item{[27]}
A. M. Polyakov,
{\sl Int. J. Mod. Phys. } {\bf A5} (1990) 833.
\item{[28]}
I. Bakas and D. Depireux, {\sl Mod. Phys. Lett.} {\bf A6} (1991) 1561;
\item{}
P. Mathieu and W. Oevel, {\sl Mod. Phys. Lett.} {\bf A6} (1991) 2397;
\item{}
D. A. Depireux and P. Mathieu,
{\it On the classical $W_n^l$ algebras},
preprint LAVAL PHY-27/91.
\item{[29]}
L. J. Romans, {\sl Nucl. Phys.} {\bf B357} (1991) 549;
\item{}
J. Fuchs, {\sl Phys. Lett.} {\bf 262B} (1991) 249.
\item{[30]}
E. Witten,
{\sl Commun. Math. Phys.} {\bf 92} (1984) 483.
\item{[31]}
P. Goddard and D. Olive,
{\sl Int. J. Mod. Phys.} {\bf A1} (1986) 303.
\item{[32]}
N. Bourbaki, \lq\lq Groupes et Alg\`ebres de Lie", Hermann, Paris, 1975;
chap. 8.
\item{[33]}
N. Jacobson,
\lq\lq Lie Algebras",
Interscience Publishers, Wiley, New York - London, 1962.
\item{[34]}
V. I. Arnold,
\lq\lq Mathematical Methods of Classical Mechanics",
Springer, Berlin-Heidelberg-New York, 1978;
\item{}
V. Guillemin and S. Sternberg, \lq\lq Symplectic techniques in physics",
Cambridge University Press, 1984.
\item{[35]}
A.M. Polyakov and P.B. Wiegmann, {\sl Phys. Lett.} {\bf 131B} (1983)
121.
\item{[36]}
S. Helgason, \lq\lq Differential Geometry, Lie Groups and Symmetric
Spaces", Academic Press, New York, 1978.
\item{[37]}
L. Feh\'er, L. O'Raifeartaigh, P. Ruelle, I. Tsutsui and A. Wipf,
{\it  Polynomial and Primary Field Character of $W_n^l$-Algebras},
Dublin preprint DIAS-STP-91-42.
\item{[38]}
B. Kostant, {\sl Amer. J. Math.} {\bf 81} (1959) 973.
\item{[39]}
A.I. Malcev, Amer. Math. Soc., Transl. 33 (1950);
\item{}
E. B. Dynkin, {\sl Amer. Math. Soc. Transl.} {\bf 6 [2]} (1957) 111.
\item{[40]}
L.Palla, {\sl Nucl. Phys.} {\bf B341} (1990) 714.
\item{[41]}
P. Bowcock, {\sl Nucl. Phys.} {\bf B316} (1989) 80;
\item{}
see also, A.P. Balachandran, G. Marmo, B.-S. Skagerstam and A. Stern,
   \lq\lq {\it Gauge Symmetries and Fibre Bundles}", Springer-Verlag
   Lecture Notes in Physics 188, (Springer-Verlag, Berlin and
   Heidelberg, 1983).
\item{[42]}
L.D. Faddeev, {\sl Theor. Math. Phys.} {\bf 1} (1970) 1.
\item{[43]}
A. Alekseev and S. Shatashvili,
   {\sl Nucl. Phys.} {\bf B323} (1989) 719.
\item{[44]}
L. O'Raifeartaigh, P. Ruelle and  I. Tsutsui,
   {\sl Phys. Lett.} {\bf 258B} (1991) 359.
\item{[45]}
See, for example, T. Kugo and I. Ojima,
   {\sl Prog. Theor. Phys. Supplement} {\bf 66} (1979) 1.
\item{[46]}
M. Bershadsky and H. Ooguri,
   {\sl Commun. Math. Phys.} {\bf 126} (1989) 49;
\item{}
N. Hayashi,
   {\sl Mod. Phys. Lett.} {\bf A6} (1991) 885;
   {\it Conformal Integrable Field Theory from WZNW via Quantum
   Hamiltonian Reduction},
   Osaka preprint OU-HET 149.
\item{[47]}
H. Freudenthal and H. de Vries, \lq\lq Linear Lie Groups",
Academic Press, New York and London, 1969.
\item{[48]}
M. F. De Groot, T. J. Hollowood and J. L. Miramontes,
{\it Generalized Drinfeld-Sokolov Hierachies},
preprint IASSN-HEP-91/19,  PUTP-1251,
{\sl Commun. Math. Phys.}, to appear;
\item{}
H. J. Burroughs, M. F. De Groot, T. J. Hollowood
and J. L. Miramontes,
{\it Generalized Drinfeld-Sokolov Hierachies II:
The Hamiltonian Structures},
preprint PUTP-1263, IASSN-HEP-91/42;
{\it Generalized W-algebras and Integrable Hierarchies},
preprint PUTP-1285, IASSN-HEP-91/61.
\item{[49]}
B. Spence,
{\it W-algebra Symmetries of Generalized Drinfeld-Sokolov
Hierarchies}.
preprint IMPERIAL/TP/91-92/02.
\item{[50]}
P. Bowcock and G. M. T. Watts,
{\it On the classification of quantum W-algebras},
preprint EFI 91-63, DTP-91-63.

\vfill\eject
\end